\documentclass[12pt,a4paper]{article}
\usepackage{ifthen} % for conditional statements
\newboolean{pdflatex}
\setboolean{pdflatex}{true} % False for eps figures 

\newboolean{articletitles}
\setboolean{articletitles}{true} % False removes titles in references

\newboolean{uprightparticles}
\setboolean{uprightparticles}{false} %True for upright particle symbols

\usepackage{microtype}
\usepackage{lineno}  % for line numbering during review
\usepackage{xspace} % To avoid problems with missing or double spaces after
\usepackage{graphicx}  % to include figures (can also use other packages)
\usepackage{color}
\usepackage{colortbl}
\graphicspath{{./figs/}} % Make Latex search fig subdir for figures

\usepackage{amsmath} % Adds a large collection of math symbols
\usepackage{amssymb}
\usepackage{amsfonts}
\usepackage{upgreek} % Adds in support for greek letters in roman typeset

\newcommand*\patchAmsMathEnvironmentForLineno[1]{%
\expandafter\let\csname old#1\expandafter\endcsname\csname #1\endcsname
\expandafter\let\csname oldend#1\expandafter\endcsname\csname
end#1\endcsname
 \renewenvironment{#1}%
   {\linenomath\csname old#1\endcsname}%
   {\csname oldend#1\endcsname\endlinenomath}%
}
\newcommand*\patchBothAmsMathEnvironmentsForLineno[1]{%
  \patchAmsMathEnvironmentForLineno{#1}%
  \patchAmsMathEnvironmentForLineno{#1*}%
}
\AtBeginDocument{%
\patchBothAmsMathEnvironmentsForLineno{equation}%
\patchBothAmsMathEnvironmentsForLineno{align}%
\patchBothAmsMathEnvironmentsForLineno{flalign}%
\patchBothAmsMathEnvironmentsForLineno{alignat}%
\patchBothAmsMathEnvironmentsForLineno{gather}%
\patchBothAmsMathEnvironmentsForLineno{multline}%
}

\usepackage{hyperref}    % Hyperlinks in references
\usepackage[all]{hypcap} % Internal hyperlinks to floats.

\def\sPlot{\mbox{\em sPlot}}

\def\RSens {$R$\xspace}
\def\PhiSens {$\Phi$\xspace}

\def\lhcb {LHCb\xspace}
\def\ux85 {UX85\xspace}
\def\cern {CERN\xspace}
\def\lhc {LHC\xspace}
\def\sps {SPS\xspace}

\def\velo   {VELO\xspace}

\ifthenelse{\boolean{uprightparticles}}%
{

 \def\Pmu         {\ensuremath{\upmu}\xspace}

 \def\Ppsi        {\ensuremath{\uppsi}\xspace}

 \def\PDelta      {\ensuremath{\Delta}\xspace}                 
 \def\PXi      {\ensuremath{\Xi}\xspace}                 
 \def\PLambda      {\ensuremath{\Lambda}\xspace}                 
 \def\PSigma      {\ensuremath{\Sigma}\xspace}                 
 \def\POmega      {\ensuremath{\Omega}\xspace}                 
 \def\PUpsilon      {\ensuremath{\Upsilon}\xspace}                 
 
 \def\PB      {\ensuremath{\mathrm{B}}\xspace}                 
                  
 \def\PD      {\ensuremath{\mathrm{D}}\xspace}

 \def\PJ      {\ensuremath{\mathrm{J}}\xspace}                 
 \def\PK      {\ensuremath{\mathrm{K}}\xspace}

 \def\Pb      {\ensuremath{\mathrm{b}}\xspace}

 \def\Pi      {\ensuremath{\mathrm{i}}\xspace}

}
{

 \def\Pmu         {\ensuremath{\mu}\xspace}

 \def\Ppsi        {\ensuremath{\psi}\xspace}                 
                  
 \mathchardef\PDelta="7101
 \mathchardef\PXi="7104
 \mathchardef\PLambda="7103
 \mathchardef\PSigma="7106
 \mathchardef\POmega="710A
 \mathchardef\PUpsilon="7107
                  
 \def\PB      {\ensuremath{B}\xspace}                 
                  
 \def\PD      {\ensuremath{D}\xspace}

 \def\PJ      {\ensuremath{J}\xspace}                 
 \def\PK      {\ensuremath{K}\xspace}

 \def\Pb      {\ensuremath{b}\xspace}

 \def\Pi      {\ensuremath{i}\xspace}

}

\def\mumu       {\ensuremath{\Pmu^+\Pmu^-}\xspace}

\def\b     {\ensuremath{\Pb}\xspace}
\def\bbar  {\ensuremath{\overline \b}\xspace}
\def\bbbar {\ensuremath{\b\bbar}\xspace}

\def\kaon  {\ensuremath{\PK}\xspace}
  \def\Kbar  {\kern 0.2em\overline{\kern -0.2em \PK}{}\xspace}

\def\Kz    {\ensuremath{\kaon^0}\xspace}
\def\Kzb   {\ensuremath{\Kbar^0}\xspace}
\def\KzKzb {\ensuremath{\Kz \kern -0.16em \Kzb}\xspace}
\def\Kp    {\ensuremath{\kaon^+}\xspace}
\def\Km    {\ensuremath{\kaon^-}\xspace}

\def\KpKm  {\ensuremath{\Kp \kern -0.16em \Km}\xspace}

  \def\Dbar    {\kern 0.2em\overline{\kern -0.2em \PD}{}\xspace}
\def\D       {\ensuremath{\PD}\xspace}

\def\Dz      {\ensuremath{\D^0}\xspace}
\def\Dzb     {\ensuremath{\Dbar^0}\xspace}
\def\DzDzb   {\ensuremath{\Dz {\kern -0.16em \Dzb}}\xspace}
\def\Dp      {\ensuremath{\D^+}\xspace}
\def\Dm      {\ensuremath{\D^-}\xspace}

\def\DpDm    {\ensuremath{\Dp {\kern -0.16em \Dm}}\xspace}

\def\B       {\ensuremath{\PB}\xspace}
  \def\Bbar    {\kern 0.18em\overline{\kern -0.18em \PB}{}\xspace}

\def\Bs      {\ensuremath{\B^0_s}\xspace}
\def\Bsb     {\ensuremath{\Bbar^0_s}\xspace}

\def\jpsi     {\ensuremath{{\PJ\mskip -3mu/\mskip -2mu\Ppsi\mskip 2mu}}\xspace}

  \def\Y#1S{\ensuremath{\PUpsilon{(#1S)}}\xspace}% no space before {...}!

\newcommand{\decay}[2]{\ensuremath{#1\!\to #2}\xspace}         % {\Pa}{\Pb \Pc}

\def\to                 {\ensuremath{\rightarrow}\xspace}

\newcommand{\dm}{\ensuremath{\Delta m}\xspace}

\def\BsToJPsiPhi  {\decay{\Bs}{\jpsi\phi}\xspace}

\def\AT#1     {\ensuremath{A_T^{#1}}\xspace}           % 2

\def\C#1      {\ensuremath{\mathcal{C}_{#1}}\xspace}                       % 9
\def\Cp#1     {\ensuremath{\mathcal{C}_{#1}^{'}}\xspace}                    % 7
\def\Ceff#1   {\ensuremath{\mathcal{C}_{#1}^{\mathrm{(eff)}}}\xspace}        % 9  
\def\Cpeff#1  {\ensuremath{\mathcal{C}_{#1}^{'\mathrm{(eff)}}}\xspace}       % 7
\def\Ope#1    {\ensuremath{\mathcal{O}_{#1}}\xspace}                       % 2
\def\Opep#1   {\ensuremath{\mathcal{O}_{#1}^{'}}\xspace}                    % 7

\newcommand{\unit}[1]{\ensuremath{\rm\,#1}\xspace}          % {kg}

\newcommand{\tev}{\ensuremath{\mathrm{\,Te\kern -0.1em V}}\xspace}
\newcommand{\gev}{\ensuremath{\mathrm{\,Ge\kern -0.1em V}}\xspace}
\newcommand{\mev}{\ensuremath{\mathrm{\,Me\kern -0.1em V}}\xspace}
\newcommand{\kev}{\ensuremath{\mathrm{\,ke\kern -0.1em V}}\xspace}
\newcommand{\ev}{\ensuremath{\mathrm{\,e\kern -0.1em V}}\xspace}
\newcommand{\tevc}{\ensuremath{{\mathrm{\,Te\kern -0.1em V\!/}c}}\xspace}
\newcommand{\gevc}{\ensuremath{{\mathrm{\,Ge\kern -0.1em V\!/}c}}\xspace}
\newcommand{\mevc}{\ensuremath{{\mathrm{\,Me\kern -0.1em V\!/}c}}\xspace}
\newcommand{\gevcc}{\ensuremath{{\mathrm{\,Ge\kern -0.1em V\!/}c^2}}\xspace}
\newcommand{\gevgevcccc}{\ensuremath{{\mathrm{\,Ge\kern -0.1em V^2\!/}c^4}}\xspace}
\newcommand{\mevcc}{\ensuremath{{\mathrm{\,Me\kern -0.1em V\!/}c^2}}\xspace}

\def\m    {\ensuremath{\rm \,m}\xspace}
\def\cm   {\ensuremath{\rm \,cm}\xspace}

\def\mm   {\ensuremath{\rm \,mm}\xspace}

\def\mum  {\ensuremath{\,\upmu\rm m}\xspace}

\def\invpb {\ensuremath{\mbox{\,pb}^{-1}}\xspace}

\def\invfb   {\ensuremath{\mbox{\,fb}^{-1}}\xspace}

\def\mus  {\ensuremath{\,\upmu{\rm s}}\xspace}
\def\ns   {\ensuremath{{\rm \,ns}}\xspace}

\def\fs   {\ensuremath{\rm \,fs}\xspace}

\def\ghz  {\ensuremath{{\rm \,GHz}}\xspace}
\def\mhz  {\ensuremath{{\rm \,MHz}}\xspace}

\def\hz   {\ensuremath{{\rm \,Hz}}\xspace}

\def\invps{\ensuremath{{\rm \,ps^{-1}}}\xspace}

\def\Xrad {\ensuremath{X_0}\xspace}

\def\neutroneq {\ensuremath{\rm \,n_{eq}}\xspace}

\def\gsim{{~\raise.15em\hbox{$>$}\kern-.85em
          \lower.35em\hbox{$\sim$}~}\xspace}
\def\lsim{{~\raise.15em\hbox{$<$}\kern-.85em
          \lower.35em\hbox{$\sim$}~}\xspace}

\def\pt         {\mbox{$p_T$}\xspace}
\def\invpt      {\mbox{$1/p_T$}\xspace}

\def\degrees{\ensuremath{^{\circ}}\xspace}

\def\murad  {\ensuremath{\,\upmu\rm rad}\xspace}
\def\mrad{\ensuremath{\rm \,mrad}\xspace}

\def\degreesC{\ensuremath{^{\circ}}C\xspace}

\def\mbar {\ensuremath{\rm \,mbar}\xspace}

\def\geantfour      {\mbox{\textsc{Geant4}}\xspace}

\def\nonn {\ensuremath{\rm {\it{n^+}}\mbox{-}on\mbox{-}{\it{n}}}\xspace}

\def\nonp {\ensuremath{\rm {\it{n^+}}\mbox{-}on\mbox{-}{\it{p}}}\xspace}

\def\tell1  {TELL1\xspace}
\def\ukl1   {UKL1\xspace}
\def\beetle {Beetle\xspace}

\newcommand{\eg}{\mbox{\itshape e.g.}\xspace}
\newcommand{\ie}{\mbox{\itshape i.e.}}

\usepackage{cite} % Allows for ranges in citations
\usepackage{mciteplus}

\begin{document}

%%%%%%%%%%%%%%%%%%%%%%%%%
%%%%% Title     %%%%%%%%%
%%%%%%%%%%%%%%%%%%%%%%%%%
\renewcommand{\thefootnote}{\fnsymbol{footnote}}
\setcounter{footnote}{1}

% %%%%%%% CHOOSE TITLE PAGE--------
%\onecolumn
% \input{title-LHCb-ANA}
% \input{title-LHCb-CONF}
% $Id: title-LHCb-PAPER.tex 54939 2014-05-29 11:19:01Z parkesb $
% ===============================================================================
% Purpose: LHCb-PAPER journal paper title page template
% Author: 
% Created on: 2010-09-25
% ===============================================================================

%%%%%%%%%%%%%%%%%%%%%%%%%
%%%%%  TITLE PAGE  %%%%%%
%%%%%%%%%%%%%%%%%%%%%%%%%
\begin{titlepage}
\pagenumbering{roman}

% Header ---------------------------------------------------
\vspace*{-1.5cm}
\centerline{\large EUROPEAN ORGANIZATION FOR NUCLEAR RESEARCH (CERN)}
\vspace*{1.5cm}
\hspace*{-0.5cm}
\begin{tabular*}{\linewidth}{lc@{\extracolsep{\fill}}r}
\ifthenelse{\boolean{pdflatex}}% Logo format choice
{\vspace*{-2.7cm}\mbox{\!\!\!\includegraphics[width=.14\textwidth]{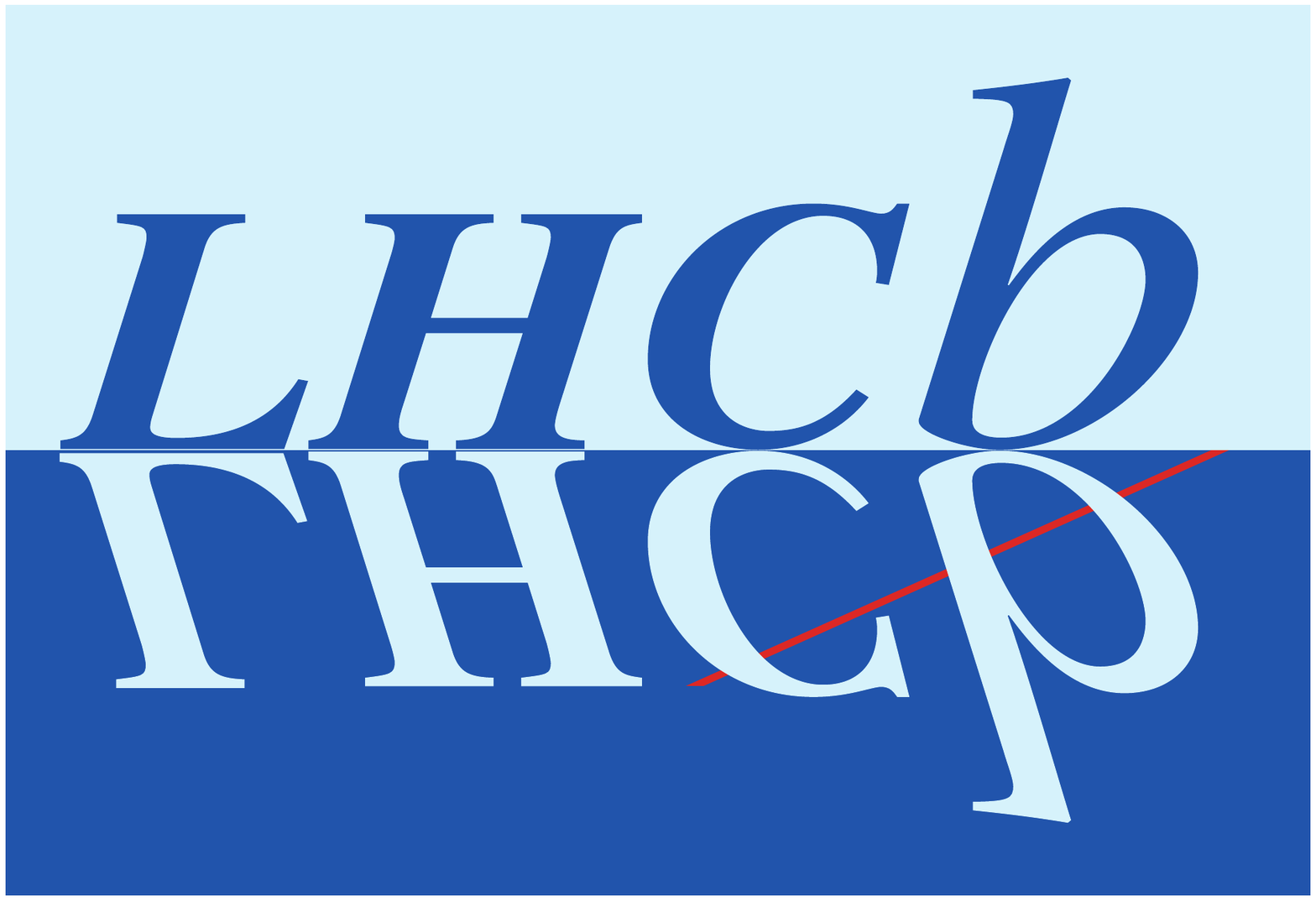}} & &}
{\vspace*{-1.2cm}\mbox{\!\!\!\includegraphics[width=.12\textwidth]{figs/lhcb-logo.eps}} & &}%
\\
 & & CERN-LHCb-DP-2014-001 \\  % ID 
 & & \today \\ % Date - Can also hardwire e.g.: 23 March 2010
% & & Draft 7 \\
% not in paper \hline
\end{tabular*}

\vspace*{1.0cm}

% Title --------------------------------------------------
{\bf\boldmath\huge
\begin{center}
Performance of the \lhcb Vertex Locator  
\end{center}
}

\vspace*{2.0cm}

% Authors -------------------------------------------------
\begin{center}
\lhcb \velo Group\footnote{Authors are listed on the following page.}
\end{center}

\vspace{\fill}

% Abstract -----------------------------------------------
\begin{abstract}
  \noindent
%[Chris to provide 0.3 page]
The Vertex Locator (\velo) is a silicon microstrip detector that surrounds the proton-proton interaction region in the \lhcb experiment. The performance of the detector during
% the collection of 1.2~\invfb of data in 
the first years of its physics operation is reviewed. The system is
operated in vacuum, uses a bi-phase $\rm CO_2$ cooling system, and the
sensors are moved to 7\mm from the LHC beam for physics data taking. The performance and stability of these characteristic features of the detector are described, and details of the material budget are given. The calibration of the timing and the data processing algorithms that are implemented in FPGAs are described.
%, and the data quality monitoring reviewed.  
The system performance is fully characterised. The sensors have a
signal to noise ratio of approximately 20 and a best hit resolution of
4\mum is achieved at the optimal track angle. The typical detector
occupancy for minimum bias events in standard operating conditions in
2011 is around 0.5\%, and the detector has less than 1\% of faulty
strips. The proximity of the detector to the beam means that the inner
regions of the \nonn sensors have undergone space-charge sign
inversion due to radiation damage. The \velo performance parameters
that drive the experiment's physics sensitivity are also given. The
track finding efficiency of the \velo is typically above 98\% and the
modules have been aligned to a precision of 1\mum for translations in
the plane transverse to the beam. A primary vertex resolution of 13\mum in the
transverse plane and 71\mum along the beam axis is achieved for
vertices with 25 tracks. An impact parameter resolution of
less than 35\mum is achieved for particles with transverse momentum
greater than 1\gevc. 
%The consequent decay time resolution for the \BsToJPsiPhi{} decay channel is 50~\fs.

%---------------------------------

%Length approx. 40 pages: max total length of sections - sub-system: 8 pages; calibration 10 pages; overall 13 pages; physics 12 pages.

%Provide a reference on system performance. Emphasis on use for LHCb physics papers, and for construction/operation of future Si detectors.

%Not intended as a system description or as internal documentation for velo group use. 

%Please use, and add to, the latex definitions in lhcb-symbols-def.tex

%Please recommit (as long as it compiles) to svn after each editing session to prevent conflicts.

\end{abstract}

\begin{center}
To be submitted to Journal of Instrumentation
\end{center}
\vspace{\fill}

\end{titlepage}

%%%%%%%%%%%%%%%%%%%%%%%%%%%%%%%%
%%%%%  EOD OF TITLE PAGE  %%%%%%
%%%%%%%%%%%%%%%%%%%%%%%%%%%%%%%%

%  empty page follows the title page ----
%\newpage
%\setcounter{page}{2}
%\mbox{~}
%\newpage

% Author List ----------------------------
%  You need to get a new author list!
\begin{center}
\textbf{\large LHCb VELO group}
\end{center}
\begin{flushleft}
%-- Velo Author List - veloPerfPaperAuthors2014
R.~Aaij$^{1}$,
A.~Affolder$^{2}$,
K.~Akiba$^{3}$,
M.~Alexander$^{4}$,
S.~Ali$^{1}$,
R.B.~Appleby$^{5}$,
M.~Artuso$^{6}$,
A.~Bates$^{4}$,
A.~Bay$^{7}$,
O.~Behrendt$^{8}$
% ...

\bigskip

\footnotesize

\it{
$^{1}$Nikhef National Institute for Subatomic Physics, Amsterdam, Netherlands
\newline$^{2}$Oliver Lodge Laboratory, University of Liverpool, Liverpool, United Kingdom
\newline$^{3}$Universidade Federal do Rio de Janeiro (UFRJ), Rio de Janeiro, Brazil
\newline$^{4}$School of Physics and Astronomy, University of Glasgow, Glasgow, United Kingdom
\newline$^{5}$School of Physics and Astronomy, University of Manchester, Manchester, United Kingdom
\newline$^{6}$Syracuse University, Syracuse, NY, United States
\newline$^{7}$Ecole Polytechnique F\'{e}d\'{e}rale de Lausanne (EPFL), Lausanne, Switzerland
\newline$^{8}$European Organization for Nuclear Research (CERN), Geneva, Switzerland
\newline$^{9}$H.H. Wills Physics Laboratory, University of Bristol, Bristol, United Kingdom
\newline$^{10}$Institute of Nuclear Physics, Moscow State University (SINP MSU), Moscow, Russia
\newline$^{11}$LAL, Universit\'{e} Paris-Sud, CNRS/IN2P3, Orsay, France
\newline$^{12}$Department of Physics, University of Warwick, Coventry, United Kingdom
\newline$^{13}$Deutsches Elektronen-Synchrotron, Hamburg, Germany
\newline$^{14}$Center for High Energy Physics, Tsinghua University, Beijing, China
\newline$^{15}$Department of Physics, University of Oslo, Oslo, Norway
\newline$^{16}$Physikalisches Institut, Ruprecht-Karls-Universit\"{a}t Heidelberg, Heidelberg, Germany
\newline$^{17}$Department of Physics, University of Oxford, Oxford, United Kingdom
\newline$^{18}$Vrije Universiteit Brussel, Brussel, Belgium
\newline$^{19}$Institute of Physics, Academy of Sciences of the Czech Republic, Prague, Czech Republic
\newline$^{20}$Universidade Federal do Rio de Janeiro, Campus UFRJ - Maca\'{e}, Rio de Janeiro, Brazil
\newline$^{21}$School of Physics, University College Dublin, Dublin, Ireland
\newline$^{22}$Universidad Aut\'{o}noma de Madrid, Spain
\newline$^{23}$School of Physics and Astronomy, University of Edinburgh, Edinburgh, United Kingdom
\newline$^{24}$AGH - University of Science and Technology, Faculty of Physics and Applied Computer Science, Krak\'{o}w, Poland
\newline$^{25}$School of Physics, University of Sydney, Sydney, Australia
\newline$^{26}$Universit\'{e}~de Lyon, Universit\'{e}~Claude Bernard Lyon 1, CNRS-IN2P3, Institut de Physique Nucl\'{e}aire de Lyon, Villeurbanne, France
\newline}
\end{flushleft}
%-- Number of authors: 103
%-- Number of institutes: 26

% for ArXiv submission
%LHCb VELO Group: R. Aaij, A. Affolder, K. Akiba, M. Alexander,
%S. Ali, R.B. Appleby, M. Artuso, A. Bates, A. Bay, O. Behrendt, J. Benton, M. van Beuzekom, P.M. Bj{\o}rnstad, G. Bogdanova, S. Borghi, A. Borgia, T.J.V. Bowcock, J. van den Brand, H. Brown, J. Buytaert, O. Callot, J. Carroll, G. Casse, P. Collins, S. De Capua, M. Doets, S. Donleavy, D. Dossett, R. Dumps, D. Eckstein, L. Eklund, C. Farinelli, S. Farry, M. Ferro-Luzzi, R. Frei, J. Garofoli, M. Gersabeck, T. Gershon, A. Gong, H. Gong, H. Gordon, G. Haefeli, J. Harrison, V. Heijne, K. Hennessy, W. Hulsbergen, T. Huse, D. Hutchcroft, A. Jaeger, P. Jalocha, E. Jans, M. John, J. Keaveney, T. Ketel, M. Korolev, M. Kraan, T. La\v{s}tovi\v{c}ka, G. Lafferty, T. Latham, G. Lefeuvre, A. Leflat, M. Liles, A. van Lysebetten, G. MacGregor, F. Marinho, R. McNulty, M. Merkin, D. Moran, R. Mountain, I. Mous, J. Mylroie-Smith, M. Needham, N. Nikitin, A. Noor, A. Oblakowska-Mucha, A. Papadelis, M. Pappagallo, C. Parkes, G.D. Patel, B. Rakotomiaramanana, S. Redford, M. Reid, K. Rinnert, E. Rodrigues, A.F. Saavedra, M. Schiller, O. Schneider, T. Shears, R. Silva Coutinho, N.A. Smith, T. Szumlak, C. Thomas, J. van Tilburg, M. Tobin, J. Velthuis, B. Verlaat, S. Viret, V. Volkov, C. Wallace, J. Wang, A. Webber, M. Whitehead, E. Zverev

\tableofcontents

\cleardoublepage

%\twocolumn
% %%%%%%%%%%%%% ---------

\renewcommand{\thefootnote}{\arabic{footnote}}
\setcounter{footnote}{0}

%%%%%%%%%%%%%%%%%%%%%%%%%%%%%%%%
%%%%%  Table of Content   %%%%%%
%%%%%%%%%%%%%%%%%%%%%%%%%%%%%%%%
%%%% Uncomment next 2 lines if desired
%\tableofcontents
%\cleardoublepage

%%%%%%%%%%%%%%%%%%%%%%%%%
%%%%% Main text %%%%%%%%%
%%%%%%%%%%%%%%%%%%%%%%%%%

\pagestyle{plain} % restore page numbers for the main text
\setcounter{page}{1}
\pagenumbering{arabic}

% %%%%%%% CHOOSE --------
%% ----------------------------------
%% Line numbering on the left margin 
%% ----------------------------------
%% Uncomment during review phase. 
%% Comment it out before a final submission.
%\linenumbers
%% --------------------------------
% %%%%%%%%%%%%% ---------

% You can include short sections directly in the main tex file.
% However, for larger papers it is desirable to split the text into
% several semiautonomous files, which can be revised independently.
% This is especially useful when developing a document in
% collaboration with several people, since then different parts can be
% edited independently.  This type of file organization is shown here.
% 

% $Id: introduction.tex 54939 2014-05-29 11:19:01Z parkesb $
% ===============================================================================
% Purpose: introduction to the standard template
% Author: Tomasz Skwarnicki
% Created on: 2010-09-24
% ===============================================================================

\section{Introduction}
\label{sec:Introduction}
%[Chris to provide 1 page]

%[brief description of system, references to LHCb JINST paper and TDR.
%motivation phrased in terms of physics aims.]

\lhcb \cite{Alves:2008zz} is an experiment dedicated to heavy flavour physics at the
\lhc. Its primary aim is to discover new physics through precision
studies of CP violation and rare decays of beauty and charm hadrons. The
detector is comprised of the Vertex Locator (\velo), silicon strip and straw chamber
trackers,  a warm dipole magnet, ring imaging Cherenkov particle
identification systems, calorimeters and a muon detection system.

The \lhcb Vertex Locator \cite{LHCbVELO_TDR:2001hf} is a
silicon microstrip detector positioned around the proton-proton interaction region.
The \velo provides measurements of track coordinates which are used to
identify the primary interaction vertices and the secondary vertices that are a distinctive feature of $b$-
and $c$-hadron decays. The VELO was designed to optimise the LHCb physics programme in the following ways:

\begin{itemize}

\item {\bf{Angular coverage}}. 
The \velo is designed to cover the forward region, such that all
tracks inside the nominal \lhcb acceptance of 15--300\mrad cross at
least three \velo stations.  In this way the detector fully
reconstructs roughly $27\%$ of \bbbar production for 7\tev
proton-proton centre-of-mass collisions, while covering just
$1.8\%$ of the solid angle \cite{Aaij:2010gn,Sjostrand:2006za}. 
The VELO also reconstructs tracks in the forward direction and backward directions which do not have momentum information, but are
useful to improve the primary vertex reconstruction.

%The detector is designed to
%  cover the forward region, and has measured the \bbbar cross-section to
%  be  $(75.3 \pm 5.4 \pm 13.0)$ \mub at 7\tev centre-of-mass energy \cite{Aaij:2010gn} in the range $2<\eta
%  <6$ .  This represents $27\%$ of the total cross-section (obtained from
%  Pythia \cite{Sjostrand:2006za}), while only covering $1.8\%$ of the solid
%  angle.  All tracks inside the nominal \lhcb acceptance of 15-300 \mrad
%  cross at least three \velo stations.

%cone of half-angle theta has solid angle 2pi(1-cos (theta))
% eta 2 = 15.4 deg, 270mrad, eta 6= 0.28 deg, 5 mrad
% solid angle = 0.2256 - 7.5e-5

\item {\bf{Triggering}}. The reconstruction of the primary vertex and
 the displaced secondary decay vertex of a heavy flavour hadron in the
 \velo is a key ingredient of the high level trigger which reduces
 the event rate from a 1\mhz event rate to a few kHz~\cite{Aaij:1493820}.

\item {\bf{Efficient reconstruction}}. \lhcb has studied
  decay modes with up to six charged tracks in the final
  state~\cite{Aaij:2011rj}. This relies on the highly efficient cluster reconstruction
  in the \velo  since track reconstruction efficiency
  losses are transmitted as the sixth power. The cluster
  reconstruction efficiency in the \velo is paramount, both for the
  selection of those tracks, as six measurements per track are required,
  and for efficient pattern recognition and fake track rejection.

%Measurements of the Cabibbo-favored branching fractions for B(s) → D(s)πππ and
%Λ 0b → Λ +c π π π

\item {\bf{Displaced tracks and vertices}}.  Excellent vertex resolution
  is essential to the LHCb physics programme. Most  analyses rely
  heavily on selection cuts on the distance with which tracks approach
  the primary vertex (impact parameter) and on the displaced vertex
  reconstruction with the \velo to  identify the signal channels.
 The impact parameter resolution was optimised by positioning the \velo
sensors as close to the LHC beam as permitted by safety consideration,
having a small inter-strip pitch at the inside of the sensors, and minimising the amount of material traversed by a particle
before the first measured hits in the \velo.

%\item {\bf{Precision reconstruction}}. 
%The \velo intrinsic cluster position precision and the alignment
%quality of the \velo contribute to the impact parameter resolution 
%and angular precision of the reconstructed tracks. These properties of
%the \velo 

\item {\bf{Decay time}}. The decay time of a particle is obtained from the measurement of its flight distance in the \velo.
This is required for lifetime measurements and,
critically,  for time-dependent measurements in the rapidly
oscillating \Bs--\Bsb meson system \cite{Aaij:1543533, LHCb:2011aa}. 

\end{itemize}

\begin{figure}
\begin{center}
    \resizebox{1.0\textwidth}{!}{
	\includegraphics[width=0.51\textwidth]{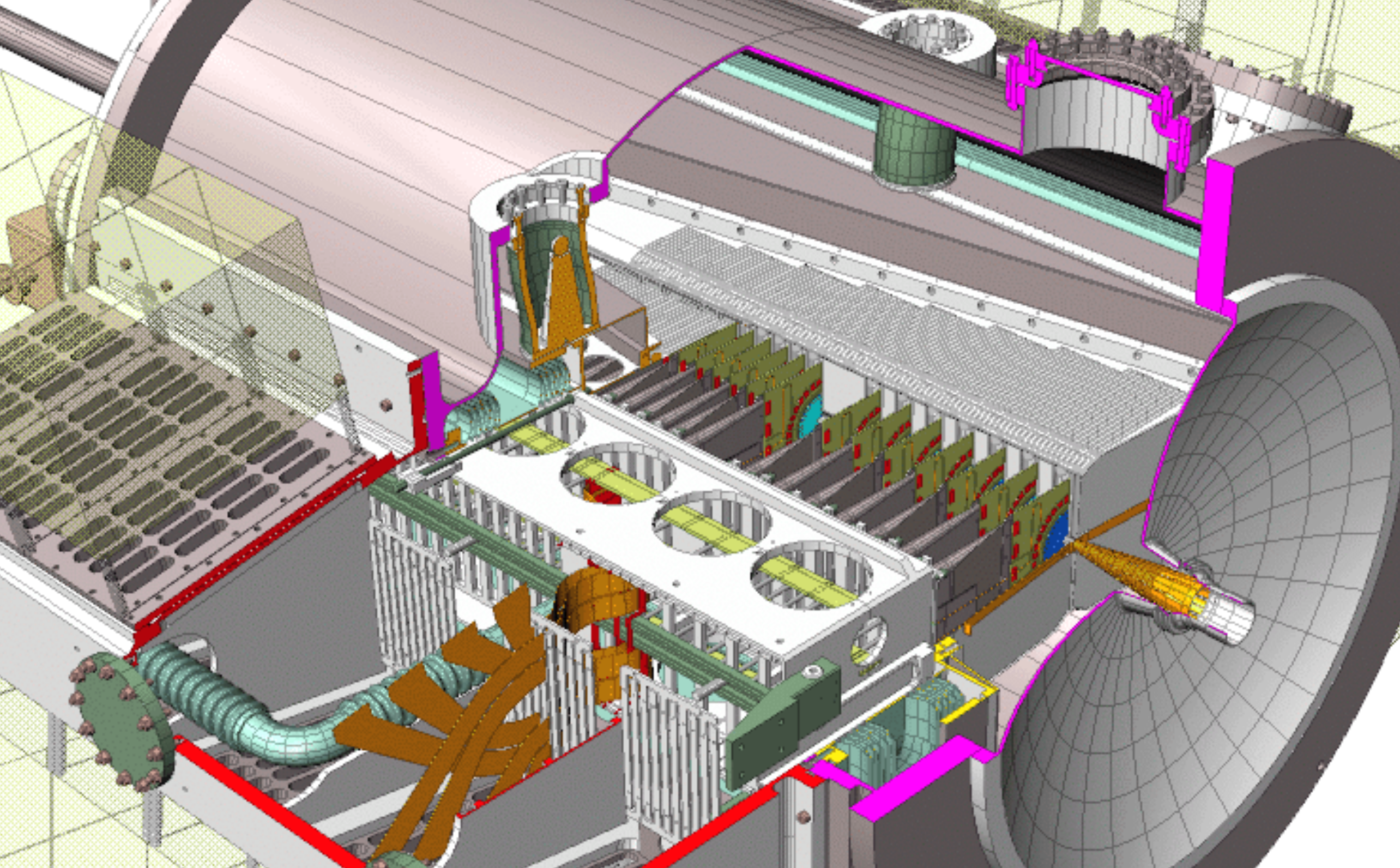}
	\includegraphics[width=0.47\textwidth]{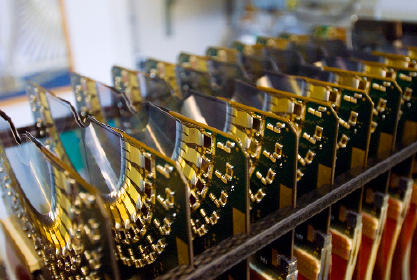}
    }
 \includegraphics[width=1.0\textwidth]{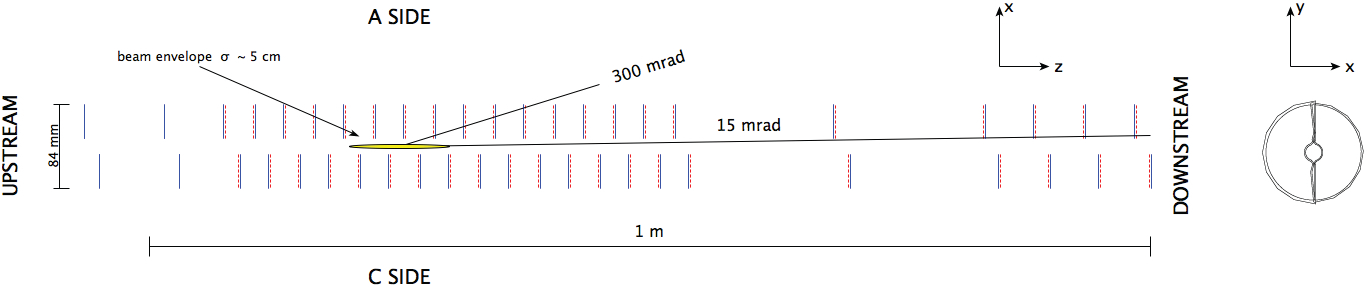}

	\caption{ {\small (top left) The LHCb \velo vacuum tank. The
            cut-away view allows the \velo sensors, hybrids and module
            support on the left-hand side to be seen. (top right) A
            photograph of one side of the \velo during assembly
            showing the silicon sensors and readout hybrids. (bottom)
            Cross-section in the $xz$ plane at $y=0$ of the sensors
            and a view of the sensors in the $xy$ plane. The detector
            is shown in its closed position. \RSens (\PhiSens) sensors are shown with solid blue (dashed red) lines. The modules at positive (negative)
            $x$ are known as the left or A-side (right or C-side). }}
          \label{fig:veloLayout}
\end{center}
\end{figure}

The \velo contains a series of silicon modules
%, each providing a
%measure of radial ($r$) and azimuthal ($\phi$) coordinates, 
arranged along the beam
direction, see Fig.~\ref{fig:veloLayout}.  A right-handed co-ordinate
system is defined with $z$ along the beam-axis into the detector, $y$
vertical and $x$ horizontal. Cylindrical polar co-ordinates
($r,\theta,\phi$) are also used. The region of the detector at
positive (negative) $z$ values is known as the forward (backward) or downstream (upstream) end. 

The sensors are positioned only 7\mm from the \lhc beams. This is smaller
than the aperture required by the \lhc beam during injection. 
Hence, the detector is produced in two retractable halves. 
There is a small overlap between the two detector halves when closed. This aids
alignment and ensures that full angular coverage is maintained. The
position of the \velo halves are moveable in $x$ and $y$ and the \velo is
closed at the beginning of each fill such that it is centred on the
interaction region.

\begin{figure}
\begin{center}
	\includegraphics[width=0.51\textwidth]{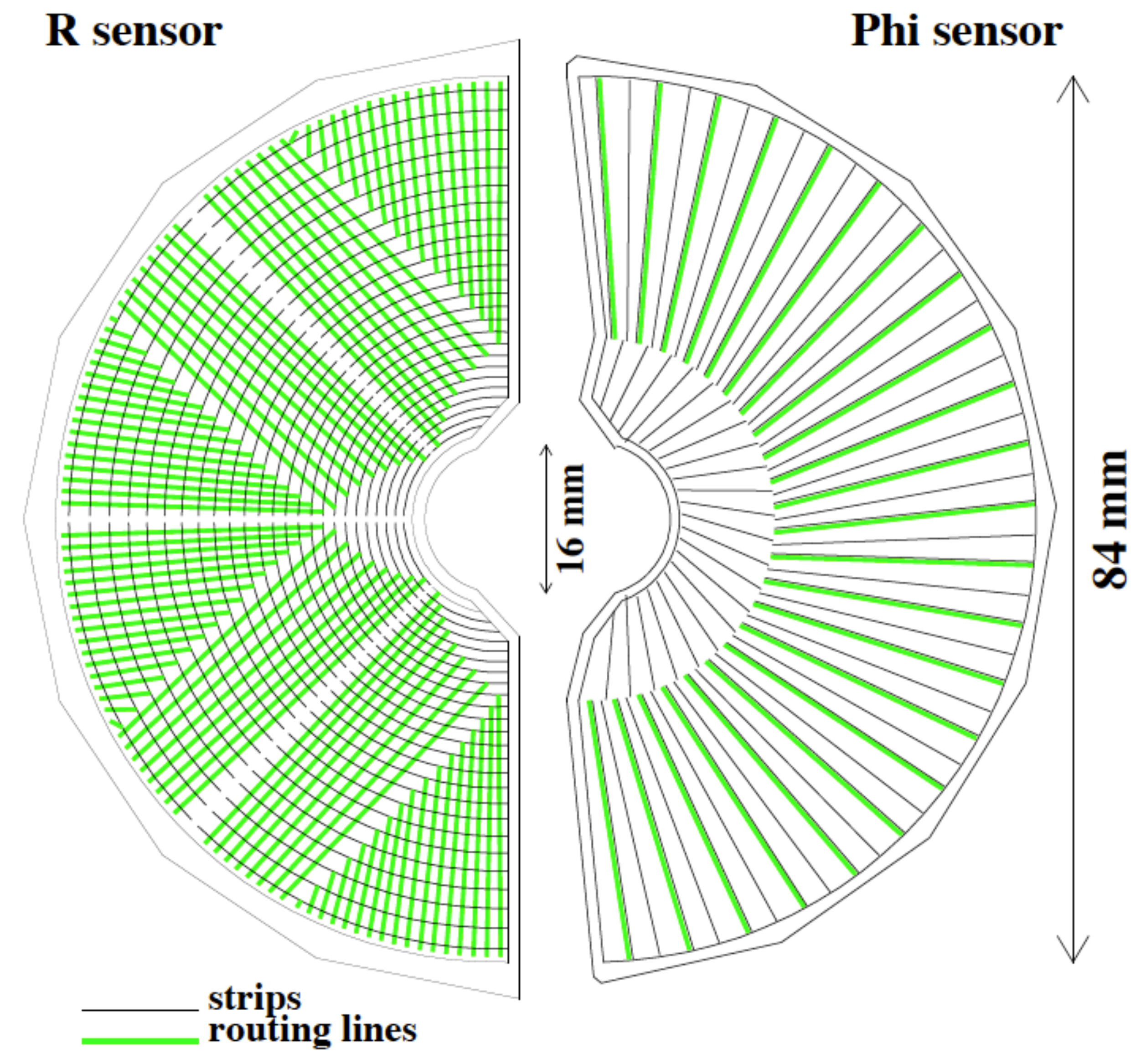}
	\caption{ {\small Schematic representation of an \RSens and a
            \PhiSens sensor. The \RSens sensor strips are arranged into four
            approximately  $45\degrees$ segments and have routing
            lines perpendicular to the strips. The $\Phi$ sensor has two
zones with inner and outer strips. The routing lines of the inner
strips are orientated parallel to the outer strips.}} \label{fig:veloSensors}          
\end{center}
\end{figure}

Approximately semi-circular silicon sensors are used. Each module
contains one $r$ and one $\phi$ coordinate measuring sensor, known as
\RSens and \PhiSens sensors and shown schematically in Fig.~\ref{fig:veloSensors}. The
inter-strip pitch varies from approximately 40 to 100\mum across the
sensor. The strips are read out from around the circumference of the
sensor through the use of routing lines on the sensor. 
The sensors are read out  using the \beetle \cite{Agari:2004ja} analogue front-end ASIC,
operated with a 40\mhz input event sampling rate.  The signals are digitised and processed 
to form clusters in an FPGA-based readout board known as the \tell1 \cite{Haefeli:2006cv},
before being passed to the high level trigger. More details on the
readout chain are given in Sect.~\ref{sec:DAQ}

There are 21 standard modules in
each \velo half. Two further modules, known as the pile-up system,
containing \RSens sensors only are
located in the most upstream positions.

Owing to the proximity of the detector to the beam, the \velo is exposed to a high radiation fluence and radiation
tolerant oxygenated \nonn sensors, consisting of an $n$-type implant
on an $n$-type bulk with a backplane $p^{+}$-type implant, are
employed. One of the most upstream modules uses \nonp silicon (one
\RSens and one \PhiSens sensor pair). These are the only \nonp sensors
in operation at the \lhc  and were installed as this technology  is a leading candidate for use in the \lhc upgrades.  All sensors were fabricated by Micron Semiconductor.\footnote{Micron Semiconductor Ltd., 1 Royal Buildings,  Marlborough Road, Lancing Business Park, Lancing, Sussex,  BN15 8SJ, UK.}

The detectors are mounted in a vacuum vessel and are located
in a secondary vacuum separated from the \lhc machine vacuum by an RF-box. The
surfaces facing the beam are 0.3\mm thick corrugated sheets, known as
the RF foil.  These RF-boxes and foil provide three functions: they provide shielding against RF pickup from the LHC
beams, guide wakefields to prevent impedance disruptions to
the LHC beams, and protect the \lhc vacuum from outgassing of the detector
modules.  The detector is located upstream of the \lhcb dipole magnet in a
region with a negligible magnetic field.

This paper reports on the performance of the \velo detector over the
first period of \lhc physics operation. The first proton-proton collisions occurred in November 2009
at beam energies of 450\gev, with the first 3.5\tev beam collisions occurring in March 2010.
The beam energy was raised further to 4\tev in April 2012. All physics data
were recorded with a 50\ns minimum bunch spacing. \lhcb recorded
integrated luminosities of 0.04\invfb in 2010, 1.11\invfb in 2011 and
2.08\invfb in 2012. The first period of LHC operations ended in
February 2013, when the LHC entered a shutdown for an upgrade to
increase the beam energy. The \lhcb \velo performance results in this
paper are primarily given on 2011 data,  the first year in which the
instantaneous luminosity reached, and exceeded, the nominal design value. 
Section \ref{sec:Subsystem} describes the performance of the component subsystems in
the \velo. The calibration of the timing, gain and processing
parameters is described in Sect.~\ref{sec:Calibration}, along with the
detector monitoring strategy and the simulation and reconstruction software.
Sections \ref{sec:Overall} and \ref{sec:Physics} provide the system performance
results, with the former section providing detector performance related results
and the latter results on quantities more directly related to physics performance.

 % $Id: introduction.tex 4475 2011-04-05 10:27:28Z uegede $
% ===============================================================================
% Purpose: introduction to the standard template
% Author: Tomasz Skwarnicki
% Created on: 2010-09-24
% ===============================================================================

\section{Subsystem performance}
%[Themis to oversee, max. 8 pages]
\label{sec:Subsystem}

The performance of individual subsystems from the \velo are described in this section. The section starts with a description of the commissioning stages of the detector and then describes the performance of the vacuum, cooling, low and high voltage and motion systems. The section ends with information on the material budget of the \velo.

\subsection{Commissioning results}
\label{sec:Commissioning}
%[to discuss ? Paula, 1 page]

%Commissioning with TED data

The subsystems and overall operations of the \velo were extensively commissioned before the first LHC beam collisions.
After production, the commissioning of the \velo consisted of four main stages: testing during assembly, operation in a test beam, commissioning post-installation at the LHCb experimental area, and  testing with beam-absorber collisions.

The detectors underwent extensive quality assurance tests at the production sites and after delivery for system assembly, utilising components of the final readout system \cite{VeloBurnin}. During detector assembly  (Fig.~\ref{fig:veloLayout} shows a photograph of the system at this stage) each module was operated with a complete readout slice, simulating as closely as possible the final vacuum, cooling, and powering conditions of the \velo.  The main mechanical difference from the final system was the absence of the RF foil. This first testing step allowed a complete ``fingerprint'' of the pedestal and noise map of each module to be produced, along with the commissioning of the zero suppression algorithms running in the TELL1 DAQ boards (see Sect.~\ref{sec:TELL1}). 

In the second step, a partially assembled \velo half was taken to the 120\gev pion/muon beam at the North area SPS test beam facility at CERN.  Complete track and vertex reconstruction was performed, using the products of interactions of the beam in lead targets placed in similar positions to that of the LHC beam interaction region \cite{Papadelis:Thesis}.  This
allowed the data acquisition and software chain to be debugged, in addition to performing a full speed test of significant components of the LHCb readout chain.  It also served as a ``dress rehearsal" for the transportation of the \velo using a truck with special traction and speed control. The \velo half was mounted on four shock
absorbers on a trailer with special suspension, and the transport motion was fully logged using accelerometers.   

The third stage of the commissioning was carried out after the installation of the \velo at the \lhcb experimental area.  Due to the proximity of the  silicon sensors ($\leq 1$\mm)
to the 1\m long corrugated RF foil, and the insertion procedure of the detector halves into the RF foil relying purely on the manufacturing and assembly tolerances, it was necessary to employ special monitoring procedures. Each module slot was tested with an oversized dummy sensor model for contact with the foil, and a test insertion was
performed with a genuine module.  Characteristic IV curves of all sensors were taken at several occasions during  the transport, movement and installation procedures, and it was verified that no damage had occurred.  Following this, the cabling of the detector was verified by applying custom test-pulse patterns to each of the 5632 readout links, and verifying the correct
pattern was read back.  Each link was tuned to select the optimal ADC sampling point using test-pulses (see Sect.~\ref{sec:timing}). The digitisation uniformity
 was confirmed by injecting sine wave pulses with an amplitude close to the full dynamic range and a period matching the pulse train from the front-end ASICs.  In addition, the noise and pedestals were measured, and compared with the ``fingerprint'' recorded during the
assembly and commissioning process.   

In the fourth and final commissioning step, a special method was found to study genuine tracks in the detector before the start of LHC collisions.  The synchronisation tests of the LHC beam were utilised in which, during the initial phase of each test, the \lhc proton beam was collided with a beam absorber at the end of the transfer line between the \cern \sps and the \lhc about 340\m from the \lhcb cavern. This allowed the \velo to reconstruct the first tracks from the \lhc machine \cite{VeloFirstTracks}. The particles produced by the proton interactions in the absorber, and by their re-interaction, were detected by the \lhcb experiment and were used to commission the detector. These injection tests were performed in 2008 before the first circulating beam in the LHC and in 2009 before the first proton-proton collisions. These data were used to set up the timing and the alignment of the system, and for commissioning of the control, reconstruction and monitoring software. They were of particular importance to the \velo as it is not possible to collect sufficient cosmic ray data for commissioning due to the horizontal geometry, and hence provided the first full system test of the \velo.   With this method about 50k tracks were reconstructed in \lhcb.  This was sufficient to determine that the modules were displaced by less than 10\mum  from their surveyed positions (see Sect.~\ref{sec:Alignment}). The alignment of the modules was determined with 5\mum precision for $x$ and $y$ translation and 200\murad for the rotations around the $z$-axis \cite{VeloTedArticle}. In addition, the tracks traversing from one \velo half to
the other could be used, by evaluating the mismatch of the two segments, to measure the distance between the halves to a precision of 100\mum.  The time alignment of the \velo was determined to a precision of 2\ns, and measurements made of the signal to noise and cluster finding efficiency.   These commissioning stages allowed \lhcb to be ready for an immediately successful start to physics data taking after the first \lhc beams collided in November 2009.

\subsection{Vacuum stability}
\label{sec:Vacuum}

\begin{figure}
\begin{center}
    \resizebox{0.6\textwidth}{!}{
	\includegraphics[width=0.96\textwidth]{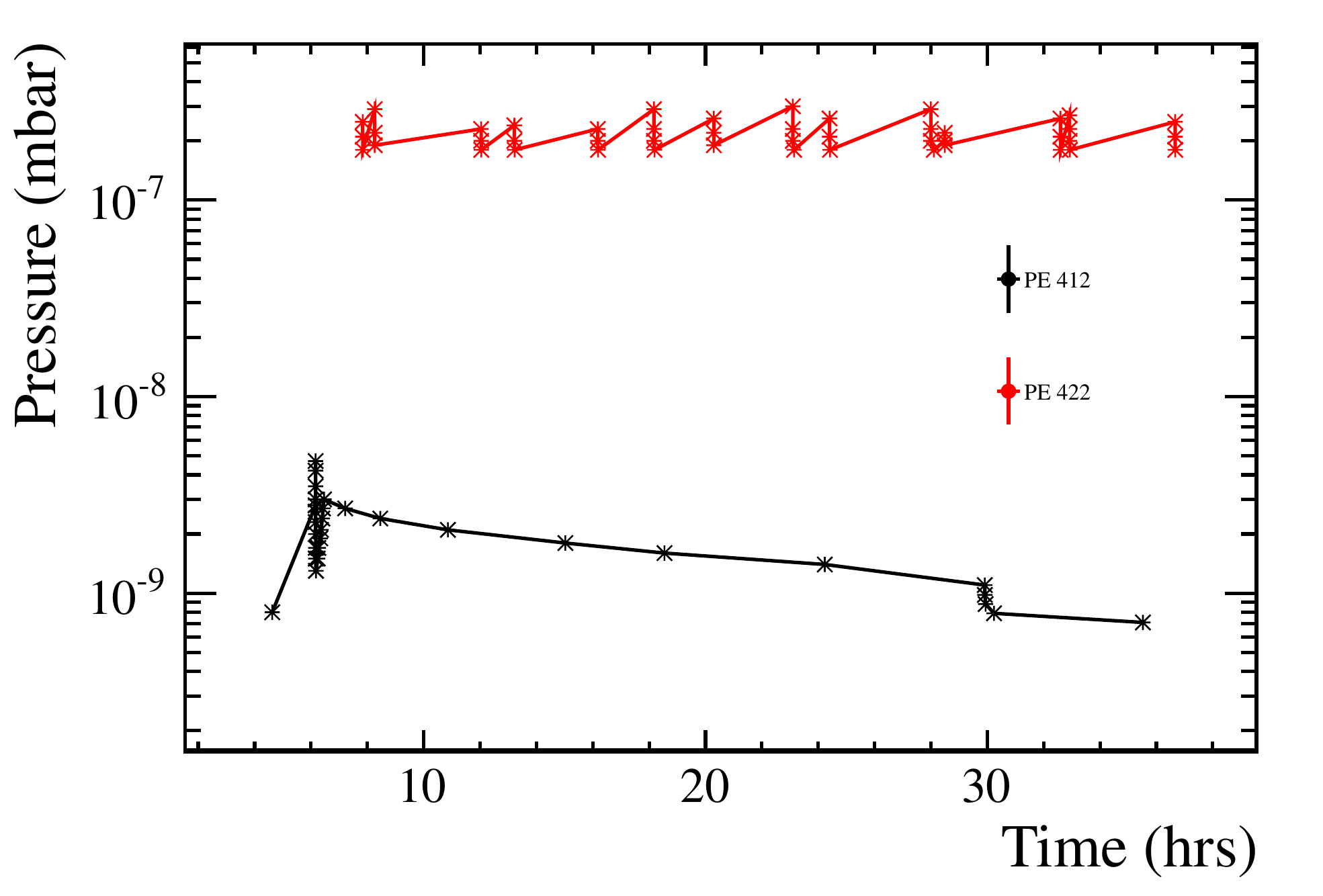}
    }
	\caption{{\small \label{fig:pressures} 
 Illustration of the influence of the proton beams on the beam volume pressure.  The readings of the beam volume (PE~412) and detector volume (PE~422) are shown, and the increase in pressure in the beam volume coinciding with the beam injections (each containing $2\times10^{14}$ protons) at $t = 6~\mathrm{hours}$ can be seen. The beams were dumped 24~hours later.  }} 
\end{center}
\end{figure}

The \velo vacuum vessel is composed of two sections: the first is part
of the beam volume of the \lhc, and the other contains the 
detector modules. The two sections are separated by a vacuum tight RF-box, welded from a
0.3\mm thick corrugated AlMg3 foil with 0.5\mm thick side walls
which encapsulates each detector half. 
The differential pressure
between the beam and detector volume should always remain below 5\mbar
to protect the foil against irreversible deformation.
Consequently, the pump-down and gas venting procedures are very delicate operations that are controlled by a programmable logic controller (PLC), based on the readings from three membrane switches that have preset trigger values.
The PLC continuously monitors the performance of the system and takes, when needed, appropriate actions.
A PVSS\footnote{PVSS is a Supervisory Control and Data Acquisition software package developed by ETM professional control GmbH.} project monitors and archives all relevant system variables in
a database, that is common to the whole \lhcb experiment~\cite{Clemencic:951971}.
For redundancy purposes, two sets of roughing, turbo-molecular and ion pumps are implemented in the system. 
The pressure in the detector volume is around $2\times10^{-7}$\mbar, 
while the beam volume is at $1\times10^{-9}$\mbar  in the absence of circulating beams.
Under the influence of 1380~bunches per beam this pressure can rise to $5\times10^{-9}$\mbar. 
The influence of the proton beams on the pressure of the beam volume
can clearly be seen in Fig.~\ref{fig:pressures}.

\subsection{Cooling performance}
\label{sec:Cooling}
The \velo modules are cooled by means of an evaporative system using CO$_{2}$ as coolant. 
The cooling plant, which is located outside the radiation zone, has two chillers filled with hydrofluorocarbon (R507a): a 2.5\,kW water-cooled main chiller and a 1\,kW air-cooled backup chiller, operating at $-40$ and $-25\degreesC$, respectively (see Ref.~\cite{vanBeuzekom:2007zz}).
The actual cooling system of the detector consists of two parallel, largely independent, loops: one for each detector half. 
The CO$_{2}$ is subcooled by the main chiller and subsequently pumped to the corresponding detector via a 55\m long transfer line.
The operation principle is that of a bi-phase accumulator controlled loop, in which the evaporative temperature in the detector is regulated by controlling the pressure of the saturated mixture of vapour and liquid in the accumulator. 
The cooling system is controlled by a PLC. 
In total 31 pressure and 192 temperature sensors are installed throughout the system. 
The PLC controls the system via 24 proportional, integrating and differentiating loops based on a subset of these readings.
Alarm handling tasks are executed continuously and if appropriate the system switches to the backup chiller or a spare pump. 
A PVSS project monitors and archives all relevant system variables in an Oracle database that is common to the whole \lhcb experiment.
A separate heating system takes care of maintaining the temperature of the module base at 20\degreesC to minimise deformations.

The default evaporator temperature is $-28$\degreesC.
Due to thermal gradients between the cooling blocks and the components of the hybrid the silicon temperatures are $(-7\pm2)$\degreesC.
From the moment the front-end electronics is switched on, it takes 3 minutes till the hybrid temperatures have stabilised.
The hybrid temperatures have been shown to be stable within 0.1\degreesC over a period of four weeks. 

The operation of the cooling system has been very smooth. The only major intervention required was the replacement of the insulation of the transfer lines in 2011 to eliminate the formation of ice. During operations in both 2009 and 2011 an increase of $\sim$1\degreesC was observed in one detector half and found to be due to clogging of filters: the filters were replaced during the following winter shutdowns. 

During injection and ramp of the beams, the temperatures of the RF-boxes increase by 0.5\degreesC. 
Subsequently the temperatures of the RF-boxes decrease by 1.5\degreesC when the \velo is closed. 

\subsection{Low voltage and high voltage}
\label{sec:Voltage}

Each silicon sensor has its own hybrid with separate low voltage (LV) and high voltage (HV) supplies. In addition, repeater boards are located directly outside the \velo vacuum tank, which have their own positive and negative supplies. Both the LV and HV power supplies are inside the counting-house shielded from the radiation zone of the detector.

The LV system is based on the A3009 EASY low voltage module, with compatible crates and mainframe controller, manufactured by CAEN.\footnote{CAEN S.p.A., Via Vetraia 11, 55049 - Viareggio (LU) - Italy.} 
%Twenty two modules are located in five crates. Three A3496 bulk power supplies provide the 48~V service and mains power to the crates, and the system is controlled by an SY1527 CAEN mainframe.  Each hybrid and repeater board are supplied with three independent voltages. Each power supply is floating, and has its own return cable and sense wires. 
The main system performance issue has been with the LV Anderson connectors used in the system. Due to oxidation a voltage drop across the connector can occur. This requires careful monitoring as the dissipated power caused severe thermal damage on one of the connectors during operation. The connectors were replaced by CAEN during the shutdown at the end of 2011.

The HV system is based on the EHQ F607n-F 16-channel HV module from Iseg.\footnote{Iseg Spezialektronik GmbH, Bautzner Landstr. 23, 01454 Radeberg / OT Rossendorf, Germany.} 
%Six HV modules are located in one crate ISEG ECH 238 UPS in the counting house. Every channel supplies up 700~V at 4 mA maximum output. Maximum output current and voltages are set for each module, and the system is connected to the interlock safety system.  
Careful monitoring of the HV parameters is a key requirement as the current and depletion voltage change with radiation damage (see Sect.~\ref{sec:Radiation}).  The main performance issue was with the control of the HV supplies. The supplies occasionally get stuck when ramping in voltage, requiring the software commands to be resent or control software restarted.

%The repeater board is primarily for the transmission of data signals, timing and fast control signals and front-end chip configurations signals. Monitoring signals, including those of temperatures, are also send out from the repeater board to the detector slow control system. The repeater boards carry the LV regulators required by the front-end electronics and the first level trigger electronic service system.

Voltage and current monitoring are implemented for both LV and HV systems and they are connected to an FPGA based interlock safety system.  All HV and LV modules can have their channels switched off quickly ($<1$~s)  by the interlock system in case of problems. The primary uses of the interlock system have been in the case of power cuts or disruptions, and interruptions to the chilled water supply that is used by the cooling system.

\subsection{Motion performance}
\label{sec:Motion}

The first active silicon strips are brought to within 8.2\mm of the LHC beams, with the inner surface of the RF foil at a 5.5\mm radius. To ensure detector safety during beam injection and adjustments, each \velo half is retracted $\sim29\mm$ in the horizontal plane and is only closed once stable-beam conditions are declared. The VELO halves are moved using radiation hard stepper motors, and the motion towards the final position is always done from the same direction in order to minimise the effects of mechanical backlash.
The position is read from resolvers mounted on the motor axes. The reproducibility of the position has been measured to be better than 10\mum. The motion system is controlled by a PLC which performs many safety checks to prevent unwanted movements due to hardware failures or corruption of the destination position calculated by the closing procedure.

%The two VELO halves can be moved individually in the horizontal direction, while the vertical movements is common for both halves. Radiation hard steppermotors and gear boxes drive the spindles which take care of the actual movement of the detectors. Moving towards the final position is always done from the same direction in order to minimize the effects of mechanical backlash.
%Readback of the position is obtained from resolvers, mounted on the motor axes, whose resolution is much better than the mechanical precision.
%During installation, the reproducibility of the position has been measured to be better than 10 $\mu$m. The motion system is controlled by a PLC which performs many safety checks to prevent unwanted movements due to hardware failures or corruption of the destination position calculated by the closing manager, which is described below.
%[Martin to add two sentences on motion system mechanics]

%[Martin]

%Each VELO half contains 44 sensors, their front-end readout, their cooling, the support RF guard and the necessary mechanical support structure~\cite{velotdr}.
%The halves a are mounted on...\\
%Motion is achieved using...\\
%The accuracy ...\\

\begin{figure}
\begin{center}
\includegraphics[width=0.475\textwidth]{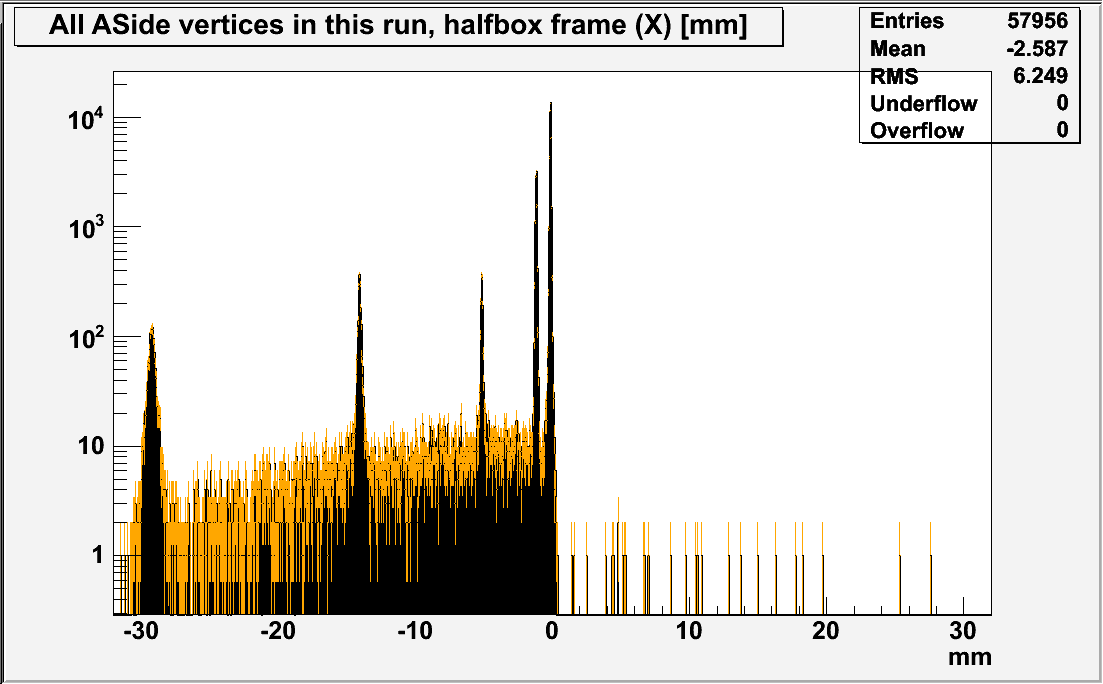}
$\ \  $
\includegraphics[width=0.475\textwidth]{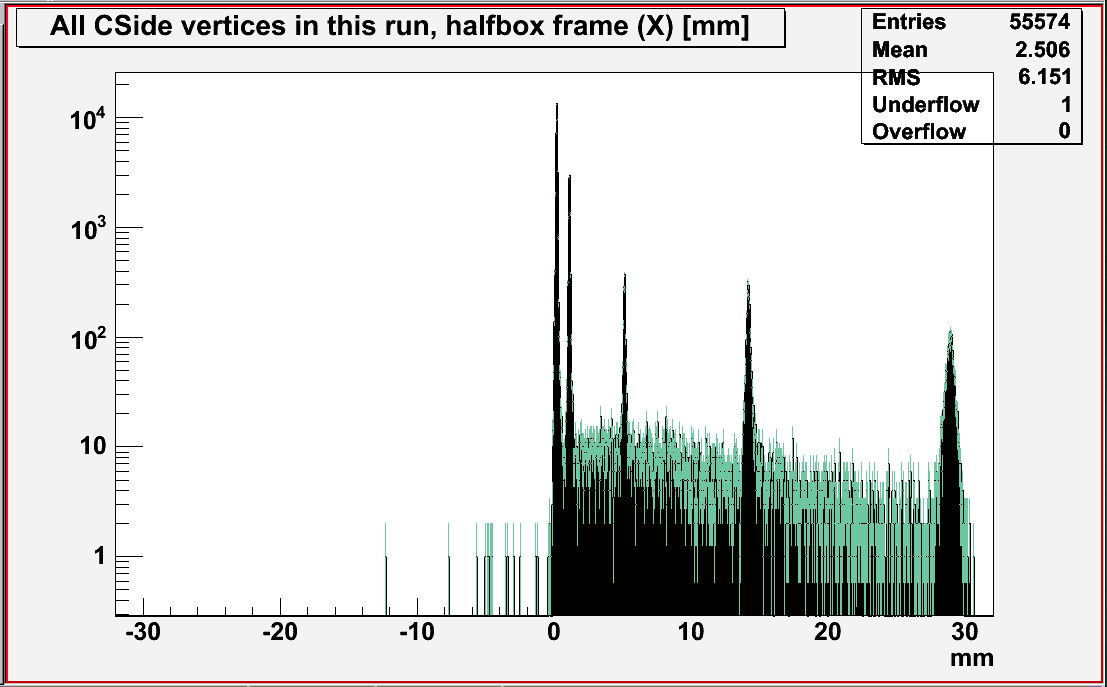}
\caption{{\small The horizontal position of \lhc collision vertices as reconstructed by the A-Side (left) and C-Side (right). 
These online screenshots of data illustrate the ``stop-measure-move'' cycle of the closing procedure, where each peak in the distribution corresponds to a ``stop'', as well as the degradation of vertex resolution with opening distance.
\label{fig:closing}}}
\end{center}
\end{figure}

An automated closure procedure has been developed to position the \velo halves around the beams, whilst taking into account the current beam conditions. It uses information about the background level, the response of the hardware and the positions of the beams to make informed decisions.
During closure, the \lhcb trigger accepts 500\hz of events which are required to contain at least one track in the \velo.
This rate ensures that a new beam-position measurement, consisting of 400 reconstructed vertices, can be acquired in 1--2\,s, even in the most open position. Figure~\ref{fig:closing} shows an example of the distribution of vertices during a typical closure.

Upon receiving notification of stable beams, the two \velo halves are moved in a series of ``stop-measure-move'' steps from open to closed. 
%The steps are: open, $\pm20~\mm$, $\pm10~\mm$, $\pm5~\mm$, $\pm1~\mm$ and closed. 
A vertical adjustment is also made at each of the steps, if the movement required to centre the \velo around the beams is greater than 20\mum.
Three criteria are checked at all times during and after closure:
\begin{enumerate}
\item the total silicon bias current from the 44 sensors in each half of the \velo is less than 1000\unit{\mu A} above the dark current. 
This limit corresponds to 10\% occupancy of ionising particles in all sensors for several seconds;
\item the Beam Conditions Monitors~\cite{Ilgner:2010vu} --- each consisting of eight CVD-diamond radiation monitors placed adjacent to the beam pipe --- are all functional and reporting fluences less than 5\% of the threshold which triggers an LHC beam dump;
\item the reconstructed beam position in $x$ and $y$, as measured by the two halves, agree given the known opening distance at any given step.
The observed beam width must be acceptable given the opening distance and the expected resolution thereof.
\end{enumerate}
By considering the two independent beam profiles compiled by each half, the \velo is observed to close symmetrically to an accuracy of better than 4\mum.

Once closed the monitoring continues throughout the data taking. 
Apart from requiring the above conditions, a slow drift of the beams is protected against by prompting the \velo to open if the beams move by more than 300\mum. The \lhc beam orbit has shown excellent stability. Consequently, the \velo is permitted to remain closed for up to 20 minutes if the reconstruction of the beam profiles is interrupted. This happens, for example, if the LHCb data acquisition is paused (e.g. for a run change or reconfiguration).  During this grace period, movement of the beam is monitored by the \lhc beam position monitors which are horizontal and vertical wire pickups located at $z=\pm22\m$ of the collision region.  A deviation of 200\mum in any of these readings will trigger the \velo to open.

The initial use of the system was performed with careful manual checking and control of the closings. The automated closing procedure was then put in place. 
From the declaration of stable beams the \velo takes, on average, 210\,s to close. Of this 160 seconds is due to the motion from the open to closed positions.  
%Around 25 seconds is spent on the beam reconstruction (2-5 seconds at each step) and a similar period is needed for the bias voltage to ramp. 
During the operations in 2010--2012 approximately 750 closing procedures were performed and 0.9\% of the stable-beam integrated luminosity delivered by the LHC was lost due to this detector-safety procedure. This number is in line with the expectation.  Only minor performance problems have been encountered and these have been addressed with changes to the closing procedure, safety checks and motion PLC code.

\subsection{Material description}
\label{sec:Material}
%[Tom - 2 pages]
The minimisation of the material budget of the detector is important for the physics performance of the experiment in order to reduce the amount of multiple scattering and particle interactions with material. An accurate description of the material is also required for the simulation of the experiment and for estimating the amount of multiple scattering that particles undergo when performing track reconstruction. The material description is implemented using the basic volumes in \geantfour~\cite{geant4} and there are limits on the accuracy that is achievable, both from the range of basic volumes (and combinations thereof)
that are available and from the CPU time taken to process very complex composite volumes.  The appropriate balance must therefore be obtained between
accuracy and simplicity for each element of the description. This is particularly an issue for the description of the complex RF foil shape (see Fig.~\ref{fig:mat-int-full} below).

%The LHCb detector description database (DDDB) holds the description of the
%detector for use in both the simulation and reconstruction software.
%A Document Type Definition is used to define a syntax for the xml files that
%build up the database.  This allows the description of each element of the
%detector structure in terms of combinations of basic volume elements and also
%provides the mechanisms to build the hierachy of elements into a complete
%description of the entire detector.
%When writing the detector description 

A comparison has been made between the measured masses of the various
elements of the detector, as determined at production, and their simulated
counterparts, which in general shows very good agreement, \eg the average mass of
the sensors is $(2.17\pm0.03)\unit{g}$, compared with the simulated mass of $2.14\unit{g}$.
Similarly the mass of the dominant component of the material, the RF foil, is
reproduced to within 2\%.
The RF foil was manufactured as a single sheet of aluminium magnesium alloy and then pressed
into its rather complex shape. So while the agreement of the total material may
be good, the distribution of that material might not be so well described.
Tests were performed by measuring the thickness of an RF foil and it was
found that the amount of material was too low in regions around the beam in the
description used up till 2012 and this has now been improved.
While this has only a small effect on the total material it does contribute
significantly to the material before the first measured point, which is
important for the impact parameter performance with lower momentum tracks
(see Sect.~\ref{sec:ImpactParameters}). 

\begin{figure}[htb]
\begin{center}
\includegraphics[width=0.49\textwidth]{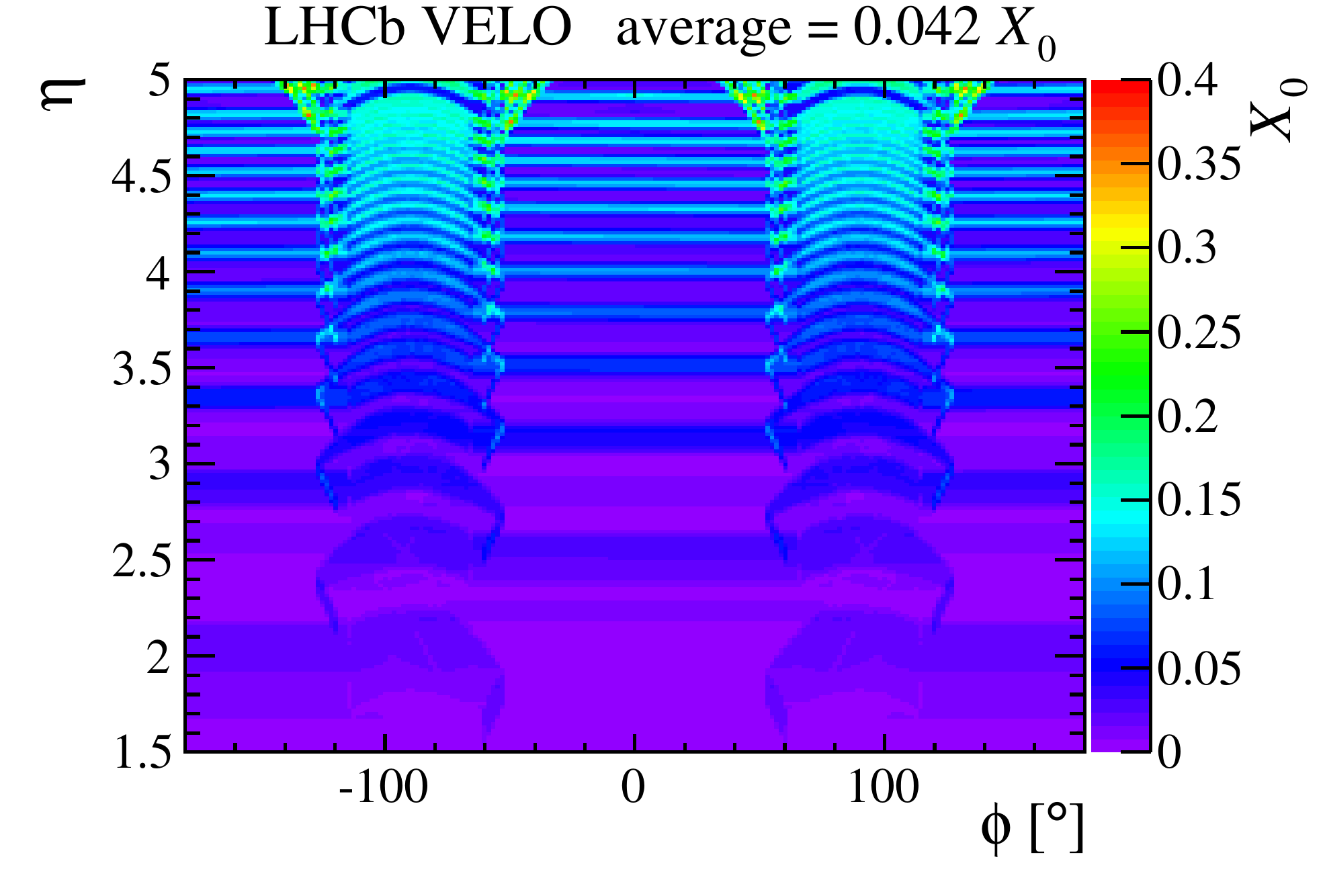}
\includegraphics[width=0.49\textwidth]{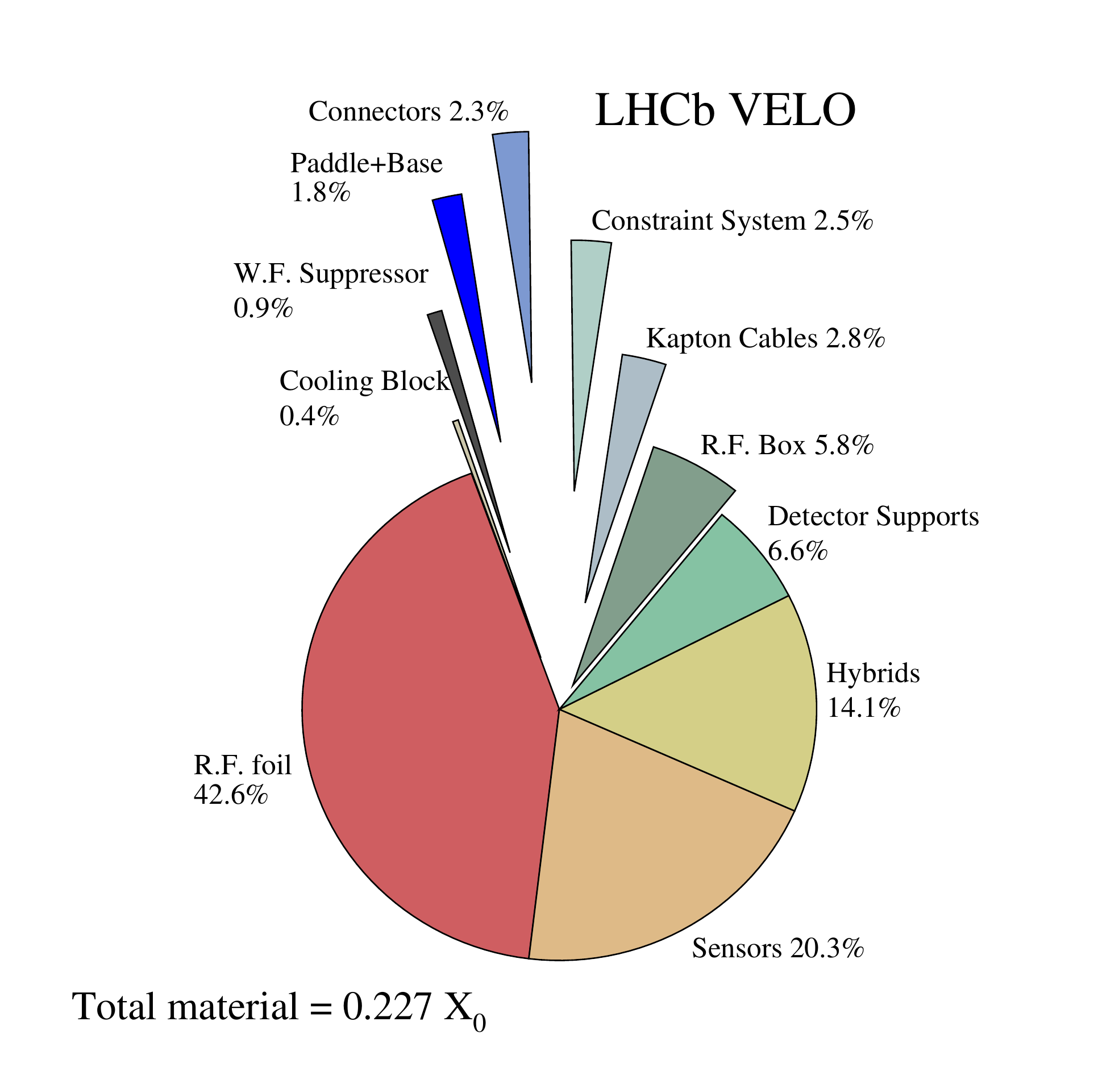}
\caption{\small
(left) Material passed through on trajectories originating from the interaction point
and terminating at the radius of the first active strip in the detector expressed in terms of the fraction of a radiation
length (\Xrad) and as a function of pseudorapidity ($\eta$) and azimuthal
angle ($\phi$). The material is higher in the regions around $\pm90\degrees$ where the two halves of the detector overlap. (right) Breakdown of the total material budget by component of the \velo.  The number given
for each component is the percentage of the total average \velo material budget
($0.227\,\Xrad$).
}
\label{fig:etaphiscanfmp}
\label{fig:piechart}
\end{center}
\end{figure}

The simulated detector description can be used to estimate the material
traversed by a particle as it moves through the \velo. Figure~\ref{fig:etaphiscanfmp} (left) shows the amount of material along the
trajectory of a particle, which originates from the interaction point, before
it reaches the radius of the first active strip on the detector ($8.2\mm$),
%it reaches a VELO sensor,
as a function of pseudorapidity and azimuthal angle.
The amount of material is expressed in terms of the fraction of a radiation
length (\Xrad) and the average material in the \velo acceptance passed through before
reaching this radius is $0.042\,\Xrad$.
The average amount of material traversed  before a particle leaves the \velo at $z=835\mm$ is $0.227\,\Xrad$. 
The breakdown of this total material budget into the components of the \velo
system inside the acceptance is shown in Fig.~\ref{fig:piechart} (right).
The \velo design means that much of the electronics and services are outside
the detector acceptance.
The RF foil dominates with $\sim43\%$ of the material, and the next largest
component is the active silicon sensors. Considering the uncertainty on all components 
of the \velo, the material traversed by a typical particle is estimated as being known to an accuracy of $\pm6\%$.

%\begin{figure}[htb]
%\begin{center}
%\includegraphics[width=0.5\textwidth]{figs/velo_pie_blacktext.pdf}
%\caption{
%Breakdown of the material budget by component of the VELO.  The number given
%for each component is the percentage of the total VELO material budget
%($0.227\,{\rm X_0}$).
%}
%\label{fig:piechart}
%\end{center}
%\end{figure}

A particle traversing the \velo can interact with the material, potentially
producing a number of other particles in the interaction.  If some or all of
these particles are charged and their trajectories are within the \velo
acceptance they can be tracked and the vertex of the interaction reconstructed.  
By examining the density distribution of these reconstructed vertices the material distribution can be studied.
The procedure is applied to a sample of data arising from interactions of the beam with gas molecules in the beam pipe as this provides a more uniformly distributed flux of interactions than in collision events.

%Ideally one would like to have coverage of the whole VELO so that the material
%can be compared in all regions.  However, $\proton\proton$ collision events do
%not provide this since the initial particles all originate from the interaction
%point.  The regions far downstream or upstream can not be covered since any
%particles from interactions in these regions will leave the VELO and cannot be
%tracked.  Instead the procedure is applied to samples of events where one of
%the beams has interacted with a gas molecule in the beampipe.  Such
%interactions can occur at any point along the beampipe and so the particles can
%travel through the VELO in any direction.  This also provides a more uniformly
%distributed flux of interactions.

Figure~\ref{fig:mat-int-full} shows the distribution of vertices in a
cross-section of the \velo between $y=-5\mm$ and $y=+5\mm$.
The vertices are plotted in the $r^{\prime}z$ plane, where $r^{\prime}$ is the
radial distance from the beam axis of the vertex multiplied by either $1$ or $-1$ depending on
the sign of the $x$ coordinate to separate the left and right halves of the detector.  
Vertices with a radius less than $5\mm$ are not plotted.
The top right and bottom left components of the figure focus on a more restricted region in $z$, just
downstream of the interaction point in data and simulation.  The main features of the \RSens and \PhiSens
sensors in each module and the undulating form of the RF foil can clearly be
seen.  The most striking difference between data and simulation is the more angular
form of the simulated RF foil.
%To make quantitative comparisons of the material using these distributions is
%not trivial.  Many systematic effects can be eliminated by forming a double
%ratio that uses the VELO sensors as normalisation:
%\begin{equation}
%R = \frac{N^{\rm data}_{\rm foil}/N^{\rm MC}_{\rm foil}}{N^{\rm data}_{\rm sensor}/N^{\rm MC}_{\rm sensor}} \,.
%\end{equation}
%However, different distributions of the fluxes of incident particles in the
%data and simulation could still give rise to a false discrepancy.  However,
%methods are being developed that should allow such comparisions in the near
%future.

\begin{figure}[htb]
\begin{center}
\resizebox{\textwidth}{!}{
\includegraphics[width=0.49\textwidth]{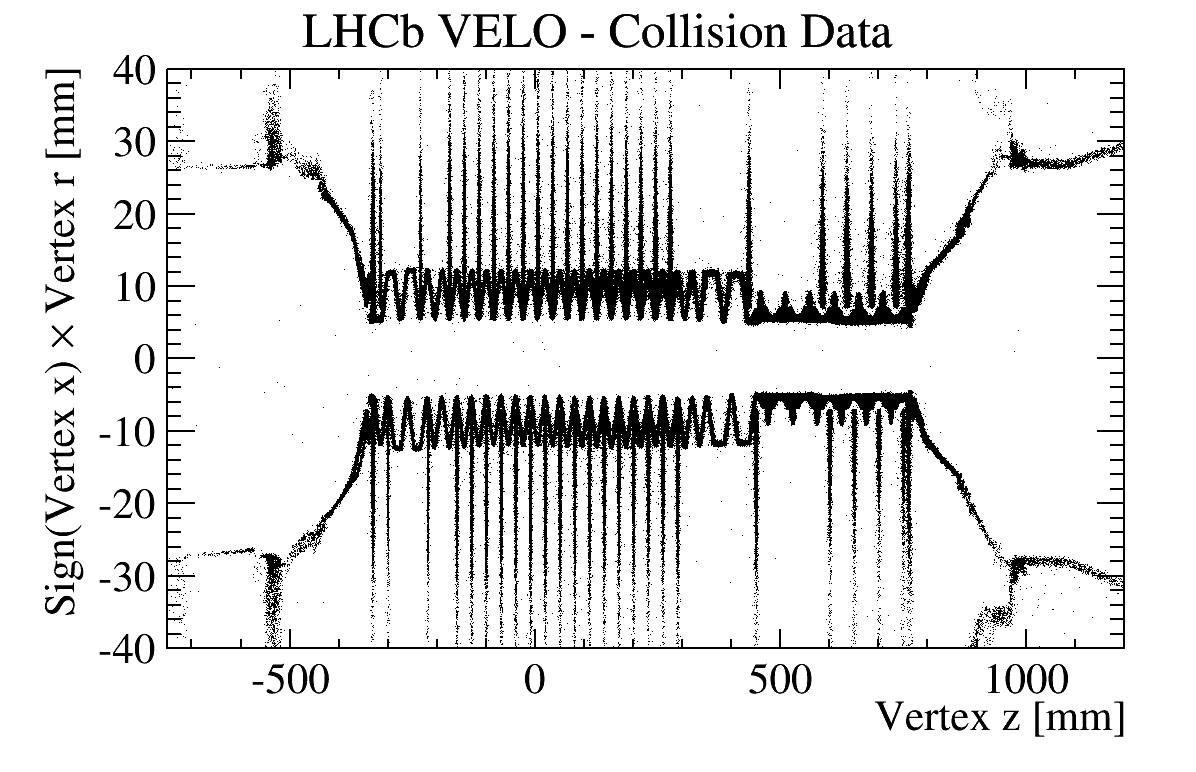}
\includegraphics[width=0.49\textwidth]{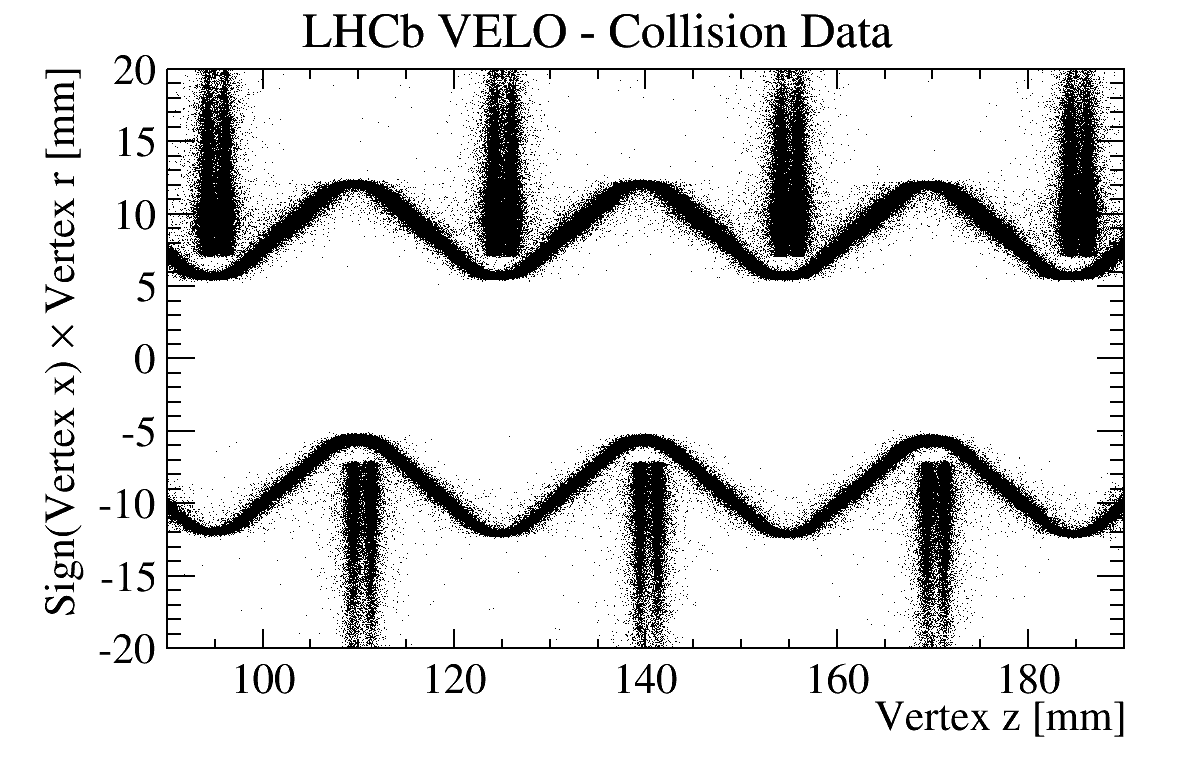}
}
\resizebox{\textwidth}{!}{
\includegraphics[width=0.49\textwidth]{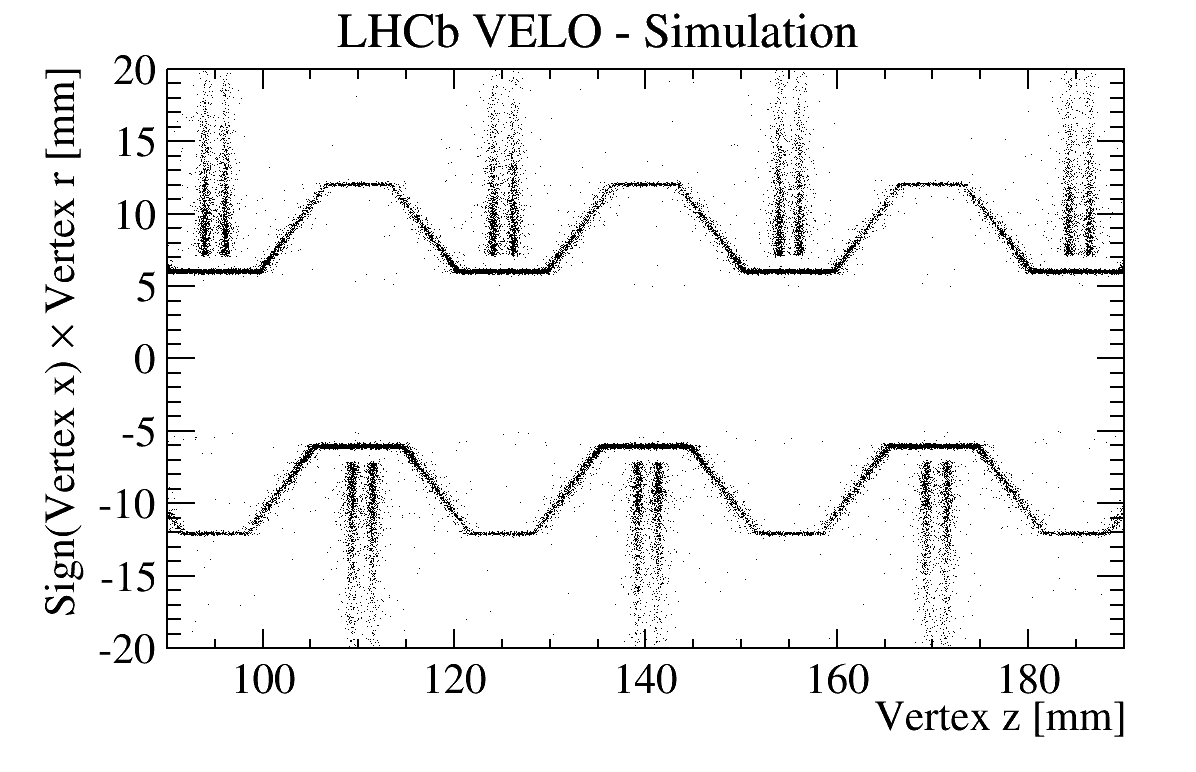}
\includegraphics[width=0.49\textwidth]{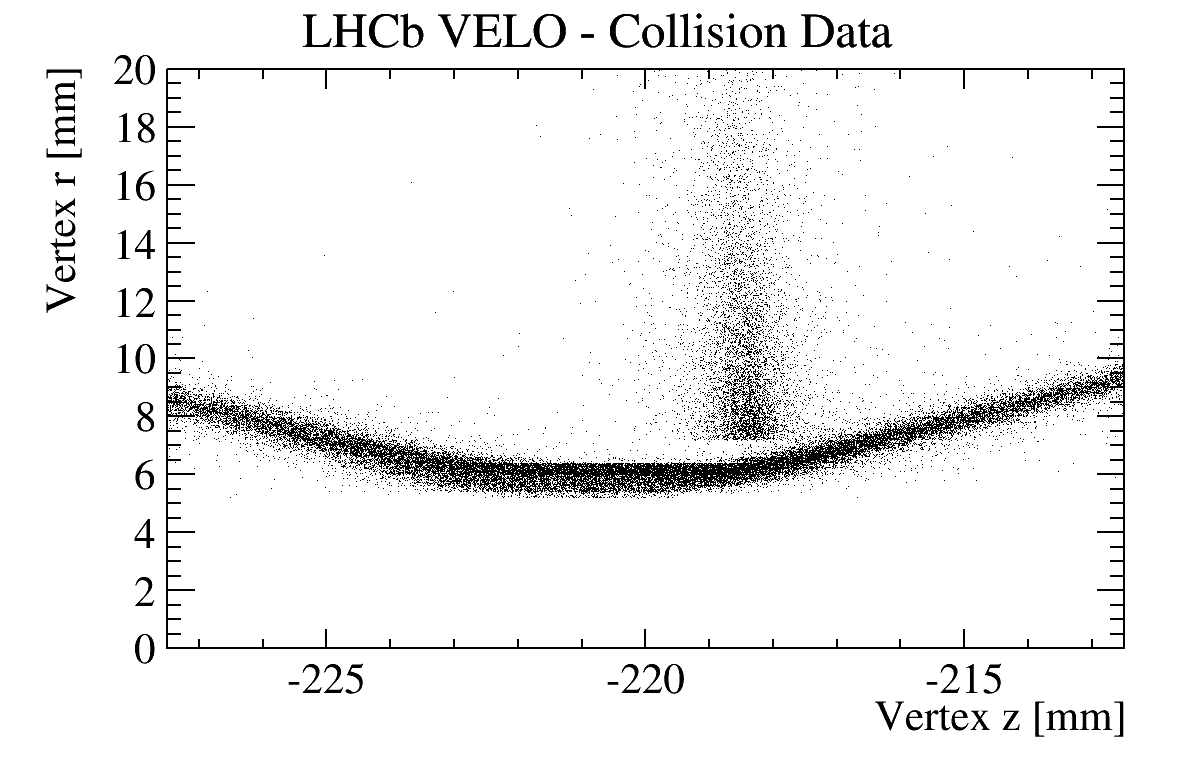}
}
\caption{\small Vertices of hadronic interactions in the LHCb \velo material. (top left) Full \velo system with entrance and exit regions visible in data. (top right) Zoom in to a group of sensors downstream from the interaction point in data. (bottom left) the same region reconstructed using simulation events. (bottom right) Zoom onto an pile-up module consisting of a single R-sensor to check the distance between the sensor and foil.}
\label{fig:mat-int-full}
\end{center}
\end{figure}

%\begin{figure}[htb]
%\begin{center}
%\resizebox{\textwidth}{!}{
%\includegraphics{figs/scan-zoom-MC.png}
%\includegraphics{figs/scan-zoom-data.png}
%}
%\caption{Vertices of hadronic interactions in the LHCb VELO material for MC (left) and data (right).
%         Zoom in to group of sensors downstream from the interaction point.}
%\label{fig:mat-int-zoom}
%\end{center}
%\end{figure}

Another useful feature of these vertex distributions is that they allow checks
to be performed of the relative positions of the various \velo components.
The proximity of the sensors to the RF foil is of particular interest since if
these were to touch it could lead to an electrical short or damage to the RF foil.  
Figure~\ref{fig:mat-int-full} (bottom right) shows the material interactions in the
region around one of the pile-up veto stations which have a single sensor.  It is clear that
in this pile-up station the sensor is quite close to, though not actually touching,
the foil.  All of the \velo stations have a greater clearance from the foil.

%\begin{figure}[htb]
%\begin{center}
%\resizebox{\textwidth}{!}{
%\includegraphics{figs/VL07_A.png}
%\includegraphics{figs/PU02_C.png}
%}
%\caption{Zoom onto individual stations to check tolerance between sensors and foil.}
%\label{fig:mat-int-stations}
%\end{center}
%\end{figure}

These data can also be used to make a determination of the aperture
available to the beam due to the mechanical tolerances of the RF foil
construction and positioning.  This information is of particular interest
for the upgrade of the \velo in which the radius of the detector, and hence of
the RF foil, is to be reduced~\cite{LHCb:1624070}.
The vertices attributed to interactions in the RF foil in the $z$ regions
around the sensors where the foil is at its smallest radius are selected. Information can be
obtained on the available beam aperture by fitting the data with a circle and extracting the variations~\cite{VeloNoteAperture}.
Conservatively assuming a foil thickness of 300\mum,  an aperture of 4.9\mm is
obtained, to be compared with the nominal value of 5.5\mm.
This is further reduced to 4.5\mm by the weld of the foil to the RF box but is
still well within the tolerance of 2.4\mm that was reserved for mechanical
imperfections of the foil with regard to the beam aperture.

\section{Calibration, monitoring and simulation}
%[Eddy Jans to oversee, max. 10 pages]
\label{sec:Calibration}

The performance of the \velo depends critically on its
calibration. This section describes the calibration of the timing and
gain of the detector, and the data processing algorithms that are
executed in FPGAs. The determination and verification of the parameters of
these algorithms are discussed. The section starts with a brief description of the data acquisition system (DAQ). The simulation of the detector is also described at the end of the section.

%The full calibration procedure for the LHCb VELO silicon detector will be presented
%in this section. First a brief description of the data acquisition system (DAQ) of the VELO
%will be given, while focusing on the data flow and processing. Next, each step of the calibration
%process will be described. Also, the software platform, called Vetra, that is used to process
%the raw - non-zero suppressed (NZS) data and determine the calibration parameters will be introduced.

\subsection{Data acquisition system}
\label{sec:DAQ}

The silicon sensor's strips are connected to the \beetle
front-end ASICs \cite{Agari:2004ja} on the hybrid circuit boards of
the \velo modules. The ASIC samples data with the LHC
bunch crossing frequency and stores them in an analogue pipeline, with
a length of 160 events, awaiting the decision of the first
level trigger. Both the shape of
the \beetle front-end amplifier pulse response and the sampling time can be
tuned in order to get the best performance; the tuning procedure for
the sampling time is described in Sect.~\ref{sec:beetleTiming}.

After a positive trigger decision is obtained, the data from the
\beetle ASIC are read out on analogue links~\cite{Bay:2010zz}. Each 128 channel \beetle
ASIC provides four-output ports. The signals of 32 channels of the sensor and four channels of header information are
sent serially at 40\mhz rate.  The system has a maximum readout rate of
1.1\mhz. The data are sent out from the
hybrid circuit board of the module on kapton cables, via vacuum
feedthroughs to the repeater boards that are located outside the \velo tank. 
The analogue data are then transmitted over a 60\m long differential
link (twisted pair) to the TELL1 DAQ boards. The TELL1 boards are located in the counting rooms outside
the radiation zone.

The TELL1 DAQ boards digitise the analogue signals and process the
data in FPGAs. The adjustment of the sampling phase is described in
Sect.~\ref{sec:timing}.  The processing algorithms and the tuning of their parameters
are described in Sect.~\ref{sec:TELL1}. The TELL1 boards output the processed data to the
high level trigger system~\cite{Aaij:1493820}.  The primary output from the TELL1
boards is zero-suppressed (ZS) data. For monitoring and tuning purposes, a number
of special output data types can also be transmitted. The raw ADC
values (non-zero suppressed, NZS) are sent out at a low rate of
approximately 1\hz. The data formats also contain the header data from the \beetle ASICs. The headers are used for the gain and error bank studies (described in Sects. \ref{sec:Gain} and 
\ref{sec:verrors} respectively).

\subsection{Timing and gain}
%The calibration of the VELO detector is performed following two basics steps. 
%Firstly the best possible digitisation parameters are determined by means of NZS data. 
%Secondly parameters of a series of processes running in the FPGAs on the TELL1 boards,
%are determined by means of algorithms that use ZS data as input.

%\subsection{Optimisation of the Digitised Data}
The ADC sampling time for the digitisation of the analogue data needs
to be determined, as does the input gain of the digitisation boards.
The ADC sampling time is set for each link to account, for example, for the
slight differences in cable lengths. 
%The timing also has a strong influence on the electronic cross-talk between neighbouring channels in the same analogue link.
The gain in the ADCs is particularly important for the uniformity of
the noise and signal levels measured subsequently. Then, the timing of
the pulse sampling in the \beetle ASIC with respect to the LHC beam collisions needs to be setup. These three calibration procedures are described below.

\subsubsection{ADC sampling time}
\label{sec:timing}
\begin{figure}[tp]
  \begin{center}
    \resizebox{\textwidth}{!}{
      \includegraphics*[width=0.02\columnwidth]{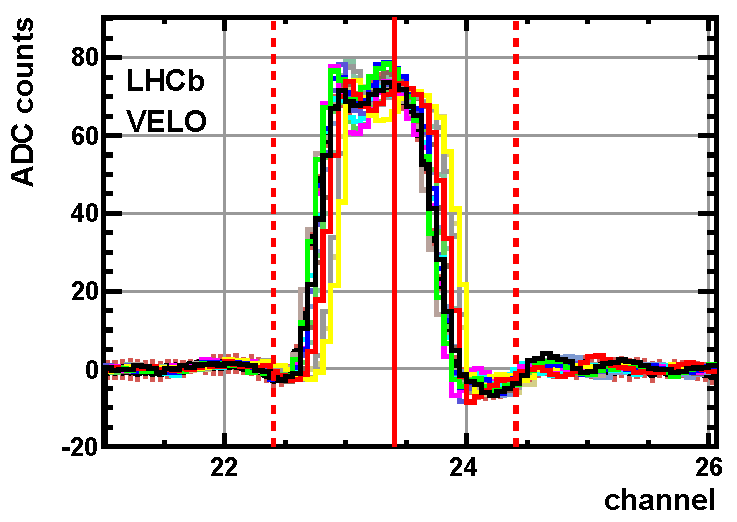}
    }
    \caption{\small An example of the scan over the digitisation
      phases (see text). Data for 16~analogue links are shown. The horizontal axis represents the 
sampling time, where the integer numbers indicate the location of the
serially transmitted \beetle channels, and between two channels there are 16 bins. The solid red vertical line indicates the 
chosen sampling point and the dashed red lines the position of the previous and next samples.} 
    \label{fig:delayscan}
  \end{center}
\end{figure}

As described in Sect.~\ref{sec:DAQ}, one triggered event of a \velo sensor is fully read out through  64 analogue links,
that each carry 36 analogue voltage levels spaced by 25\ns each.
The first four levels in this readout block are encoded
header bits, which have their heights calibrated to be slightly ($\sim30\%$) larger than one MIP.
On the receiving end, the TELL1 boards are equipped with cards that digitise
the 36 consecutive levels, sampling every 25\ns. The sampling can be
delayed by several clock cycles (25\ns)  and fine time adjustments made with steps 1/16$^{th}$ of 
a clock cycle.  The optimisation procedure first consists of roughly aligning the 
readout of the front-end signals to the digitising window. This is
performed by sending internally generated signals, test-pulses, on known channels in the \beetle
ASICs and adjusting the sampling so that the header bits are obtained in the correct positions. A scan is then performed over the fine time steps in order to get the final setting for each analogue link. The analysis of the scan data finds the optimal sampling point based on two
conditions: the best signal to noise ratio and the minimal
inter-symbol cross-talk. The inter-symbol cross-talk
 is defined as the fraction of the signal spilling over to the
 channel transmitted in the previous or next clock cycle on the same
 analogue link.  A plot of the delay scan 
analysis is shown in Fig.~\ref{fig:delayscan}, where the measured signal at
each timing step is shown. The selected sampling point of the channels
is indicated by the solid vertical line, while the small inter-symbol
cross-talk at the sampling point of the previous and next channels is indicated with the dashed lines.

\subsubsection{Gain}
\label{sec:Gain}
Each strip has a different capacitance, depending on its size and
state of depletion, and the length of its routing line.
This, in addition to the range of variances in the \beetle pipeline, cause links to have a range of signal sizes for a given deposited charge. 
The \emph{gain} is normalised with a method that is independent of these variations and is stable as the \velo undergoes radiation damage.
Here, gain refers to the conversion factor that relates a certain
\beetle output voltage to an ADC value. The gain factor can be varied using a hardware setting on each
digitisation card. This setting controls the upper limit of the voltage range digitised by the ADC and hence the proportionality of an input voltage
to ADC counts.

The four bits of header information, which precede the 32 strip
signals output on each link, encode the pipeline column number of the
event. The headers are added to the link output in the \beetle, so
they are not affected by strip capacitance or by non-uniformities in the \beetle
pipeline. They are digital bits, which are subsequently sent over an analogue link, so can act as standard candles during
the gain normalisation.

\begin{figure}[tp]
  \begin{center}
    \resizebox{\textwidth}{!}{
      \includegraphics*[width=0.02\columnwidth]{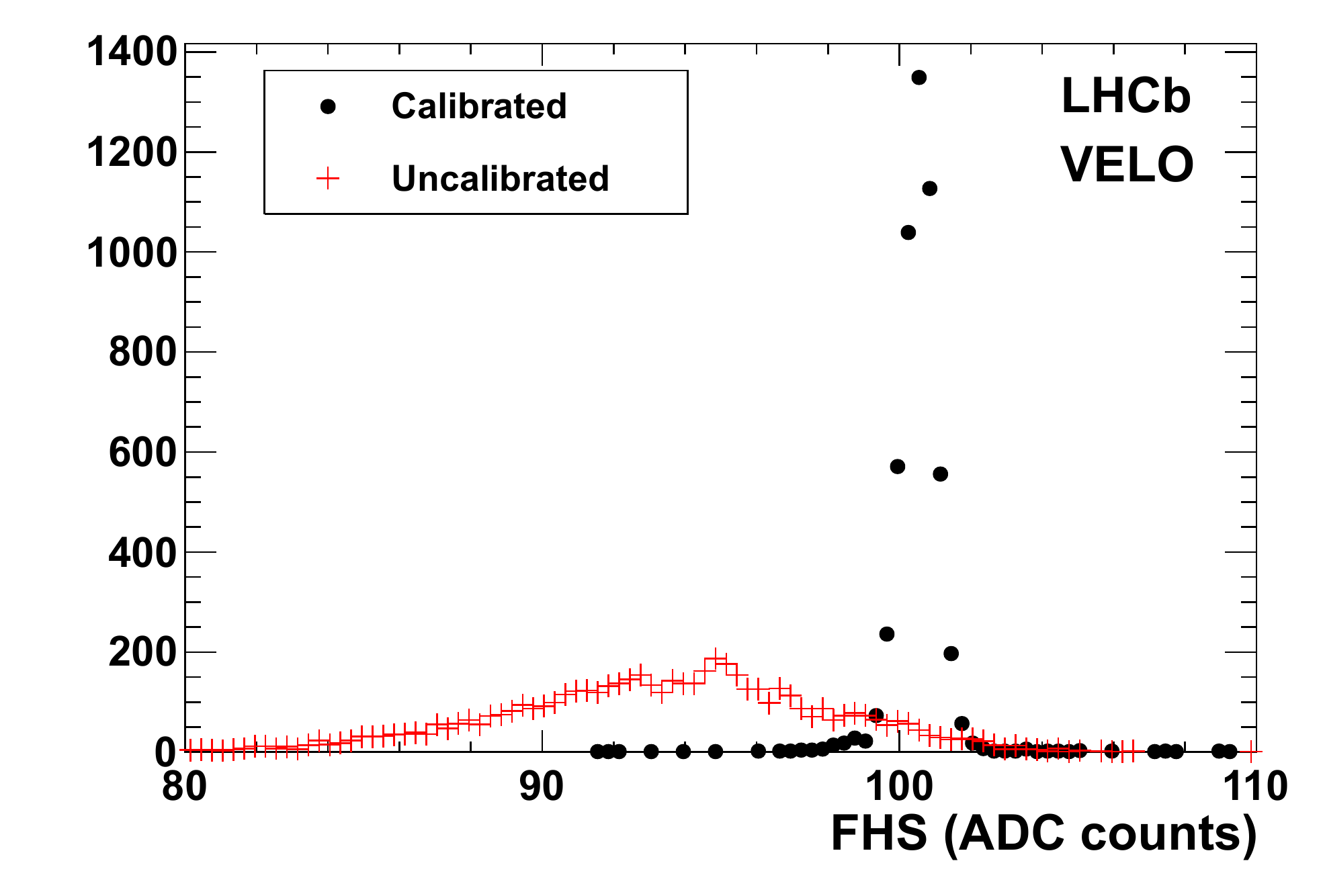}
    }
    \caption{\small The FHS distribution before and after the first gain calibration. The
    data were taken two months apart in late 2009. One entry is made in the histogram for each link.}
    \label{fig:gaincalib}
  \end{center}
\end{figure}

Depending on whether these digital bits corresponds to a `0' or `1', a header is classified as `Header High' or `Header Low', which
when uncalibrated, typically have output values of around 560 and 460~ADC counts respectively. The distributions of
high and low headers for each link are Gaussian in shape, and we define the full header swing (FHS) as the difference
between the two means. To calibrate the \velo, the gain of each link
is set to a value where FHS is equal to 100 ADC counts.
The effect of a gain calibration on the FHS is shown in Fig.~\ref{fig:gaincalib}. 

%The necessary gain settings on each TELL1 board are determined using a gain scan calibration procedure.
%NZS data is recorded at each of 23 gain settings when there are no beams in the machine.
%For each step, the FHS of each link is calculated and the resulting distribution is fitted with an exponential function.
%The range of raw FHS before any calibration was typically between 130 and 90 ADC counts.
%The result of the fit is used to determine the necessary gain setting.

With the gain normalised, the most probable value of the Landau distributions 
and average noise values of each link are more uniform than for the uncalibrated \velo.
%The gain calibration described here uses the FHS and thus is insensitive to variations in the noise and signal arising from radiation damage.
The gain has been found to be fairly stable and the calibration procedure is repeated every six months
 or after the replacement of TELL1 digitisation cards or changes to
 the FPGA firmware.

%Since the gain calibration is designed to be independent of strip capacitance and depletion,
%these will diverge as the \velo ages because the received radiation dose depends very strongly on the proximity of strips to the interaction point.

\subsubsection{Timing to the beam}
\label{sec:beetleTiming}
After performing the digitisation and gain determination, a tuning is
required to synchronise the front-end amplifier sampling to the
signals generated in the silicon by particles from beam collisions.
The signal left by a particle passing through a strip is pre-amplified and shaped in the \beetle
ASIC. The level of this pulse is sampled every 25\ns and stored in a pipeline position.
When a trigger command is received the \beetle outputs the data from the amplified
signals that were sampled 4\mus earlier; this trigger latency corresponds to 160 clock cycles.
The time alignment adjusts the \beetle sampling time to the time that the particles pass through the detector, which
is synchronous to the LHC bunch crossing time plus the time of flight
of the particle to that sensor. 

\begin{figure}[htbp]
  \begin{center}
    \resizebox{\textwidth}{!}{
       \includegraphics*[width=0.08\columnwidth]{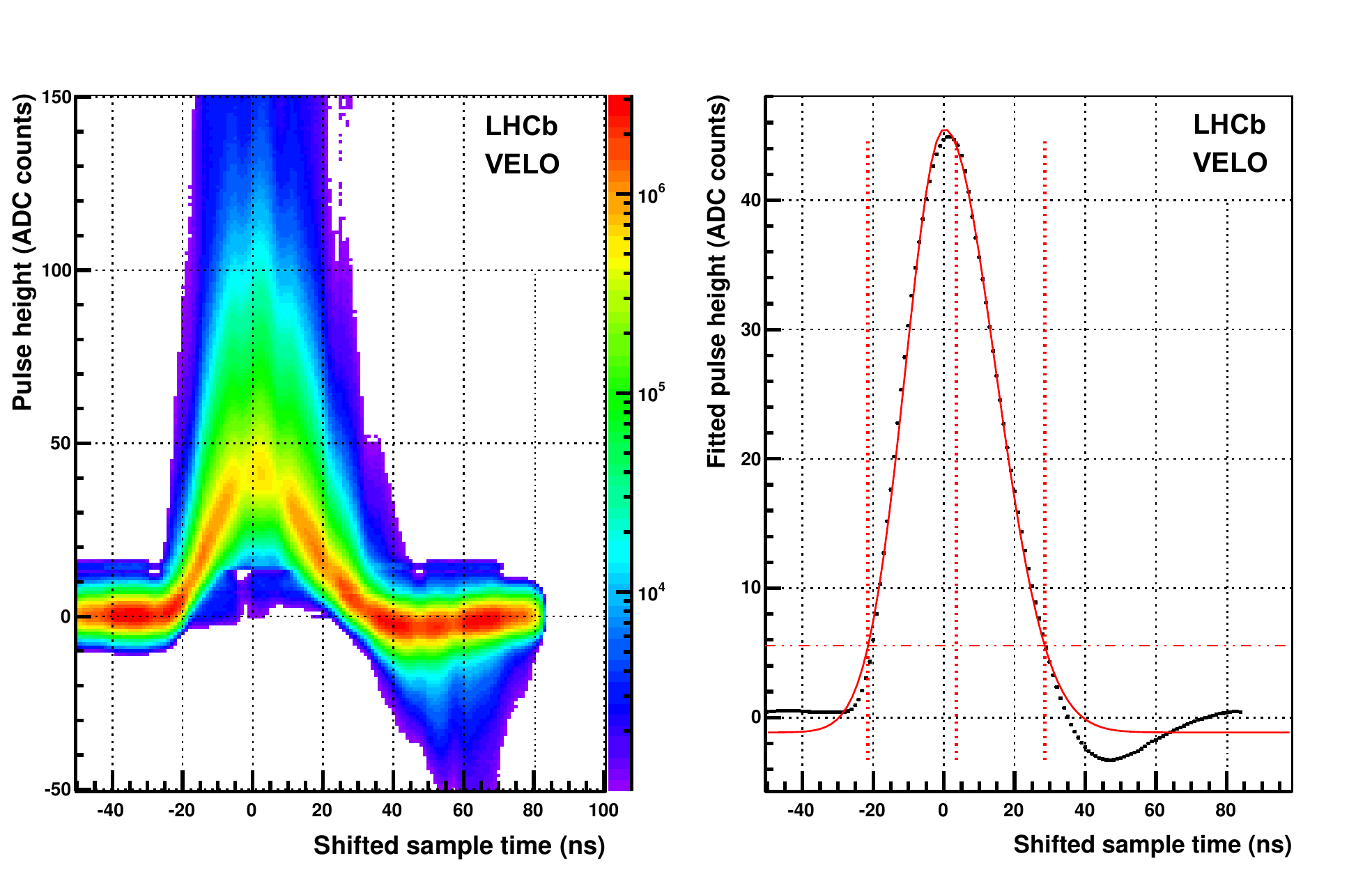}
    }
    \caption{\small Pulse-shape obtained by combining data from all sensors, with a 1\ns time step and after applying for each sensor the time shift determined by the procedure. 
             The plot on the left shows the ADC spectra versus time. 
             The plot on the right shows the most probable value for each time bin obtained from a fit of a Landau convolved with a Gaussian function. 
            A bifurcated Gaussian function is fitted to this
            distribution from which the rise and the fall widths are obtained.
             The fitted function also predicts the amount of pre-spill (25\ns before the chosen point) and the spillover (25~ns after), 
             which are marked with red vertical dotted lines. 
             The chosen sampling time is also marked with a red vertical dotted line.} 
    \label{fig:pulseshape}
  \end{center}
\end{figure}

The time alignment is performed by scanning the clocks of the front-end readout
in steps of 1\ns. Data from a few thousand collision events are stored for each step and the pulse-shape is reconstructed 
offline. The pulse-shape is fitted with a bifurcated Gaussian function, where
the two halves have different widths but are constrained to have the
same amplitude at the peak; this accounts for the difference in the
rise and fall time of the pulses. This simple function does not fully
describe the peak or the tails of the distribution, most notably the
undershoot after the pulse, but gives good agreement in the rising and
falling edges, and hence allows a quick and reliable way to optimise
the sampling time. Conventionally the sampling time would be set to
sample on the peak of the distribution, which would maximise the
signal to noise distribution.  However, we choose instead to minimise
the contamination into the previous (pre-spill) or next
(spillover)  bunch crossing. This is obtained by setting a sampling
time that equalises the spillover and pre-spill. As the rise time of
the pulse is faster than the fall time this corresponds to sampling a
few nanoseconds after the peak, and corresponds to a loss of about 4\% in the
optimal signal size. Figure~\ref{fig:pulseshape} shows the result of a
scan, with the simple function that is fitted shown and the optimal sampling time marked.

\subsection{FPGA data processing algorithms}
\label{sec:TELL1}

After digitisation the data are processed in a series of FPGA algorithms performed on the TELL1 board. 
Each TELL1 processes the output from one sensor, and contains four FPGAs each of which independently processes 512 channels of data.

A full bit-perfect emulation of the FPGA processing algorithms was
implemented in C-code at the same time as the VHDL code for the FPGAs
was developed. Extensive checks of the bit-perfectness are performed
before any firmware changes, running both the TELL1 and C-code
software on the same data samples and ensuring that bit-perfect agreement is obtained.
 The C-code emulation is then encapsulated, in a software project
 known as Vetra~\cite{Szumlak:1103717} used by both the \velo and
 \lhcb Silicon Tracker, and run inside the main software framework of
 the experiment. This allows the algorithms to be fully tested with
 data before being deployed in operation for collision data. This
 also allowed the algorithms to be fully tested and debugged during the 
 commissioning period. In operation, the ability to fully emulate the processing steps allows the output of any processing stage to be monitored, which is used for assessing the data quality and fully understanding the  system performance.
  
The algorithms described below require a set of tuneable parameters to be determined and uploaded to the FPGAs. The parameter values used are stored in XML files and recorded in a database. The parameters are primarily based on the analysis of NZS data using the Vetra package. For example, the pedestal values are determined and the cluster thresholds are optimised. In total  550k parameters are stored in the XML files for operation of the system.

%In this section a description of all TELL1 algorithms that are presently used
%(see \ref{sec:functionality}) is given together with short introduction of the input
%NZS data format. The latter is essential for understanding the whole processing chain.

%\subsubsection{NZS data format}
%\label{sec:dataformat}
%The uncompressed data from each VELO sensor consist of 2048 ADC samples. Before the input data
%can be processed the VELO dummy channels need to be added to the real data.
%This is necessary because the number of inner and outer strips on each VELO $\Phi$ sensor is an
%odd number and the FPGA processors are constructed to operate on packets of 32 channels.
%The addition of the dummy channels makes it possible to process the data from the $\Phi$ type sensors
%(for technical reasons the dummy strips are also inserted in the data from the R type sensors).  
%The input 2048 raw samples can be divided logically into 64 analogue links. Each analogue link
%consists of 32 read out (electronic) channels. The input data for each processing unit of the TELL1
%board is made of 16 analogue links. The FPGA processors can process data in a number of parallel threads
%called processing channels each of which is responsible for the handling of two analogue links of data
%(64 samples). In order to conform properly to this hardware data processing model within the VETRA emulation
%64 dummy channels need to be added at the end of the data stream for each FPGA processor. 
%The correct formatting of the input data is critical for the behaviour of the reordering and
%clusterisation algorithms. 

\subsubsection{Pedestal subtraction}
\label{sec:ped}

\begin{figure}[tb]
  \begin{center}
    \resizebox{\textwidth}{!}{
      \includegraphics*[width=0.45\columnwidth]{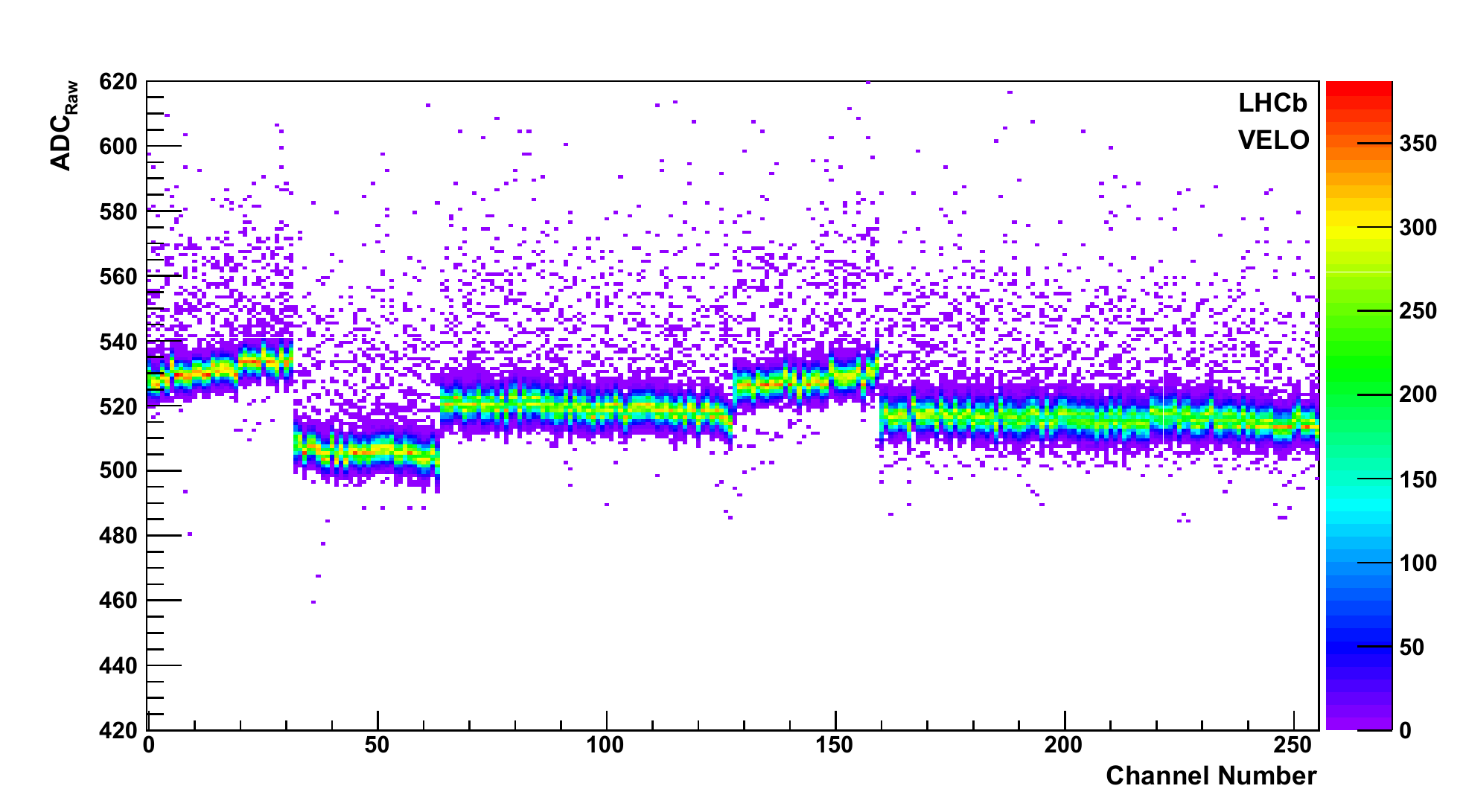}
    }
    \caption{\small Typical raw (Non-Zero Suppressed) data ADC values. Data from two ASICs, each
              with four analogue links, is shown. Variations in the pedestal values
              at the ASIC, analogue link, and channel level are seen.}
    \label{fig:rawnzs}
  \end{center}
\end{figure}

Significant variation in the average raw ADC values (pedestals) are observed for each
\beetle ASIC. Variations in pedestal from link-to-link and channel-to-channel for the raw NZS data
are also seen, as shown in Fig.~\ref{fig:rawnzs}. Hence, the pedestal values must be determined 
for each channel of the system individually, and thus 180k 10-bit pedestal values are required. 

%  The full algorithm consists of two step: the pedestal value update (using
%the pedestal value calculated for the previous event and the raw ADC value for the current event)
%and the pedestal subtraction. The full algorithm is executed only when the calibration run is taken.

The pedestal values are determined offline using NZS data collected when there are no collisions. 
%these values are stored in an XML file and kept track of in a
%database. 
%The raw data after pedestal subtraction are presented in Fig.~\ref{fig:subnzs}. 
The values are then uploaded to the FPGAs, and are subtracted from the
raw ADC when processing each event. 
%A pedestal update algorithm was also implemented in the FPGAs that
%would allow the pedestals to be followed online but has been found
%not to be needed. 
Operational experience has shown that the pedestal values remain relatively stable and a pedestal retuning is made once the pedestal subtracted ADC values of more than a few percent of channels are more than 2~ADC counts away from 0.  A retuning typically happens every two months, and is usually performed during technical stops of the LHC. A retuning is also required after the replacement of a TELL1 digitisation card or a change to the firmware. In addition to the pedestal subtraction, channels in the system that are known to be faulty are masked at this stage of the data processing.

%\begin{figure}[htbp]
%  \begin{center}
%    \resizebox{\textwidth}{!}{
%      \includegraphics*[width=0.95\columnwidth]{figs/ped_sub_data.pdf}
%    }
%    \caption{\small (left) An example of pedestal subtracted ADC values. All values are well centred around
%              the zero ADC count line. (right) The projection of this plot on the vertical axis shows a typical
%              noise Gaussian-like distribution.}
%    \label{fig:subnzs}
%  \end{center}
%\end{figure} 

\subsubsection{Mean common mode suppression}
\label{sec:mcms}

The 128 channels of the sensor that share the same \beetle ASIC, and
the 32 channels of one analogue link  
may be subject to the same sources of signal fluctuation, known as
common mode (CM) noise. The CM noise is suppressed in the FPGAs with a Mean Common Mode Suppression (MCMS) algorithm which operates on each analogue link, correcting any baseline offset in each event. 

In the algorithm, first the average pedestal-subtracted ADC value of the channels in a link is calculated for that event. Using
this mean value a search for particle hits is performed for each channel.  All channels with hits are masked, and a new mean value is calculated for each link.
This value is then used to correct the ADC values in all channels of the link. 

%The electronic signals from the VELO strips are amplified, sampled, and stored in an analog pipeline on 
%Beetle chips at the front end hybrid, as explained in previous sections. The analog 
%signals are sent to TELL1 boards via 60 meter long cables for digitisation and 
%further processing. Each beetle chip handles 128 channels and groups them
%into 4 links of 32 channels each. Differential signals of the 32 channels from one link
%are sent serially via the same pair of wires, and digitized by the same AD converter.

%The 128 channels that share the same Beetle chip, and the 32 channels of one link  
%may have the same sources of signal fluctuation. 
%These fluctuations are called common mode (CM) noise in general, although they may 
%be produced from not limited to pure noise collection. The CM noise is suppressed digitally 
%on the TELL1 FGPAs for each link. 

This algorithm would not fully correct common mode sources from the
silicon sensor, for example from noise pickup from the proton
beams. This is because consecutive channels on the sensor are not
read out on consecutive inputs of the front-end ASIC (see Sect.~\ref{sec:reorder}). A
second common mode algorithm was originally foreseen for use after
channel reordering. However,  studies performed by monitoring the noise level as a function of the \velo opening distance, showed that this source of noise pickup was small, and this second common mode algorithm has not been used in operation.

\begin{figure}
  \begin{center}
    \includegraphics[width=0.5\textwidth]{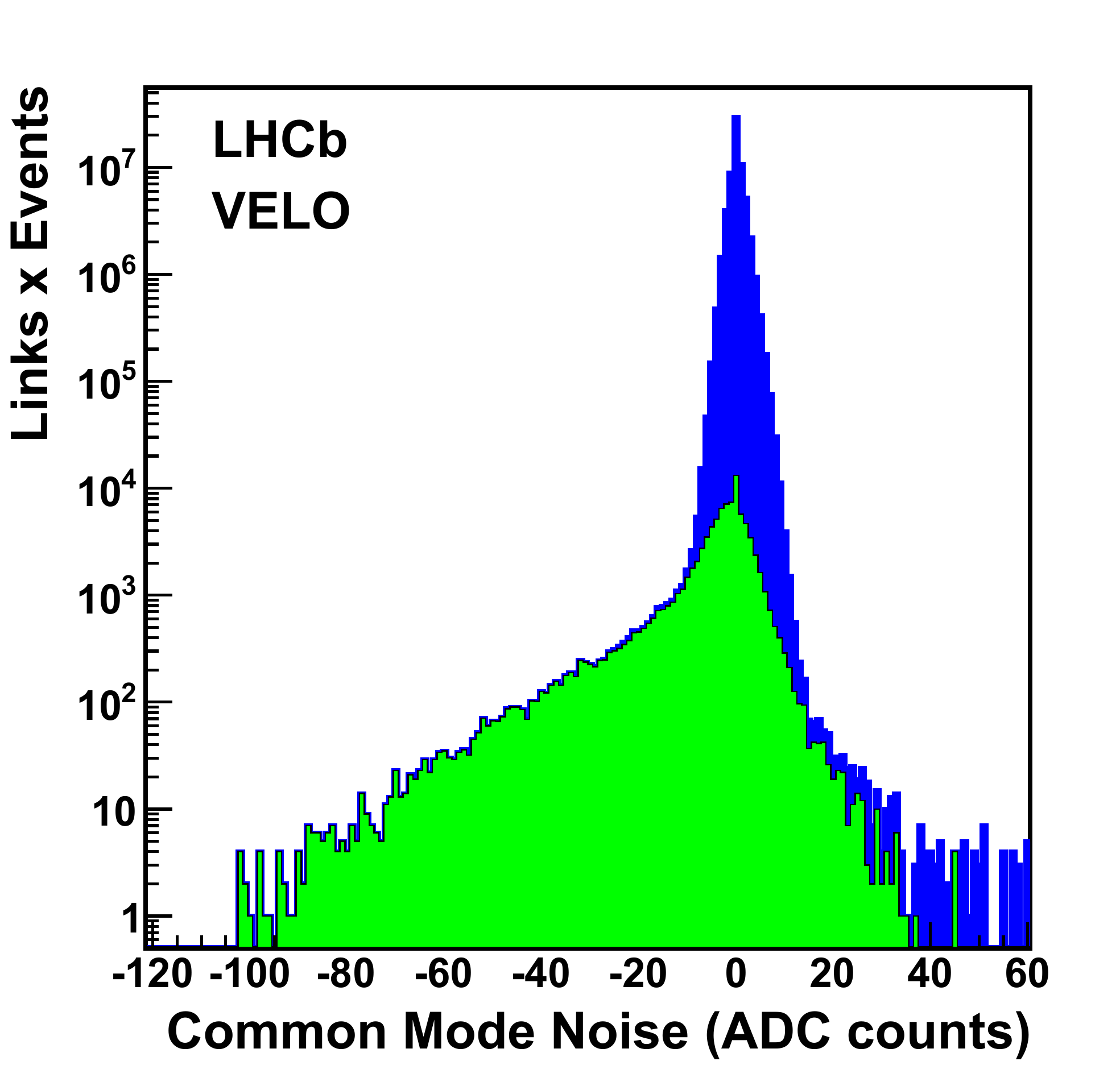}
    \caption{\small Distribution of common mode noise per link. The blue histogram is
    for all links. The green histogram is for links in a ASIC that has
    large signals (see text).}
    \label{fig:commonModeDist}
  \end{center}
\end{figure}

Figure~\ref{fig:commonModeDist} shows the distribution of the CM noise measured in the system.
It peaks at 0 and has a sigma of 1.74~ADC counts. Furthermore, out of the 1.74~ADC counts link
CM noise, there is a common fluctuation among all 5376 links, introduced by the common 
power supply and environment. This global common mode noise is 1.47~ADC counts, 
meaning the intrinsic link common mode noise is 0.93~ADC counts. For comparison the average incoherent 
(CM suppressed) noise is 1.91~ADC counts.

\begin{figure}
  \begin{center}
    \resizebox{\textwidth}{!}{
      \includegraphics{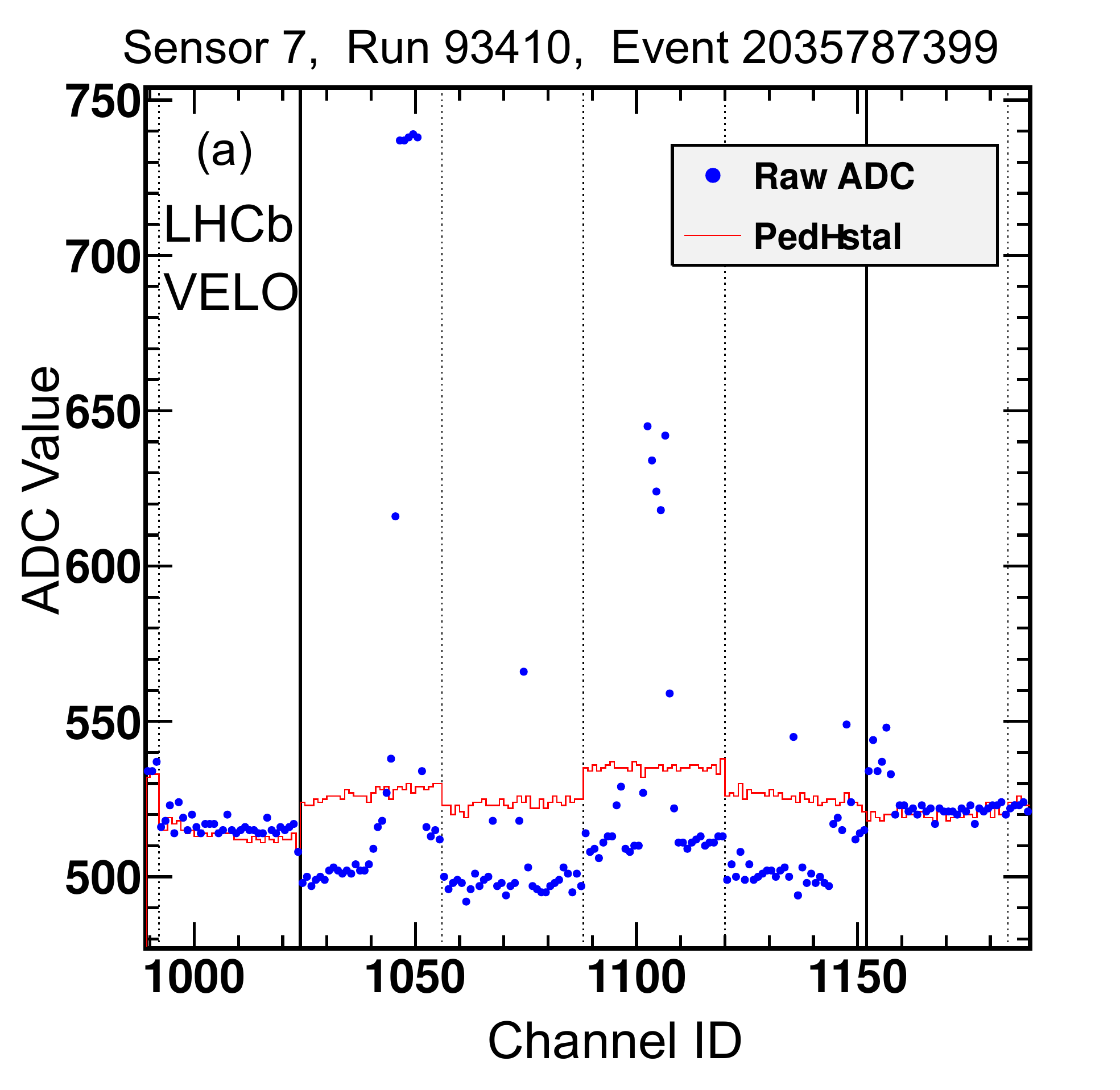}
      \includegraphics{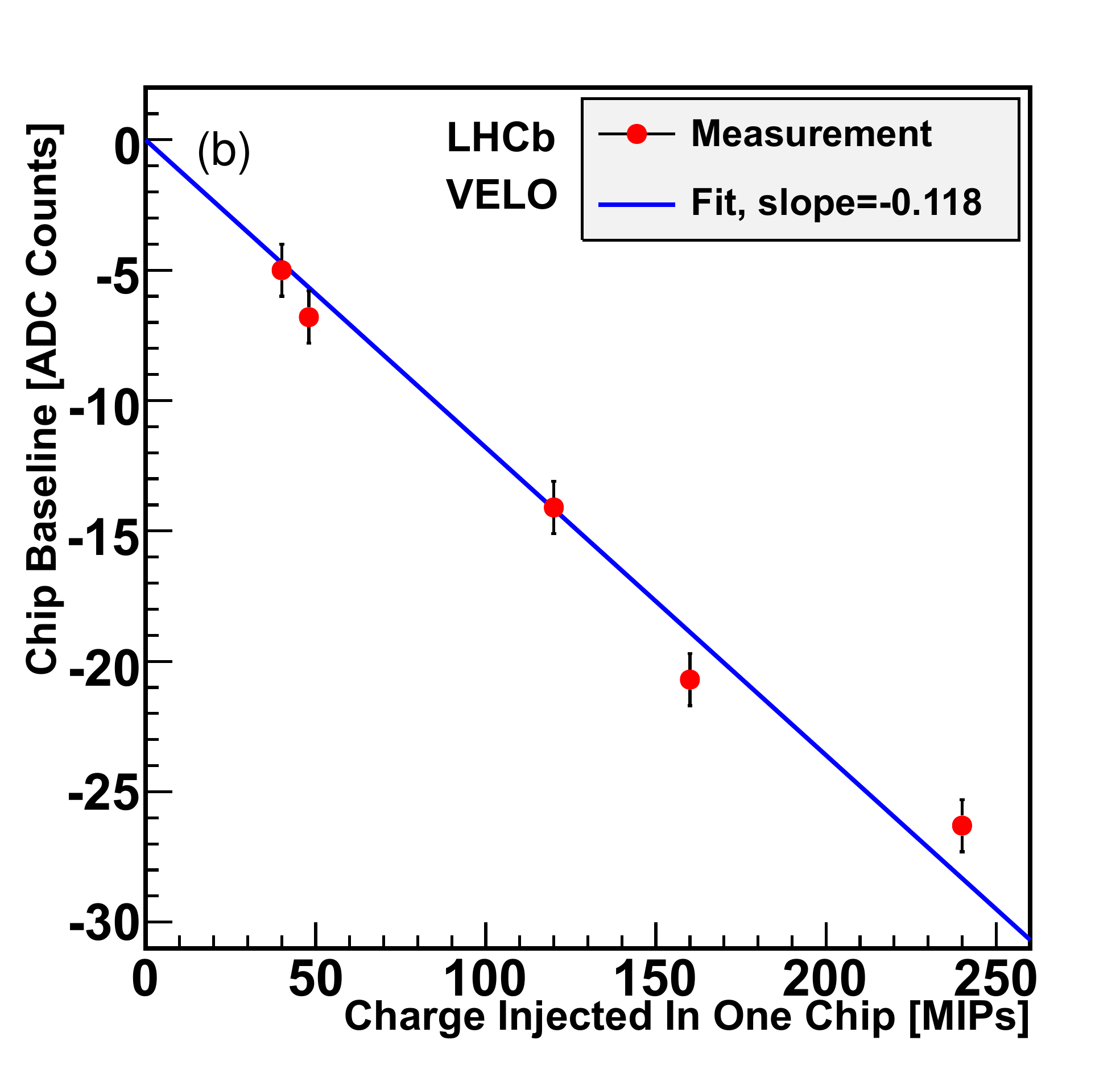}
    }
    \caption{\small (left) Event snapshot for an ASIC that has large
      signals (channels 1024-1152). The blue
    dots are the raw ADC values. The red lines are the pedestals.
    (right) ASIC baseline shift in response to injected charge in units 
    of the most probable charge deposited by a MIP in 300\mum thick silicon.}
    \label{fig:largeEnergyDepEvt}
  \end{center}
\end{figure}

The common mode distribution has a long tail on the negative side, where all four links from the same ASIC have a coherent baseline shift. 
The effect is illustrated in an event snapshot in Fig.~\ref{fig:largeEnergyDepEvt} (left).
Accompanying the baseline shift are a few channels with very large signals that reach
the saturation point of the electronics, most likely arising from soft tracks
at large angles.
In Fig.~\ref{fig:commonModeDist} the green distribution corresponds to links that have 
at least two channels with signal more than 180~ADC counts above the baseline.
The large energy deposition hypothesis was confirmed in a test-pulse calibration procedure.
A controllable amount of charge is injected to a few selected \beetle ASIC channels using test-pulse calibration capacitors. 
The baseline shift is proportional to the injected charge as shown in Fig.~\ref{fig:largeEnergyDepEvt} (right).
These baseline shifts are corrected by the CM correction algorithm.

\subsubsection{Topological strip channel reordering}
\label{sec:reorder}
As a consequence of the layout of the routing of the channels from the
\velo \RSens and \PhiSens sensors on the second metal layer,
neighbouring channels on the sensor are not necessarily read out
consecutively. In the \PhiSens sensor  strips from the inner region of
the sensor are routed out over strips in the outer region, and the
readout is thus intermingled between inner and outer region
strips, as shown in Fig.~\ref{fig:veloSensors}. Therefore the electronic readout ASIC channel numbering does not
correspond to the physical (strips) channels on the sensors. The
zero-suppression and clusterisation must be performed in terms of
physically neighbouring channels, and thus a translation algorithm is necessary.
The reordering procedure is performed separately for \RSens and
\PhiSens type sensors. The implementation of this algorithm consumes a
significant amount of the FPGA resources.

\subsubsection{Zero suppression and clusterisation}
\label{sec:zs}
The final data processing algorithm on the TELL1 produces the raw
cluster bank which is sent out to the high level trigger. This processing reduces the data size by only transmitting channels that have significant signals present. It also groups together neighbouring hits that may be due to single particles and produces a first estimate of the centre of this cluster.

%It is sensor type specific and is performed for each sensor separately. 
In the first step, hit detection is performed by finding signals above
seeding thresholds. The current seeding threshold
in each channel is set to be six times higher than the measured noise
in that channel, rounded to the nearest ADC count. This limits fake noise hits while not significantly reducing the selection efficiency of true signal clusters in fully depleted sensors (see Sect.~\ref{sec:Signal}).

In the second step, the algorithm attempts to include adjacent strips in the cluster. Strips are added to the cluster if they pass an inclusion threshold, which is currently set at 40\% of the seeding threshold. By setting a lower inclusion threshold than the seeding threshold additional charge  is added to the cluster, which helps improve the resolution on the particle position. The maximum number of strips per cluster is four. 

The centre of the cluster is calculated from a pulse-height weighted
average of the strips contributing to the cluster. This is calculated
with a precision of an eighth of a strip (3-bits). This calculation is
used to save time in the trigger; it is recalculated with floating
point precision for use in the offline tracking. The resolution and
cluster detection efficiency are discussed in Sect.~\ref{sec:Overall}.

Finally, each cluster is encoded into a bit structure that contains
information on the cluster position, cluster size, and the charge measured on each strip that contributes to this cluster. The size of the raw cluster bank varies depending on the event's occupancy, and the typical size of this bank for the full \velo is about 20\unit{kB}.

%\subsection{Noise measurement}
%\label{sec:noise}

%cross-talk??

\subsection{Error identification}
\label{sec:verrors}

Information is sent out from the \beetle  in the header bits that allow a
number of consistency checks to be performed. Error banks are then
produced by the TELL1s when the data of a link is internally
inconsistent, or there is inconsistency between data from various
links. They aim to contain the required information to trace down the
origin of the error and take appropriate measures.  The most commonly
occurring errors relate to the verification of the pipeline column
number. This number records in which column of the \beetle pipeline
the data are stored while awaiting a first level trigger
acceptance signal. The pipeline
column number is reset periodically by a broadcast command and hence
verifying that it is identical for all front-end ASICs ensures the
synchronicity of the detector.
Elements of the binary value of the pipeline column
number are transmitted as an analogue signal preceding the data from
each link of the ASICs. Gain variations and inter-symbol cross-talk can cause the analogue values to end up in between the corresponding low and high bit thresholds, and consequently make the bit assignment ambiguous. Calibration and verification procedures that are executed regularly, have managed to control these effects to a satisfactory level and errors now occur at the level of a few in every ten thousand events.
 
 \subsubsection{Single event upsets}

One error type of particular interest is a single event upset (SEU).
Since the \beetle ASIC is exposed to ionising radiation charge will be
liberated in its electronic
circuits. Most particles will however not generate enough
charge to influence the operation of the devices. Occasionally a highly ionising
particle can cause a change in the logic state of one of the flip-flops in the
\beetle ASIC. These SEU can affect the performance on the
ASIC if they occur for instance in one of the configuration registers.
Therefore all critical flip-flops are implemented in a triple redundant way,
featuring majority voting and auto-correction. To monitor the rate of SEU, all
bit-flips are counted in a dedicated counter. The two least significant bits of
this counter are sent with the data for every trigger as part of the
header bits. By analysing the transitions
of these counter bits, an average of 2.9~SEU per \invpb
of delivered integrated luminosity for all 1344 \beetle ASICs combined is observed.
% [Martin to do: express in terms
%of fluence at the \beetle, or as X-sec per flip-flop.] Martin
%attempted but wasn't
%able to do

\subsection{Monitoring}
\label{sec:Monitoring}

The \velo monitoring infrastructure is used to ensure that any degradation in data quality is observed and can be followed up quickly.
It is complementary to the \lhcb monitoring which is split in two components: online and offline monitoring.
The \lhcb online monitoring analyses a fraction of the ZS data sample that has been selected by the trigger.
The \lhcb offline monitoring uses a number of lightweight algorithms
to monitor a sample of all data that are processed for physics analyses.
The \velo monitoring adds two components: an analysis of NZS data and
a detailed analysis of the ZS data. Experience shows that many data
quality problems are spotted first in the detailed checks performed by the dedicated \velo monitoring
team.

The NZS data are recorded in a special data stream which is stored separately from the physics data at a rate of 1\hz.
Events directed to this stream contain NZS data from all \velo sensors.
This allows the analysis of pedestals, common mode suppression, cross-talk effects and noise to be made for each run.
The NZS readout has also been used to study correlations between different sensors in the same event (see Sect.~\ref{sec:mcms}).

The \velo monitoring also analyses ZS data which have not been biased
by any trigger selection. This sample of events is thus representative of the data being recorded by the detector, and allows the detector performance to be monitored independent of changes to the trigger.
These data are recorded at a rate of 10\hz and stored in a separate data stream to increase the efficiency of the data analysis.

All data in the ZS and NZS streams used by the \velo monitoring are processed automatically by monitoring algorithms shortly after they have been recorded.
The output of the monitoring algorithms are analysed daily by \velo data quality shifters.
These are people on call who analyse the incoming data and alert experts of problems if necessary.
These analyses include information on pedestals, noise, common mode,  cluster signal distributions, occupancy, tracking and alignment.
Separate monitoring is implemented for all slow control data such as voltages, currents, temperatures and pressures.

\begin{figure}[tb]
  \centering
  \includegraphics*[width=\textwidth]{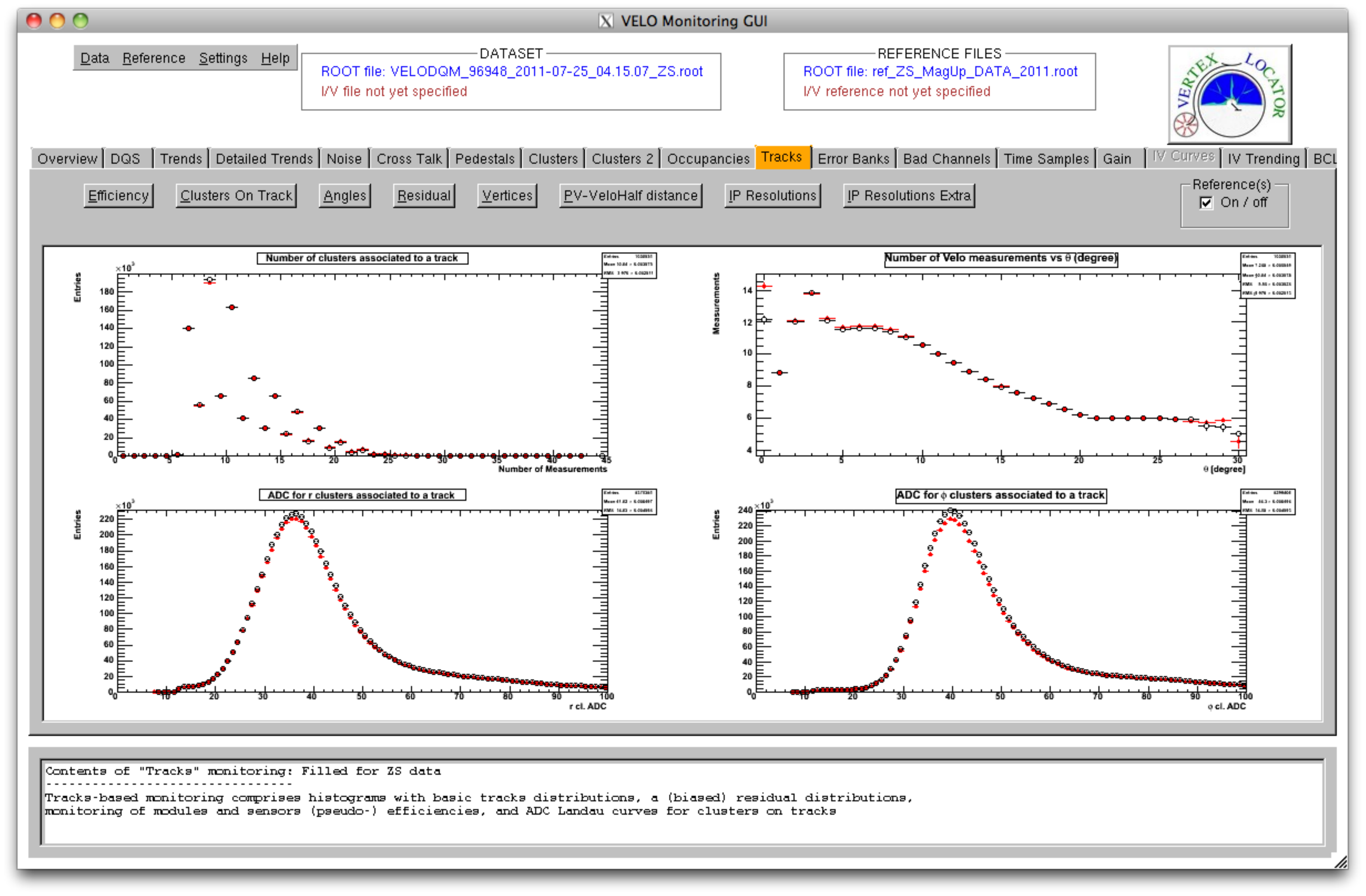}
  \caption{\small Example of the graphical user interface used in the \velo monitoring. Shown are distributions of the number of hits per \velo track segment (top left), the track angle with respect to the beam axis (top right), and the ADC counts of \RSens and \PhiSens sensor clusters associated to a track (bottom left and right, respectively). The results of the current run are shown in red filled symbols and the reference data for comparison are shown as black open symbols.}
  \label{fig:VeloMoniGUI}
\end{figure}

A graphical user interface has been developed to facilitate the analysis of the output of the regular monitoring algorithms.
An example of its usage is shown in Fig.~\ref{fig:VeloMoniGUI}.
The same tool also produces trend and correlation plots of the
relevant quantities, which are used to track the detector stability
over time. Summary tables of representative values to assess data
quality, for example numbers of high occupancy channels, for each run
are also made. The summary values and comments from the analyst of the data are stored in a dedicated electronic logbook.

Data are also taken under special conditions to cover aspects that
cannot be easily accessed with physics data. Dedicated monitoring is
in place for these periodic performance checks. These include sensor current versus voltage, current versus temperature, noise versus voltage, and charge collection efficiency  versus voltage data which are performed to monitor radiation damage (see Sect.~\ref{sec:Radiation}).

\subsection{Simulation}
\label{sec:Simulation}
%[Tomasz - 2 pages]

The \lhcb simulation is based on Monte-Carlo event generators and the use of
the \geantfour toolkit \cite{geant4}. A detailed model of the detector material in the
\velo has been produced, through which the simulated particles are
propagated (see Sect.~\ref{sec:Material}). The entry and exit points of the simulated particles in the silicon sensors are
obtained from \geantfour. A thin silicon sensor is expected to have an energy deposition distribution
that can be described by a Landau function convolved with a Gaussian distribution. The charge deposited by a particle in the silicon sensor is
calculated using the form from theory and previous experiment  \cite{Bichsel:1988if}.

The total deposited charge is distributed at uniformly spaced nodes
along the particle path in the silicon, with the energy fluctuations
along the path simulated. The diffusion of the charge as it drifts in
the electric field of the silicon is simulated through a Gaussian
smearing whose width is dependent on a bias voltage parameter.
Cross-talk between the charge attributed to the strips is then added to model capacitive coupling.
Noise is added to the strips following the measured values in the
data.  The measured pulse-shape of the front-end electronics is also simulated, and corrections are applied for the time of flight of the particles. The electronics response from the previous and the next events are also simulated by applying the pulse-shape to simulated events. 
Pedestal offsets and common mode noise are not normally simulated as they
are removed effectively in the real data by the FPGA processing algorithms.
This simulation stage results in a model of the amount of charge collected on a sensor's strips.
The model has the bias voltage and capacitive coupling as free parameters
which are tuned to obtain agreement with the measured data resolution as a function of pitch and
track angle and typically agree to within 5\%. This simple algorithm is fast and sufficiently accurate for the simulation required for
physics analyses. 

%These in turn will be processed by the same code that is used for the real data zero-suppression (see section \ref{sec:zs}). It should be stressed that The last step of the emulation procedure produces
%the raw VELO bank that is used by the track reconstruction software. The format of this bank is the same as the one
%produced by the TELL1 electronics boards.

The next stage of the simulation models the processing in the TELL1
board. First, the signal is digitised. At this point the DAQ emulation
produces an output format that corresponds to the raw non-zero suppressed
data.   The simulated data are then passed through the bit-perfect
emulation of the TELL1 clusterisation algorithm (see Sect.~\ref{sec:TELL1}). Identical clusterisation
thresholds are used to those in the data taking. The output data bank is
then produced in the same format as that produced by the TELL1
boards. A global scaling, the gain (electron to ADC conversion), is applied to normalise the observed signal in simulation, with a correction for the
variation in signal size with radius (see Sect.~\ref{sec:Signal}). The trigger, pattern recognition and track reconstruction
algorithms are applied identically for real and simulated data.

\section{Overall system performance}
\label{sec:Overall}
%[Martin to oversee, max. 13 pages]

This section characterises the \velo system performance, including the signal to noise ratio, hit resolution, occupancy, and efficiency studies. Beam backgrounds are also discussed. The results presented use data before appreciable effects of radiation damage were observed, results on the observed radiation damage are reported at the end of this section.

\subsection{Signal size and noise rate}
\label{sec:Signal}

%[David - 2 pages]

% This will be
%amplified and digitised by the Beetle, repeater board, TELL1 readout
%chain. 

\begin{figure}
  \begin{center}
    \resizebox{\textwidth}{!}{
      \includegraphics{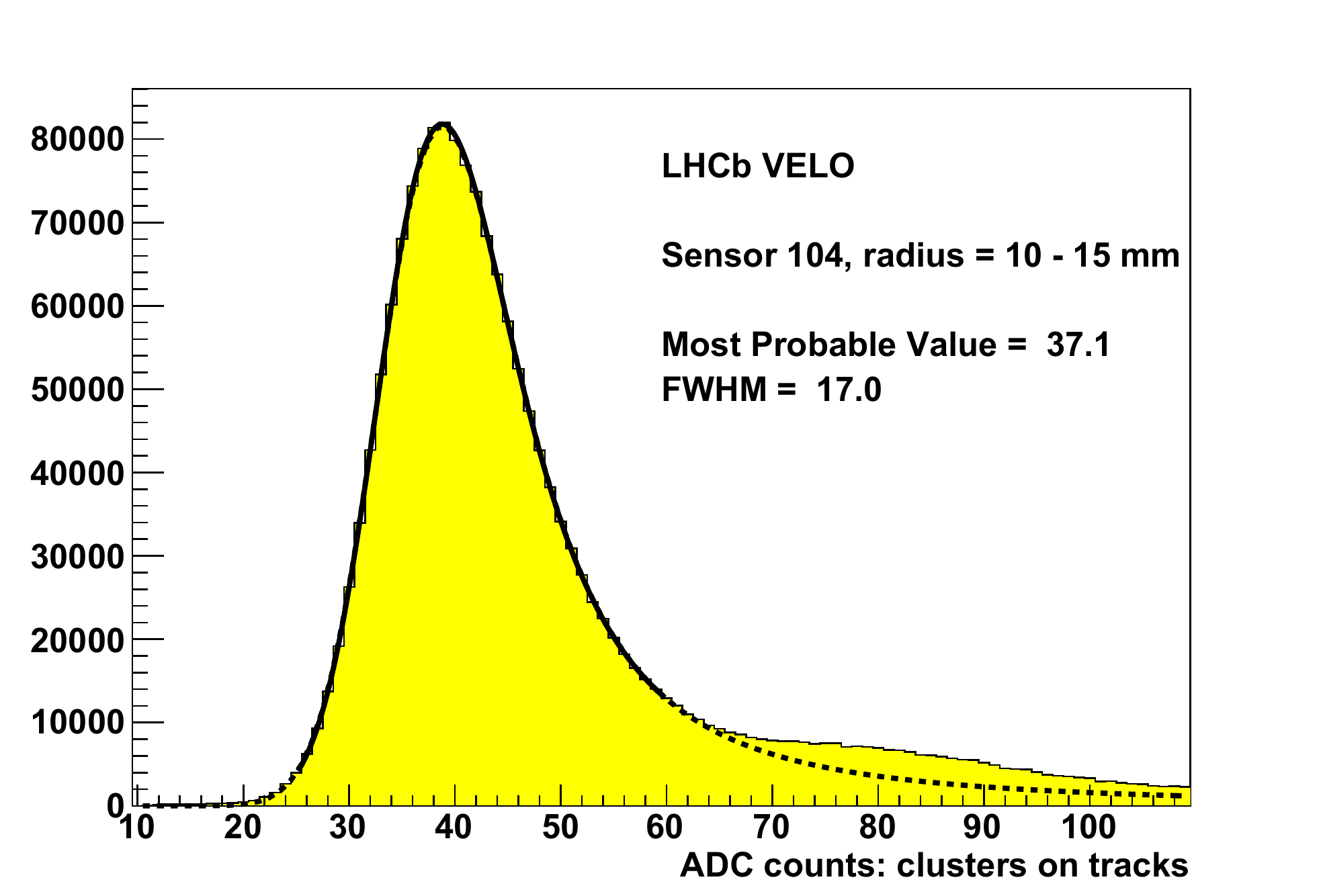}
      \includegraphics{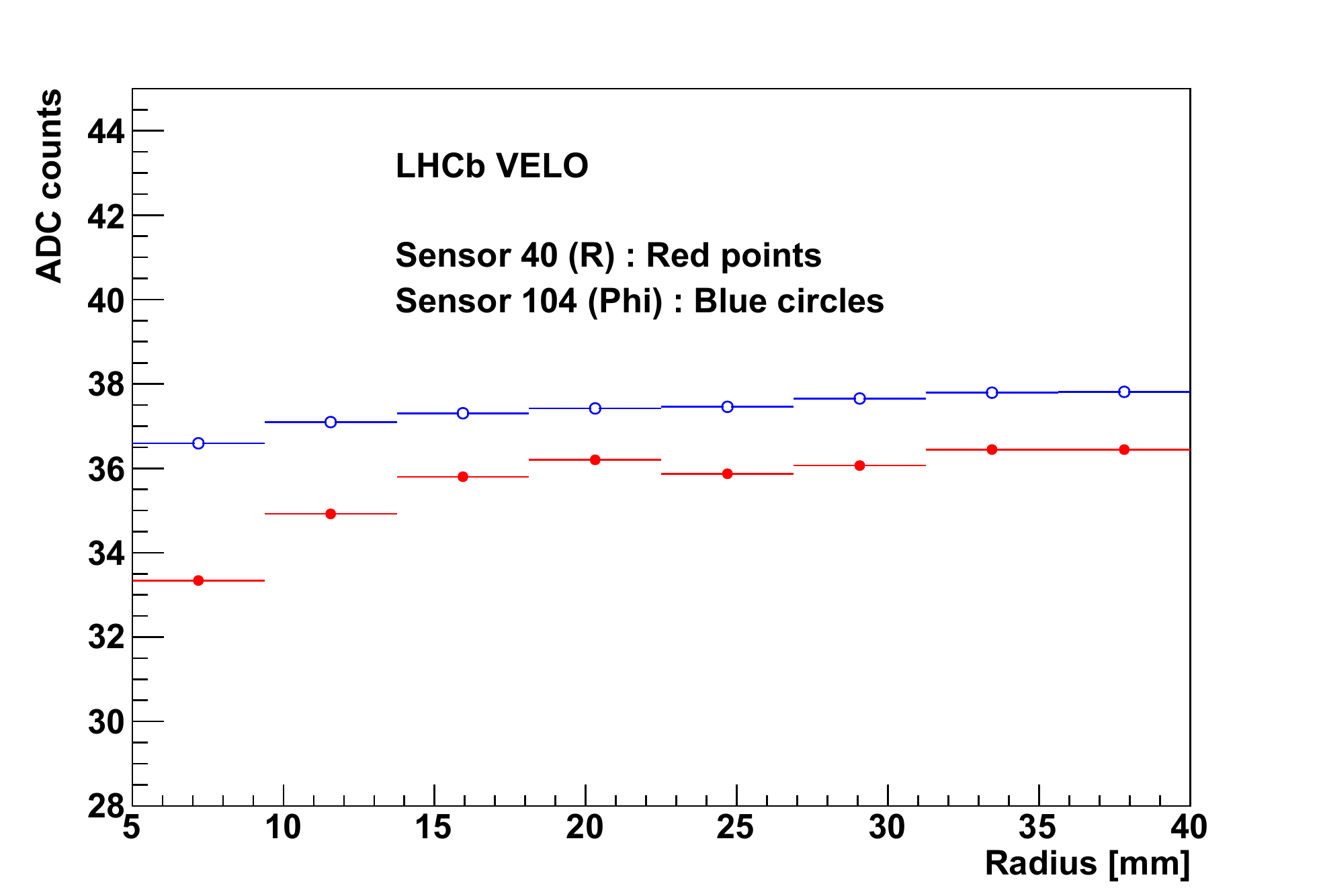}

    }
    \caption{\small (left) Fit of the signal distribution for Sensor 104 (a \PhiSens 
      sensor) for clusters on long tracks with a track intercept point on the sensor  $10 < r
      (\mm) <15$.  (right) The most probable value of the  fits to sensors 40 and 104, the \RSens and \PhiSens sensor of the same module respectively, as a function of the track intercept point radius. The cluster ADC counts were normalised to a track crossing
      300\mum of silicon. }
    \label{fig:Landau_signal}
  \end{center}
\end{figure}

The observed total signal from the clusters is determined by  fitting a Landau convolved
with a Gaussian function around the peak region. The deposited charge for each track is corrected for the track's angle to give a result corresponding to a path length in the silicon of  the nominal 300\mum thickness of the sensor. An example fit is shown in Fig.~\ref{fig:Landau_signal} (left). The function provides a good description of the rising edge and peak region, but undershoots the data in the high energy deposition region. This is in part due to the presence of photon conversions to electron positron pairs which, in the absence of an appreciable magnetic field in the \velo, can remain merged into a single cluster.  The tuned simulation is in good agreement with the data distribution, including in the high energy deposition region, and the FWHM  does not require any additional smearing of the detector response. The most probable value (MPV) of the Landau distribution varies as a function of track radius, as shown in Fig.~\ref{fig:Landau_signal} (right). 
%Also shown in Fig.~\ref{fig:Sensor104_signal} is the distribution of ADC counts from LHCb
%simulated events.

%\begin{figure}
% \begin{center}
%    \resizebox{\textwidth}{!}{
%      \includegraphics{figs/data_sens_MPV_40.pdf}
%      \includegraphics{figs/data_sens_FWHM_40.pdf}
%    }
%    \caption{The most probable value (left) and full width at half
%      maximum (right) of the fits to sensor 40 (R) and sensor 104, the
%      $\phi$ sensor of the same module, as a function of impact point
%      radius. The cluster sizes were normalised to a track crossing
%      $300\mum$ of silicon. }
%    \label{fig:Sensor40_signalVRadius}
%  \end{center}
%\end{figure}

\begin{figure}
  \begin{center}
    \resizebox{\textwidth}{!}{
      \includegraphics{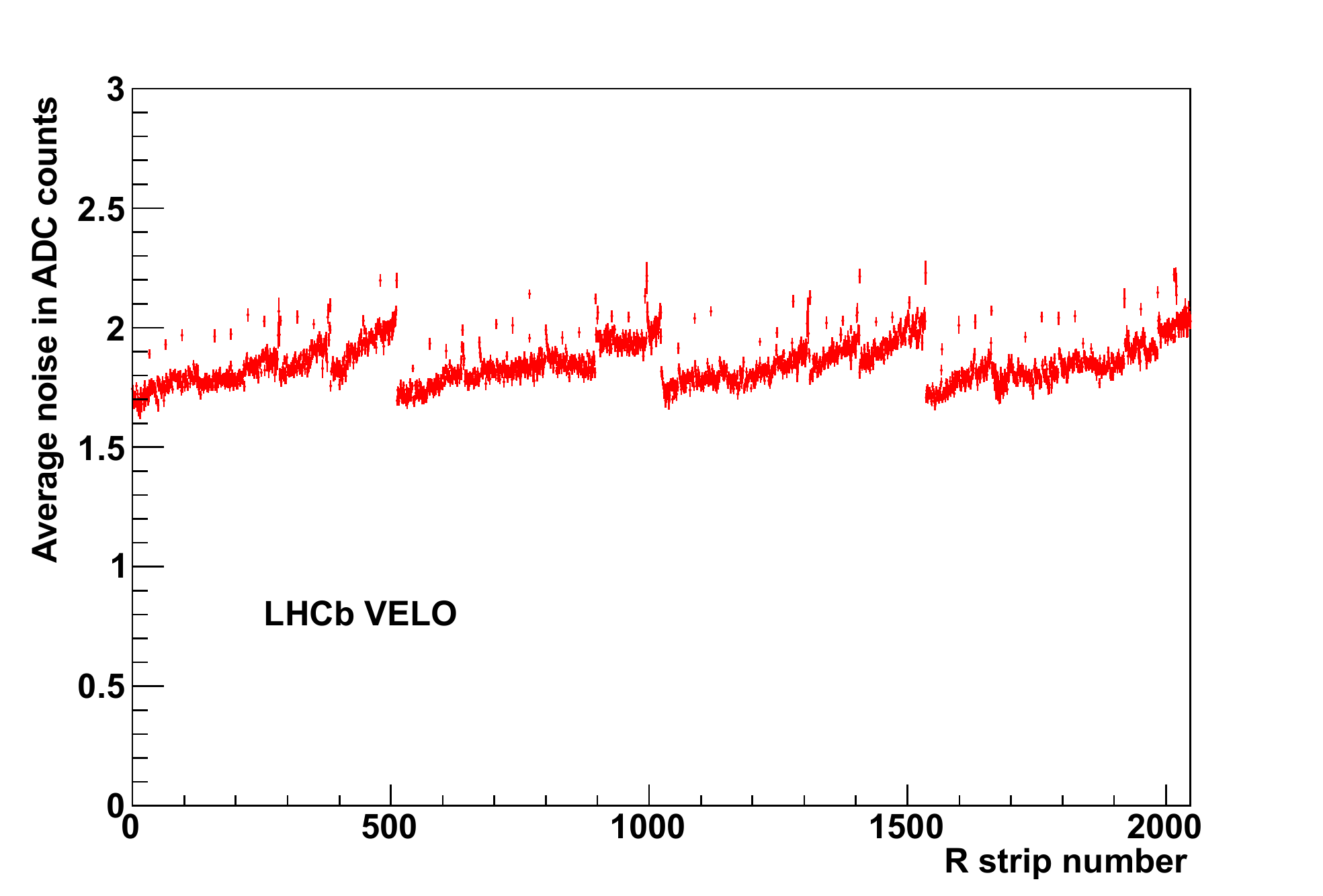}
      \includegraphics{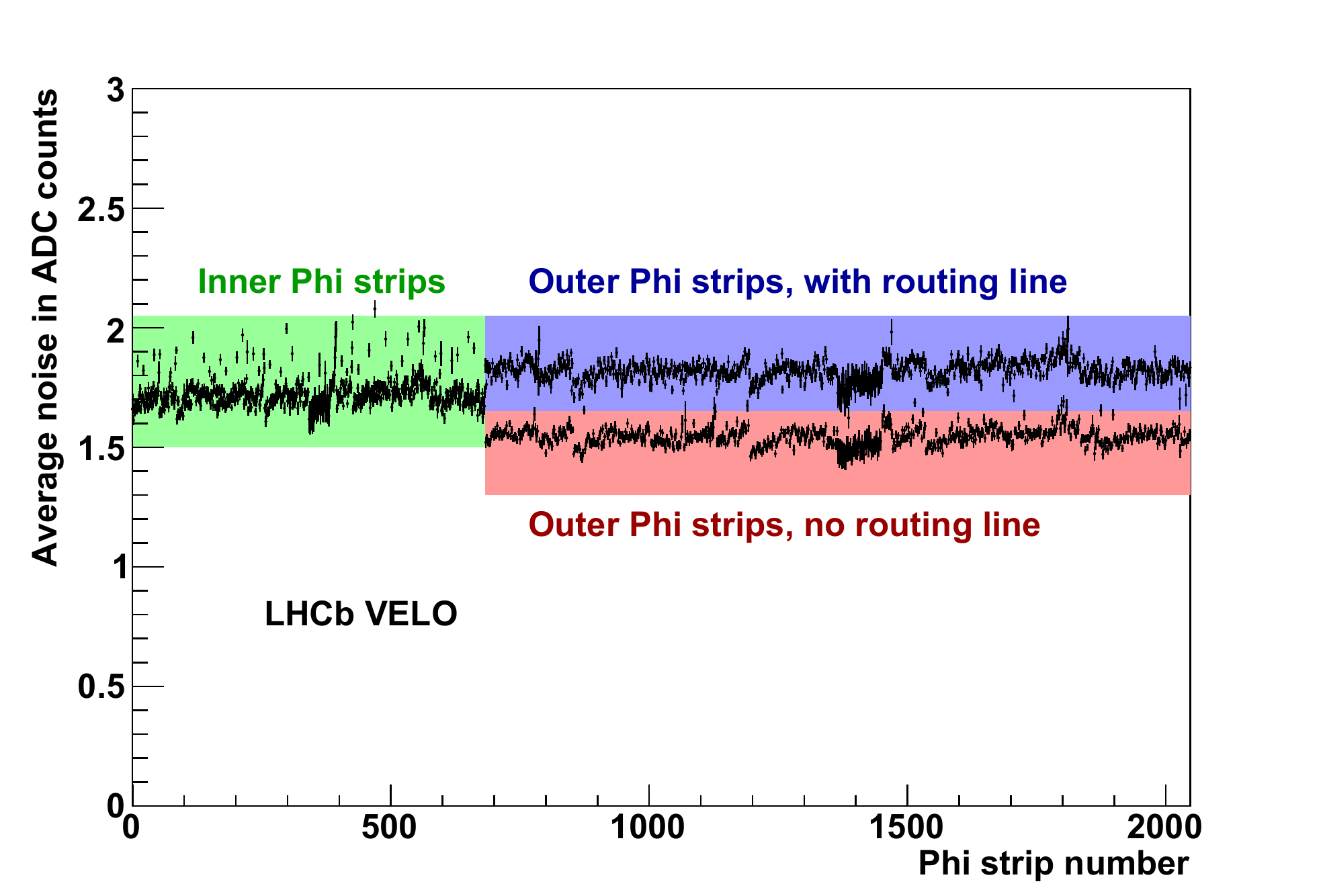}
    }
    \caption{\small Noise in ADC counts averaged across the 42 installed R
      (left) and \PhiSens (right) sensors, with the error bars indicating
      the RMS of the distribution.}
    \label{fig:Noise_StripAve}
  \end{center}
\end{figure}

The noise of the front-end ASIC depends on the strip capacitance. On the \RSens sensors the inner radius strips have the lowest capacitance due to their shorter length though this is partially compensated due to the inner strips having longer routing lines. Still, the dependence of the noise on the strip length is visible when comparing Fig.~\ref{fig:Noise_StripAve} (left) with the sensor layout shown in Fig~\ref{fig:veloSensors}. The \RSens sensor is divided into four approximately  $45\degrees$ segments, and the strip length increases with increasing strip number in each segment. The \PhiSens sensor has two
zones with inner and outer strips. The inner strips are shorter but have additional routing line contributions to their capacitance. In the outer zone every alternate strip is under the routing line for an inner strip so the capacitance for these strips is larger. The noise in these three types of \PhiSens sensor strips is shown in Fig.~\ref{fig:Noise_StripAve} (right). Larger noise is also clearly visible in both the \RSens and \PhiSens sensors every 32 channels, this is due to inter-symbol cross-talk from the digital header information into the first channel in each analogue readout link. A suppression algorithm for this inter-symbol cross-talk has been implemented in the FPGAs, but is not currently used due to the small size of this cross-talk and the large signal to noise ratio.

% , this is visible as higher noise in these channels.

\begin{figure}
  \begin{center}
    \resizebox{0.6\textwidth}{!}{
      \includegraphics{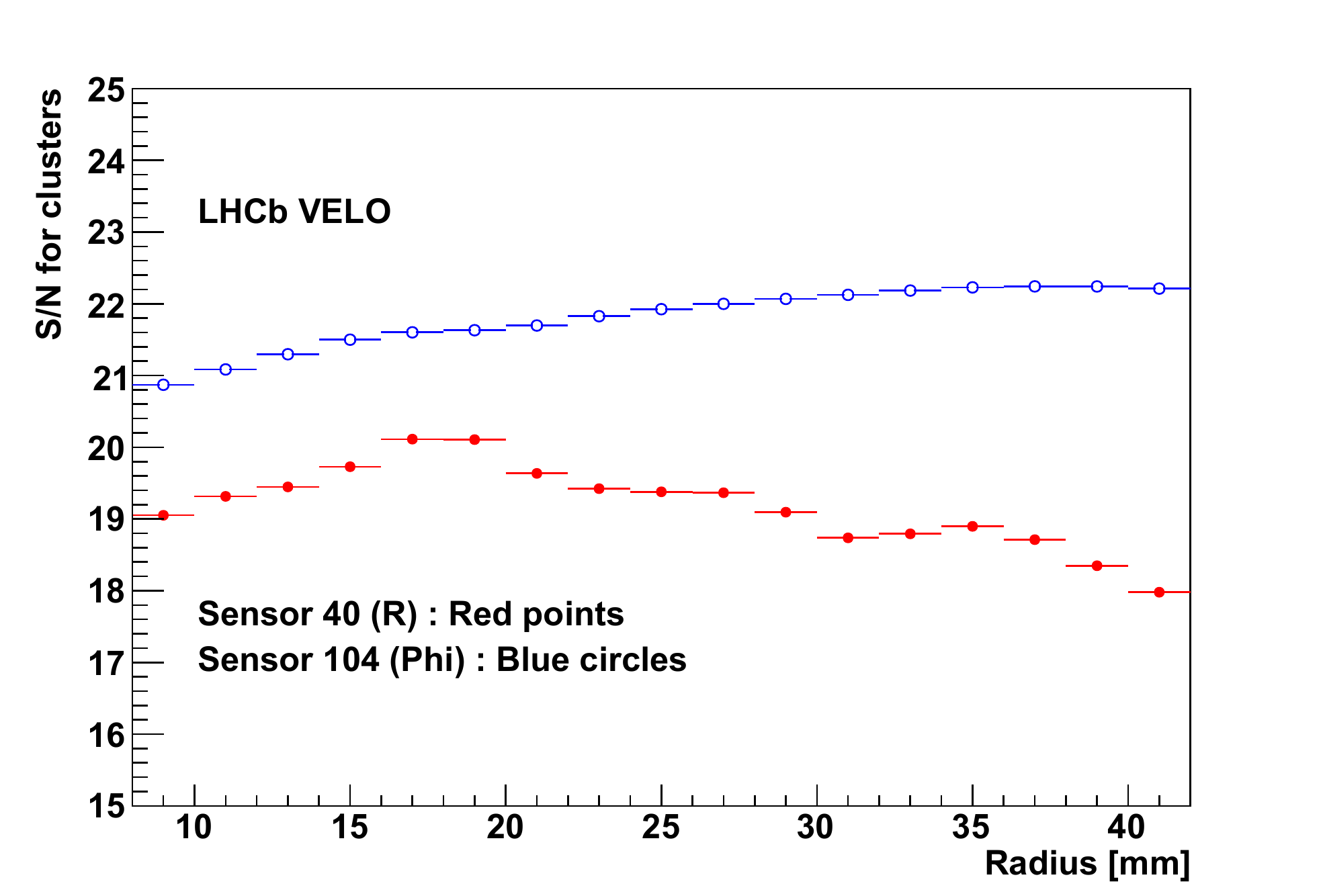}
    }
    \caption{\small Signal to noise (S/N) ratio from the MPV of the signal for
      single strip clusters on tracks divided by the noise of that strip.
      Shown are the S/N values for sensor 40 (R) and sensor 104, the
      \PhiSens sensor of the same module, as a function of impact point
      radius.}
    \label{fig:SN_ratio_sensor40}
  \end{center}
\end{figure}

The average signal to noise ratio, computed as the average MPV of single strip clusters divided by their strip noise, for the \velo is around 20:1. It is higher for the \PhiSens sensors than the \RSens sensors and shows a variation on the sensor radius as shown in Fig.~\ref{fig:SN_ratio_sensor40}. 

\subsection{Resolution}
\label{sec:Resolution}
%[Silvia - 3 pages]
% results
% show comparions with simulation
% comment on n-on-p sensor
% any change with rad. damage yet ?
% do we want to mention/explain the eta studies?
% For the results and simulation comparison we need the final numbers/plots. Work is on going

The hit resolution in silicon devices depends on the inter-strip readout pitch and the charge sharing between strips. The charge sharing varies with operational bias voltage and the projected angle of the track. The bias voltage was 150\unit{V} throughout the physics data taking in 2010--2012. The projected angle provides information on the number of strips that the particle crosses while it traverses the thickness of the silicon sensor. It is defined as the angle between the track and the perpendicular to the sensor, in the plane perpendicular to the
sensor and containing the perpendicular to the strip. Initially the resolution improves with increasing angle, due to the charge sharing between strips allowing more accurate interpolation of the hit position. The optimal resolution is obtained when the tracks cross the width of one strip
when traversing the 300\mum thickness of the sensor. For the \velo
the optimal projected angle varies between about 7\degrees at the lowest inter-strip pitch of 40\mum, to about 18\degrees for the
largest 100\mum pitch strips.  Above the optimal angle the resolution begins to deteriorate due to the fluctuations in the charge on the strips and because
the signal to noise ratio on individual strips may drop below the clustering threshold.

The clustering algorithm and charge interpolation method is described in Sect.~\ref{sec:TELL1}. The \velo reads out analogue
pulse-height information from the strips, and this information is used offline to calculate the cluster position using the weighted average of the strip ADC values. Including the track angle dependence in the clustering algorithm is found to give a small improvement in precision. The results presented here rely on the offline recalculation of the position, while the trigger relies on the lower resolution (3-bits) calculation (see Sect.~\ref{sec:zs}). The estimated resolution in the simulation is parameterised and fitted as a function of both track angle and strip pitch. This resolution estimate for each hit is then used in the Kalman fit tracking algorithm.

%The hit resolution in silicon devices mainly depends on the strip pitch and the projected
%angle\footnote{The projected angle is the angle between the track
%and the strip in the plane perpendicular to the sensor.} of the track producing the hit.  
%This angle strongly affects the charge sharing between the neighbouring strips.

%The resolution improves with increasing this angle because the charge spreads over several strips
%and reaching the optimal resolution when the tracks cross the width of one strip when traversing 
%the sensor.
%Above the optimal angle the resolution begins to deteriorate because
%the signal over noise level on individual strips may drop below threshold  
%due to the large numbers of strips.
%For the \\velo the optimal projected angle varies between 
%about 7\degrees for 40~\mum pitch strips and about 18\degrees for 100~\mum pitch strips.

The hit resolution is determined from the hit residuals which are evaluated using the LHCb  Kalman filter track fit \cite{Fruhwirth:1987fm} and include a correction for
multiple scattering and energy loss dependent on the track momentum. The residual is defined by the distance between the hit measurement and the extrapolated point of the fitted track to that sensor. As the hit for which the residual is being determined is included in the track fit this gives rise to a bias in the residual which must be corrected for.
The bias correction used to determine the residual is $\sqrt{V_M/V_R}$  \cite{Hulsbergen:2008yv}
 where $V_M$ is the variance of the measurement and $V_R$ is the variance of the residual. The evaluation of this correction is implemented in the Kalman fit.

The resolution has been determined as a function of the strip pitch and of the projected
angle. For each bin, the resolution has been determined from the sigma of the fit of a Gaussian function to the distribution of the corrected residuals. The resolution is evaluated using tracks that have hits in the tracking stations behind the magnet and hence for which the momentum measurement is available. The tracks are required to have a momentum greater than 10\gevc to reduce the dependence on the estimation of the multiple scattering effect, and a number of other track quality criteria are applied to reject fake tracks. The results are presented here for the \RSens sensor. The \PhiSens sensor results are compatible but the almost radial geometry of the strips means that tracks primarily have small projected angles.  

% and a set of quality cuts  

%The following track selection is applied:
%\begin{itemize}
%\item
%$\chi^2_{track}/ndof$ $< 5$ to reject the ghosts track and bad quality track. Only 0.2 per mille of tracks are rejected with this cut.
%\item
%Hits number between 10 and 30 (corresponding to 5 and 15 space points) to exclude ghost tracks and halo particles affected by a large multiple scattering effect.
%\item 
%Minimum momentum of 10 \gevc to reject tracks with a not negligible multiple scattering effect.
%\item
%Maximum momentum of 5 \tevc to avoid unreal tracks. 
%\end{itemize}

\begin{figure}
  \begin{center}
    \resizebox{\textwidth}{!}{
      \includegraphics{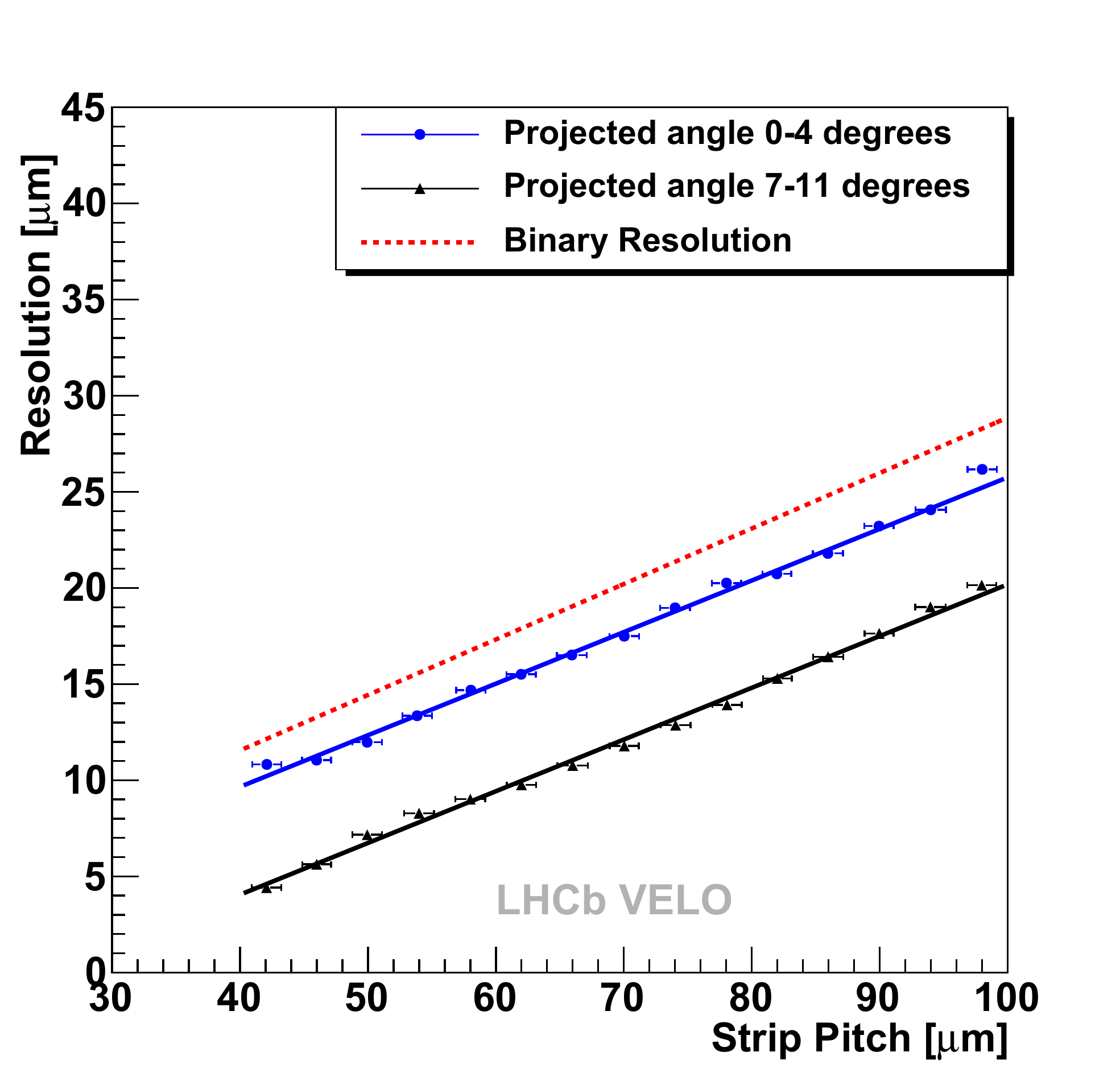}
      \includegraphics{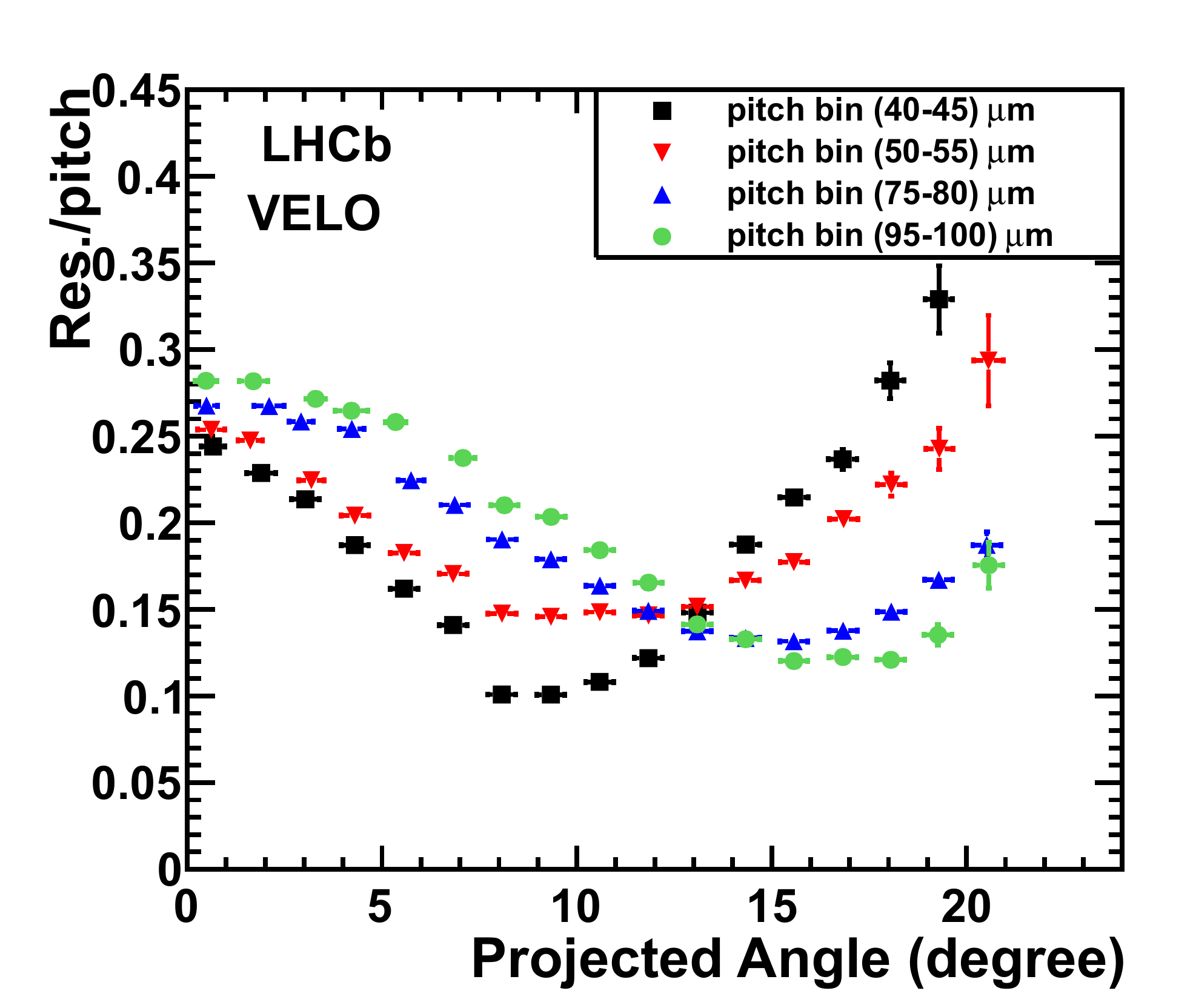}
    }
    \caption{(left) The \velo resolution for two projected angle bins for the $R$ sensors as a function of the
      readout pitch compared with binary resolution. (right) Resolution divided by pitch as function of the
      track projected angle for four different strip pitches.}
    \label{figRes}
  \end{center}
\end{figure}

\begin{figure}
  \begin{center}
    \resizebox{\textwidth}{!}{
      \includegraphics{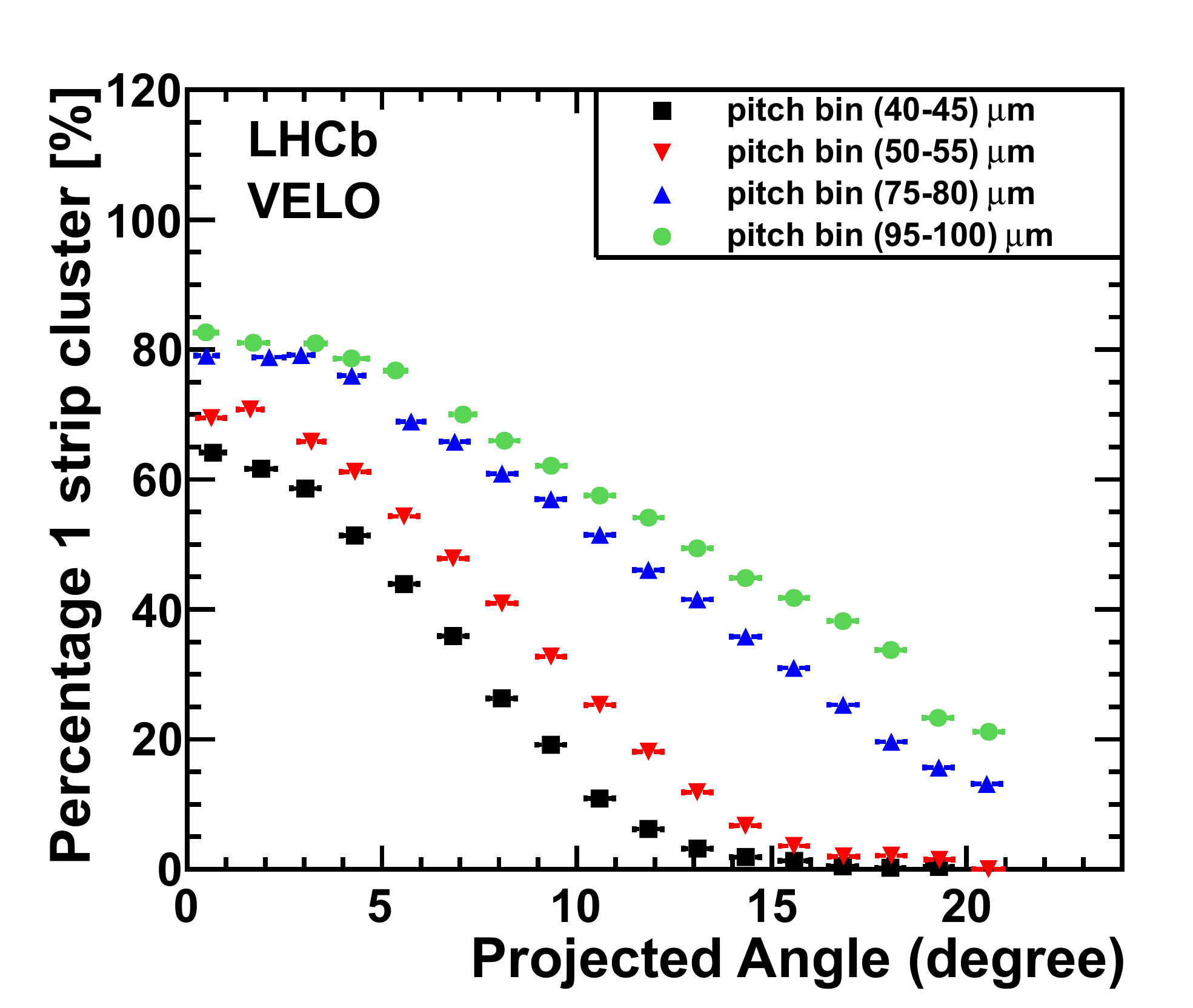}
      \includegraphics{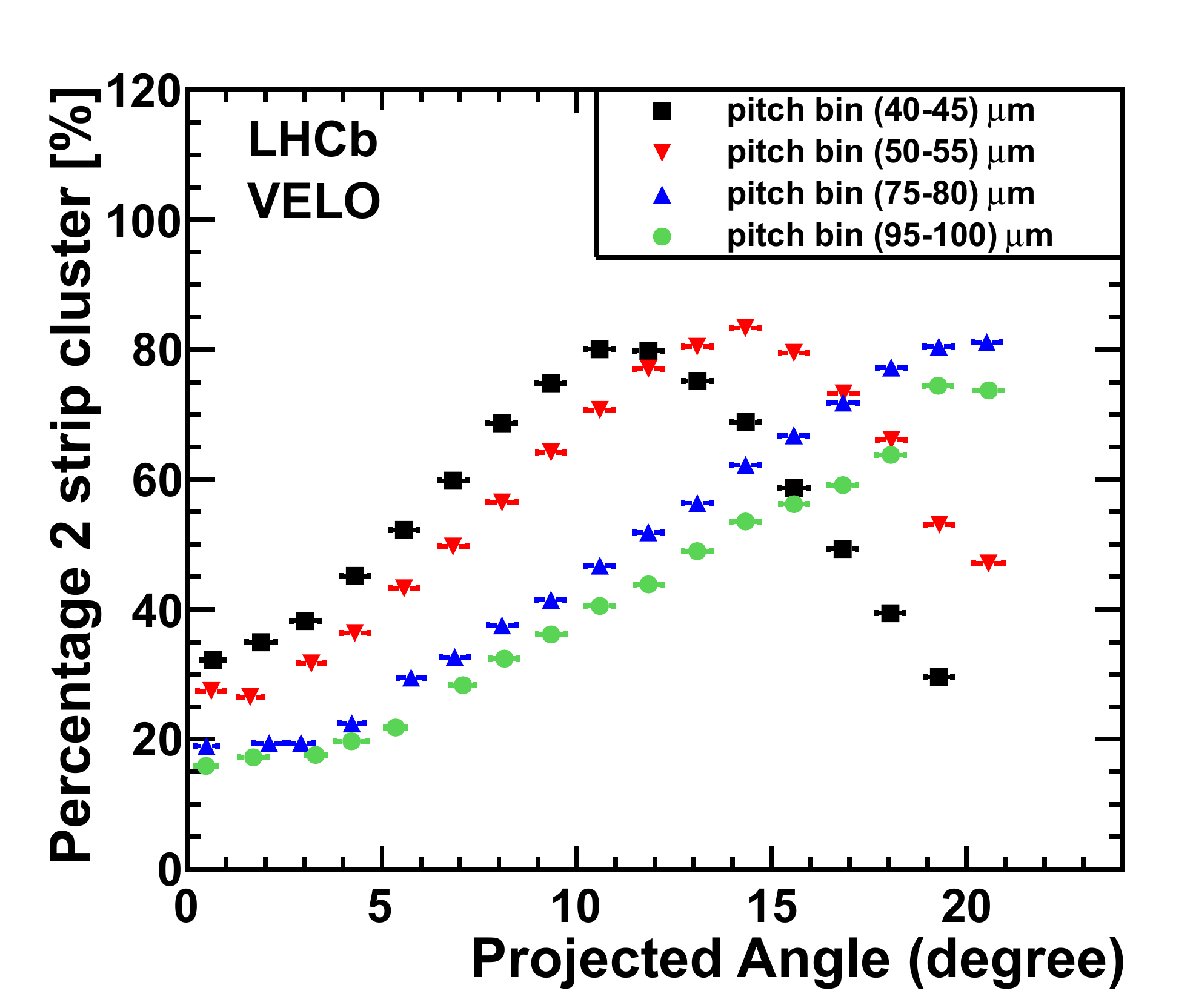}
    }
    \caption{\small The percentage of one (left) and two (right) strip clusters
      as a function of the track projected angle for four different strip pitches.}
    \label{figchargesharing}
  \end{center}
\end{figure}

%this should be re-written as soon as we have the final numbers and plots
The measured hit resolution has a linear dependence on the strip pitch in projected angle bins, as shown in Fig.~\ref{figRes} (left). 
The hit resolution at small projected angles, almost perpendicular to the sensor, has  a resolution which is close to that which would be obtained from a binary system. This is to be expected as the charge sharing between strips at this angle is minimal. A significantly better resolution is
obtained for larger projected angles, where the fraction of two strip clusters increases and the analogue readout of the pulse height in each strip is of benefit.  The hit resolution as function of the projected angle is shown in Fig.~\ref{figRes} (right) and the fraction of one and two strip clusters as a function of the projected angle and strip pitch 
are shown in Fig.~\ref{figchargesharing}. The best hit precision measured is around 4\mum for an optimal projected angle of 8\degrees and the minimum pitch of 40\mum.

%The resolution has been extracted using a simple weighted pulse height 
%algorithm for the reconstruction of the cluster position. Additional development 
%of the clustering algorithm is expected to further improve the precision.

% section written for LHCb performance paper
%The hit resolution in silicon devices depends on the inter-strip readout pitch and the charge sharing between strips. The charge sharing varies with operational bias voltage, currently $150~V$, and the projected angle of the track. The projected angle is the angle between the track
%and the perpendicular to the sensor, in the plane perpendicular to the
%sensor and containing the perpendicular to the strip. The optimal
%resolution is obtained when the tracks cross the width of one strip
%when traversing the 300~\mum thickness of the sensor. For the \velo
%the optimal projected angle varies between about 7\degrees at the
%lowest inter-strip pitch of 40~\mum, and about 18\degrees for the
%largest 100~\mum pitch strips. The \velo reads out analogue
%pulse-height information from the strips, and this information is used
%offline to perform a weighted average and improve the hit resolution. The resolution of the sensors
%is determined from the residual between the extrapolated position of
%the fitted track and the measured hit position. The use of the
%evaluated hit position in the track fit gives rise to a bias in the
%residual, for which a correction is applied. The resolution has been
%determined as a function of the strip pitch and of the projected angle using high momentum tracks and is shown in in Fig.~\ref{figVELORes}. The hit resolution has a linear dependence with the strip pitch in
%projected angle bins, and the best hit precision is $\sim 4~\mum$ at an optimal angle of 8\degrees and a pitch of 40~\mum.

\subsection{Occupancy}
\label{sec:Occupancy}
%[Kurt - 1 page]

%[Kurt computing occupancy in events passing HLT]

\begin{figure}[tp!]
  \begin{center}
  \resizebox{0.9\textwidth}{!}{
      \includegraphics{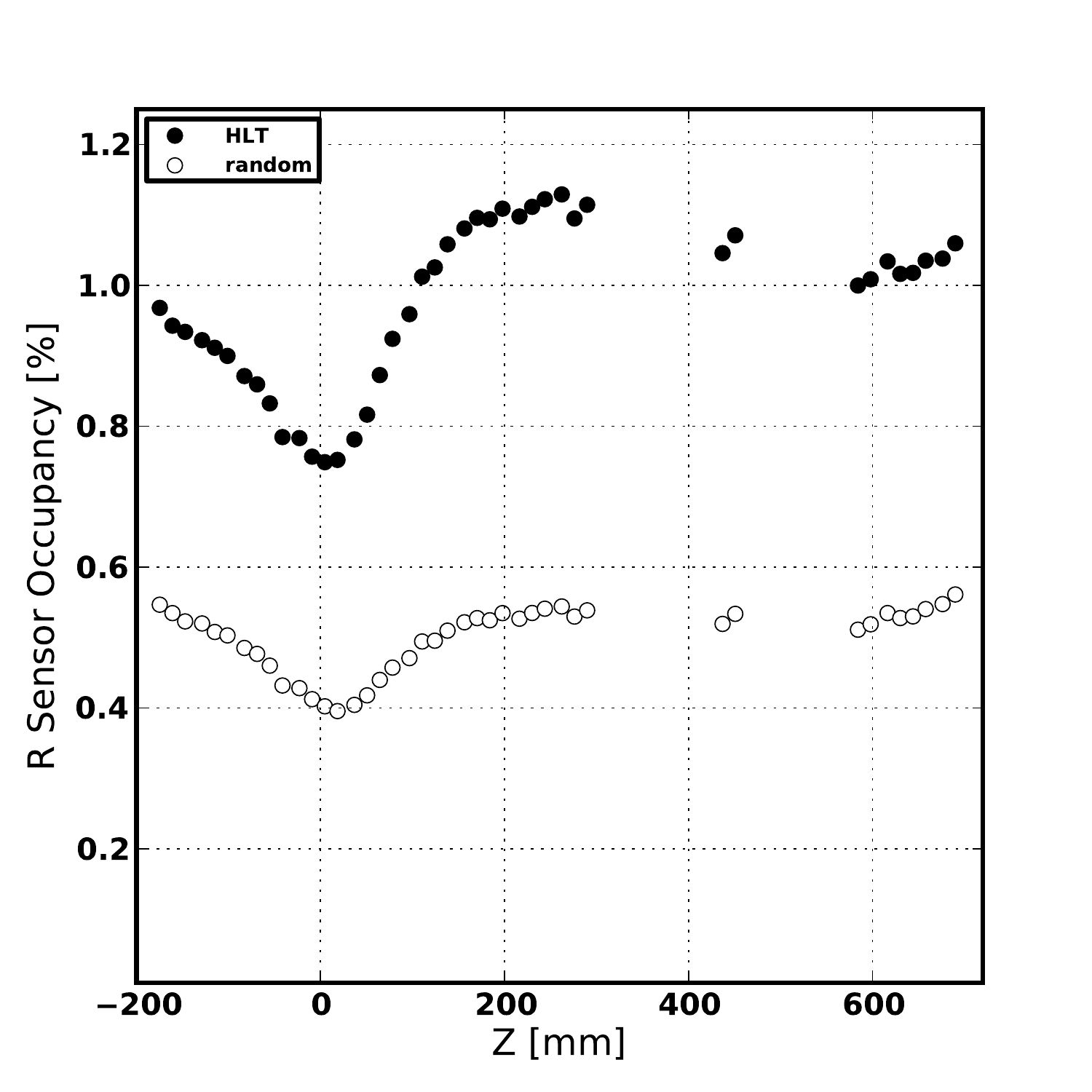}
       \includegraphics{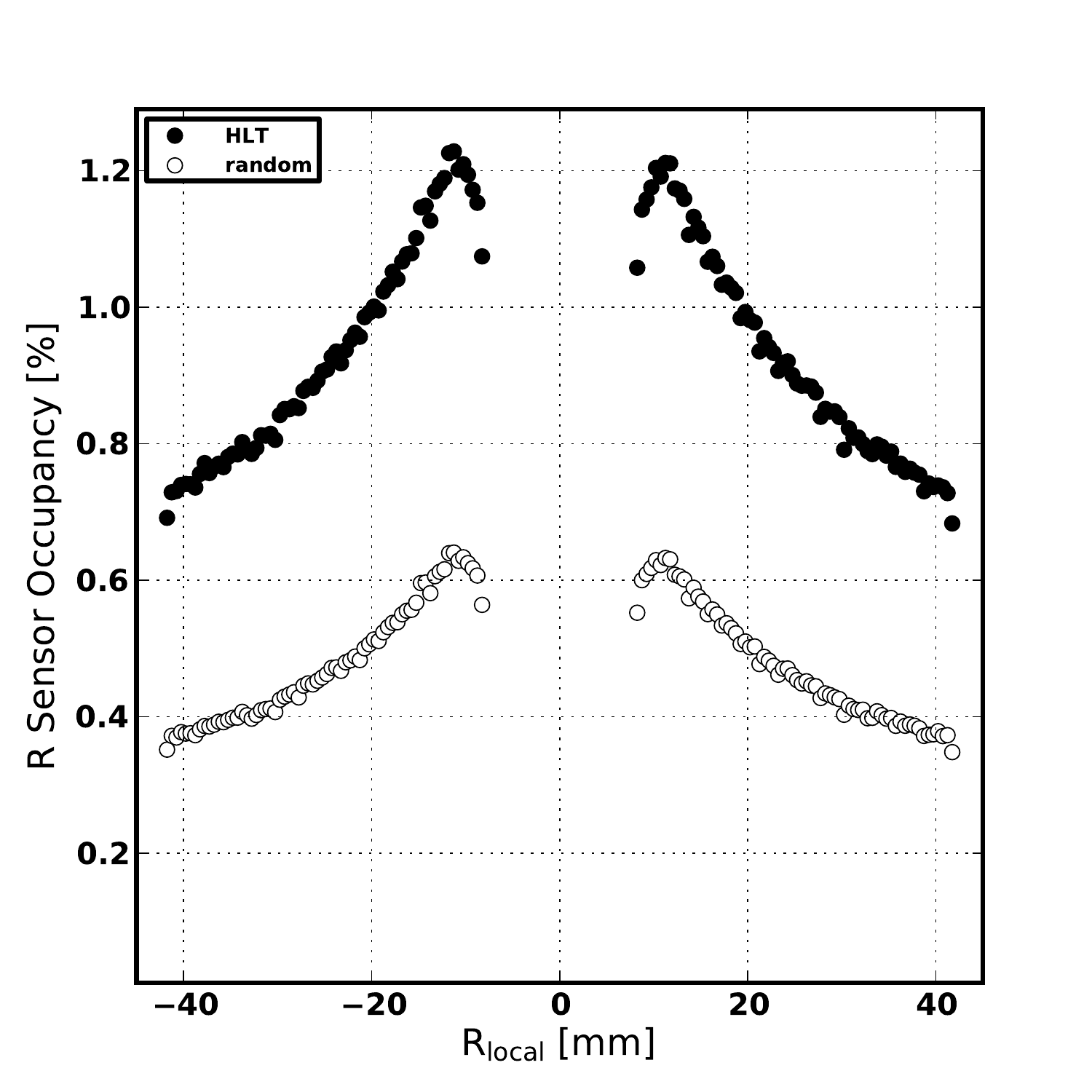}
    }  
   \resizebox{0.15\textwidth}{!}{LHCb VELO}
    \resizebox{0.9\textwidth}{!}{
      \includegraphics{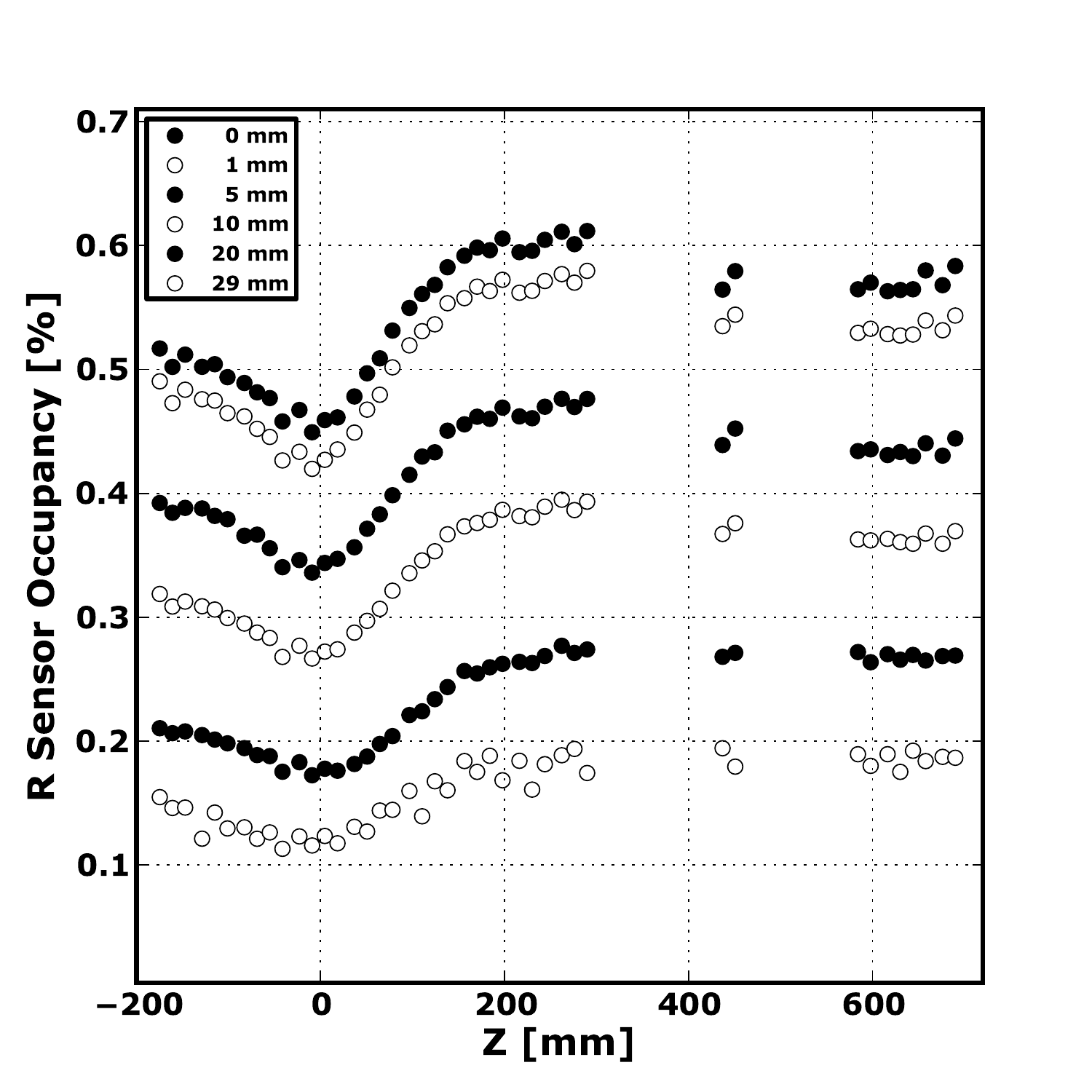}
       \includegraphics{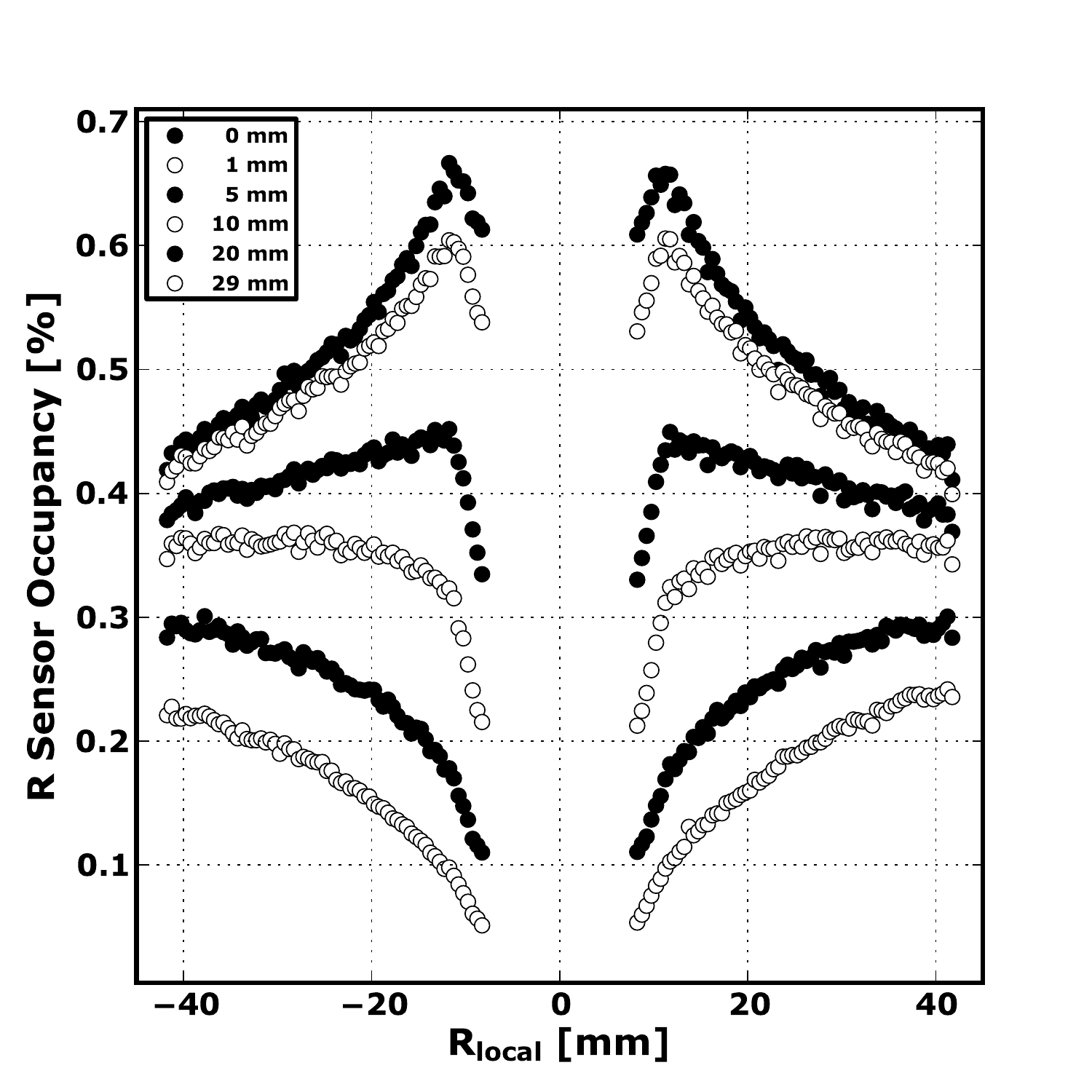}
    }
    \caption{\small Cluster occupancy in the \velo silicon strip detectors. (left plots)  Average occupancy in the sensor as a function of the position of the sensor along the beam-line. (right plots) Occupancy as a function of the local radius of strips on the \RSens sensors, with a negative sign applied to sensors on one half. The upper plots show the occupancy when fully closed using 2011 data with a $\mu$ of 1.7 and for events selected using a  random trigger or with events passing the high level trigger. The lower plots show the occupancy as a function of closing distance with fully closed labelled as 0\mm, the points below then follow in order of the retraction distance indicated in the key.}
    \label{fig:Occupancy}
  \end{center}
\end{figure}

The detector occupancy is a key parameter in the performance of the pattern recognition and tracking algorithms of the experiment.  High occupancy can lead to the mis-identification of hits on tracks and increase the number of hit combinations decreasing the speed of the algorithms. The occupancy shown here is for clusters. The cluster seeding thresholds and masking of noisy strips in the FPGA data processing algorithms ensure that the contribution to the occupancy from noise is negligible compared with that from particles; in the absence of circulating beams the observed occupancy is below 0.01\%. The typical cluster occupancy during 2011 operations is shown in Fig.~\ref{fig:Occupancy} (top plots). Only events from particle beam crossings are utilised in the computation, and this data sample has an average number of  visible interactions per beam crossing, $\mu$, of 1.7. The occupancy is shown computed with data collected using two different triggers. Data from a random trigger on beam crossings are used as this represents the average occupancy in the events observed by the detector. The occupancy for events passing the high level trigger is also given, this is higher as events with heavy flavour production are typically of higher multiplicity than the average. The distribution for events passing the high level trigger is not fully symmetric around the collision point due to the preference of selecting events in the \lhcb acceptance.

 %The occupancy shown here is for clusters, the number of strips hit is higher than this value as some particles leave hits in multiple strips in a single sensor (see Fig.~\ref{figchargesharing}).

The cluster occupancy has a dependence on the position of the sensors along the beam-line, as shown in Fig.~\ref{fig:Occupancy} (left). The location of the interaction region is clearly visible from the dip in the occupancy  distribution. The highest occupancy of 1.1\% for events passing the trigger is at around the end of the closely spaced region of sensors in the \velo, where the occupancy is 44\% larger than at its minimum. 
%In the CALIB stream (random triggers): over all sensors the occupancy varies between 0.39% and 0.56% (taken from the average sensor histogram). Over r for R sensors it varies between 0.35% and 0.64%. The PU is excluded in both cases.
%The min/max occupancies for events passing the HLT are:
%all sensors   : 0.73%/1.13%
% so mid-point in random - 0.48%, and in hlt 0.93%

The occupancy also varies across the sensor, increasing closer to the beam as seen in Fig.~\ref{fig:Occupancy} (top right). The maximum occupancy is 83\% higher than the minimum as a function of \RSens. This is a much weaker dependence than might naively be expected since the inner strips are five times closer to the beam than the outer strips. However, the semi-circular sensor geometry means that the strip length naturally reduces with decreasing radius on the \RSens sensors, and the inter-strip pitch also reduces with decreasing radius on both the \RSens and \PhiSens sensors. The occupancy reduces in the smallest radius region on the \RSens sensors as these strips are shorter due to having the corners cut off. 

Figure~\ref{fig:Occupancy} (lower plots) also shows how the occupancy varies across the sensor at a range of closing distances. These data do not have a well defined $\mu$ as the luminosity levelling (see Ref.~\cite{Evian2012}) had not yet been performed.  The minimum average sensor occupancy drops from around 0.4\% when closed to around 0.1\% when the \velo sensors are fully retracted by 29\mm from the beam. Even in the fully retracted position the number of particles that can be reconstructed in the detector allows the performance of the system to be determined and the primary vertex to be fitted to allow the detector to be safely closed (see Sect.~\ref{sec:Motion}).

\subsection{Beam backgrounds and high multiplicity events}
\label{sec:LargeEvents}

%[Marius - 0.5 page]

%[Should add splash event information to this]

%[anything from machine beam background to add ?]

%It was found that some events
%have a very high cluster multiplicity, greater than what would be expected in
%normal proton-proton interactions. 
%Some ``large'' events appear to come from interactions of beam protons
%with residual gas in the \lhc vacuum chambers outside of \lhcb (beam-gas 
%interactions). Most of the large events, however, did not have a very
%high cluster multiplicity in the other \lhcb subdetectors, and their origin is 
%not yet known. Events with very many \velo clusters
%were analysed, and the readout of such events produced data
%of poor quality, dominated by features which are clearly not due to physical
%particles. 

%Events with cluster multiplicity greater than 6000 occur at a rate of
%$10^{-4}$ relative to the total rate of collision events, and
%this has been reasonably stable over time. At this low
%rate, the large events do not degrade the physics performance of
%LHCb.

%There are ongoing studies of the machine-induced background in LHCb, including
%particle showers from beam-gas interactions. The analysis of \velo clusters is
%an important part of these studies, because the \velo sensors are very close to 
%the beam, and the \velo is most affected by machine-induced background.
%Particles from beam-gas interactions outside the \velo do not
%contribute significantly to the radiation dose received by the \velo.

Studies have been performed of beam-related backgrounds in the \velo, and a number of sources of hits have been identified: beam-gas interactions; instrumental effects triggered by the presence of beam; and beam interactions with collimators. The rate of these effects is sufficiently low to not have a significant detrimental effect on physics or to contribute significantly to radiation damage in the detector.

\begin{figure}[tb]
  \begin{center}
    \resizebox{\textwidth}{!}{
    \includegraphics[width=0.49\textwidth]{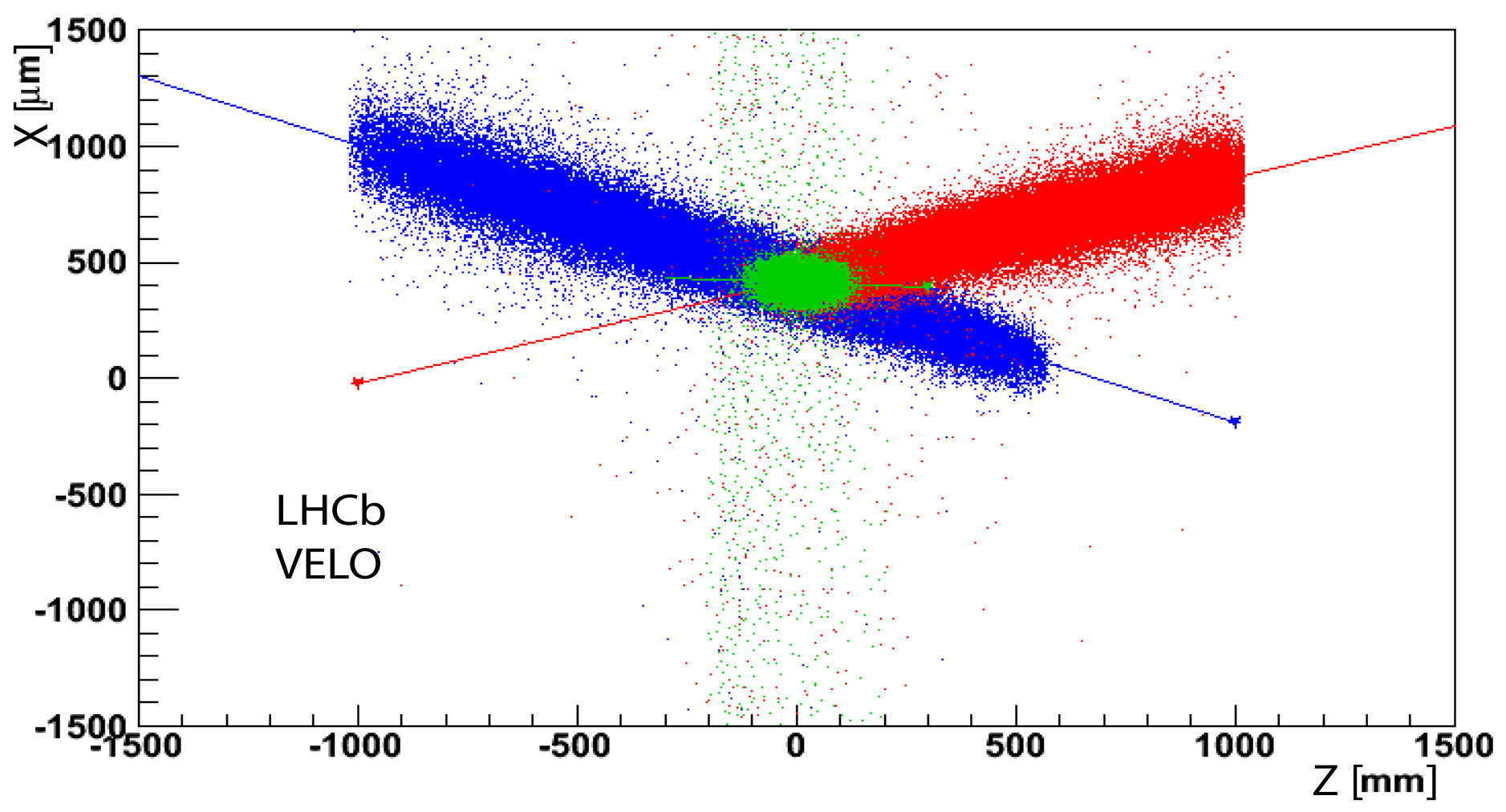}
     \includegraphics[width=0.50\textwidth]{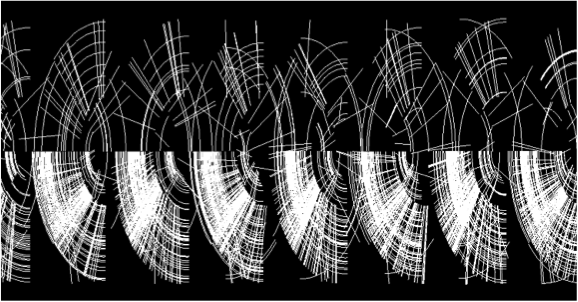}
    }
    \caption{\small (left) Imaging of the LHC beams through the reconstruction of the production vertex of tracks in 2010. The vertices from beam-gas interactions with the beam travelling in the $+z$ ($-z$) direction in LHCb are shown in blue (red). Vertices arising from beam-beam interactions are in green.  The vertical axis represents the horizontal direction ($x$) and the horizontal axis the beam direction ($z$).  (right) An event display image showing clusters in \RSens and \PhiSens sensors with a high occupancy in a localised region assumed to be due to a beam background splash event.}
    \label{fig:Background}
  \end{center}
\end{figure}

The interaction of beams with residual gas in the \lhc vacuum pipe and \velo vacuum vessel provides a useful data sample for alignment and luminosity studies. Simulations of beam-gas interactions in the \velo vacuum vessel and the long straight section of the LHC have been performed \cite{BackgroundRob}, with the latter including particle fluxes from beam-gas interactions in the long straight section and from proton interactions in collimators. Tracks arising from these beam-gas interactions provide a complementary sample to tracks from beam-collisions due to their very forward angular distribution. Many of these tracks pass through all \velo sensors and this provides useful additional constraints for alignment (see Sect.~\ref{sec:Alignment}). The reconstruction of the beam-gas interaction vertices in the \velo region is illustrated in Fig.~\ref{fig:Background} (left). The blue and red points are obtained from events in which only one beam passed through LHCb, and vertices with at least five tracks are reconstructed from interactions of the beam with residual gas. The horizontal crossing angle of the beams is apparent. The downstream beam (red) interactions are limited in their extent in the negative direction along the beam-line by the predominantly forward acceptance of \lhcb. 
This reconstruction of beam-gas interactions allows a measurement of the transverse bunch profile along the beam trajectory to be obtained and allows the beam angles, profiles and relative positions to be determined. The beam-overlap integral can then be obtained from the extracted beam profiles and has been used to obtain a measurement of the luminosity \cite{lumi}. The precision of the \velo allows this luminosity determination to be performed with a comparable precision to that obtained from the well-known van der Meer scan method. The rate of beam-gas interactions can also be increased in the \velo by the use of a gas injection system, which was commissioned in 2011 \cite{Barschel:1693671}.

Events containing a collimated spray of background hits are also observed, as shown in Fig.~\ref{fig:Background} (right). These events are characterised by high occupancy in local regions of the sensor, and they are correlated with activity in other tracking detectors. The rate of events is proportional to the beam intensity in the LHC, and typically constituted 0.03\% of events passing the high level trigger in 2011. A possible origin for these events is beam interactions with collimators. 
%[Rob to complete].

\subsection{Efficiency and faulty channel analysis}

Studies have been performed of the charge collection efficiency and cluster finding efficiency of the detector. The detector is initially operated at a bias voltage sufficient to obtain the full charge from the sensors, and the charge collection efficiency and cluster finding efficiency as a function of radiation are reported in Sect.~\ref{sec:Radiation}. In addition faulty channels have been identified through the study of the cluster finding efficiency, occupancy and noise spectrum.

%The charge collection efficiency and cluster finding efficiency measurements rely on a track extrapolation technique. 

%Collecting the charge on the test sensor around the intercept point allows the charge collection efficiency to be determined. The collected charge is fitted with a Landau convoluted with a Gaussian to obtain the most probable value of the Landau distribution. Non-zero suppressed data is collected in the test sensor, so that all charge is obtained, even if the signal is below the normal cluster threshold.

%\subsubsection{Cluster Finding Efficiency}
\label{sec:CFE}

The cluster finding efficiency is a useful measure of the performance of the detector. This is determined by excluding a sensor in the pattern recognition, and interpolating tracks to this sensor. The tracks are required to have hits in both the \RSens and \PhiSens sensors in the two modules before the test sensor and in the two modules after the test sensor. These requirements place some restrictions on the sensors that can be probed and the regions analysed within the studied sensors. Tracks that have intercept points inside the active region of the sensor are then considered.  Track quality selection cuts are applied to remove fake tracks and track isolation cuts to prevent a cluster from another track being selected. The efficiency with which clusters are found near the track intercept points is then determined.  The efficiency for finding a cluster depends on the applied bias voltage and the thresholds applied in the cluster making algorithms. The standard operational settings are used for the studies presented here.

%Clusters are used to represent the area that a particle is
%determined to have passed through a silicon sensor. A cluster may contain
%charge from several strips, where it is assumed that the charge on the strips
%was produced by a single particle. The method described in section
%\ref{sec:RDS_CCE} was developed to investigate how the most probable value
%of the charge distributed by a particle varies with sensor bias voltage. This
%method has also been used to measure the cluster finding efficiency (CFE) of a
%\emph{test} sensor. Particle tracks reconstructed using clusters on sensors
%to either side of the \emph{test} sensor are extrapolated to a position on the
%\emph{test} sensor. If a cluster is located at this extrapolated position then it
%is assumed to be associated to the particle responsible for the original track.
%The fraction of tracks for which a cluster is found is defined at the CFE.

%Cluster formation depends on the thresholds defined in the cluster making
%algorithms, which in turn are determined by the experimental signal to noise tolerance.

%Various track quality selection cuts are applied to reduce the inefficiency
%attributed due to poor track extrapolation. 

\begin{figure}
  \begin{center}
    \resizebox{\textwidth}{!}{
      \includegraphics{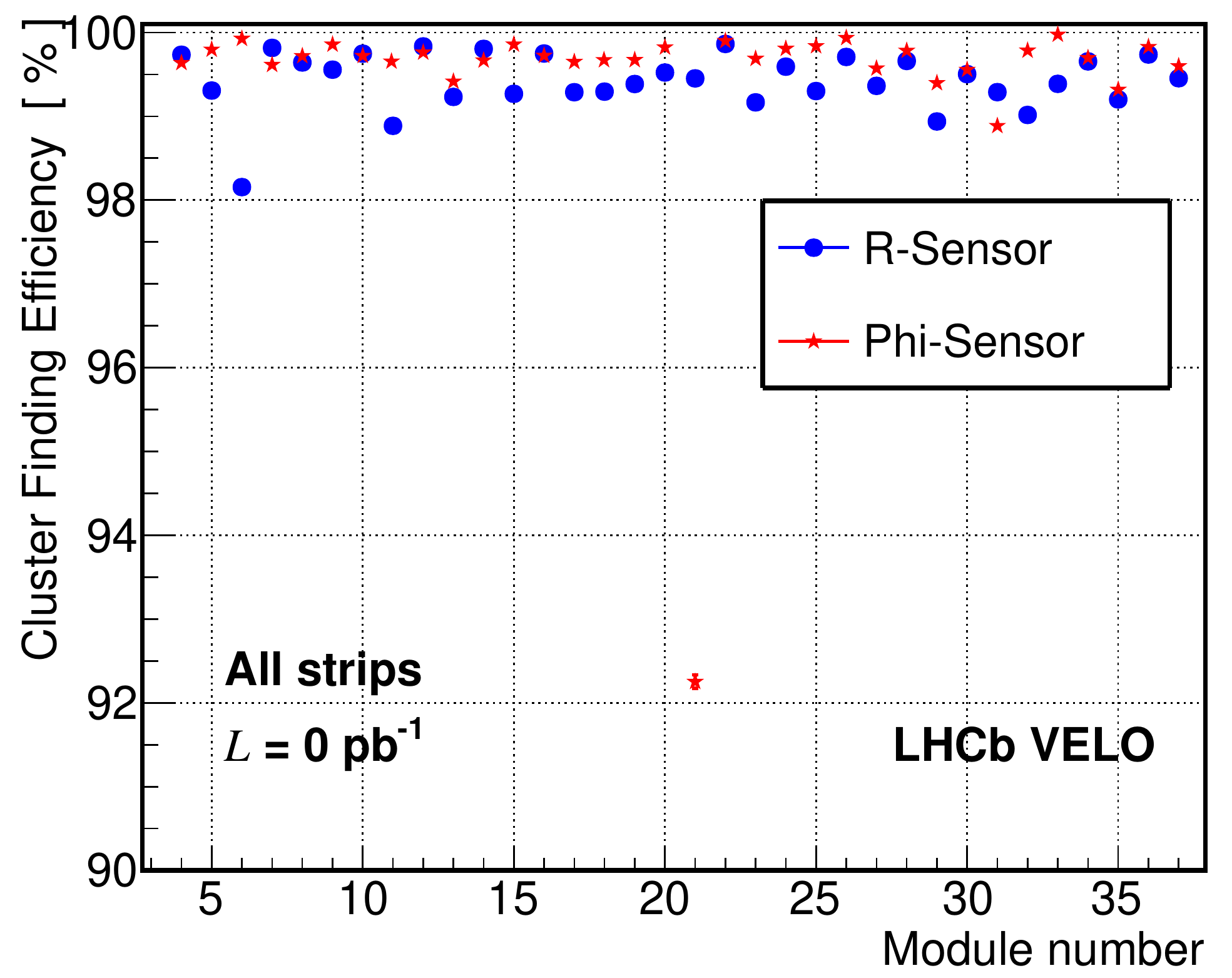}
      \includegraphics{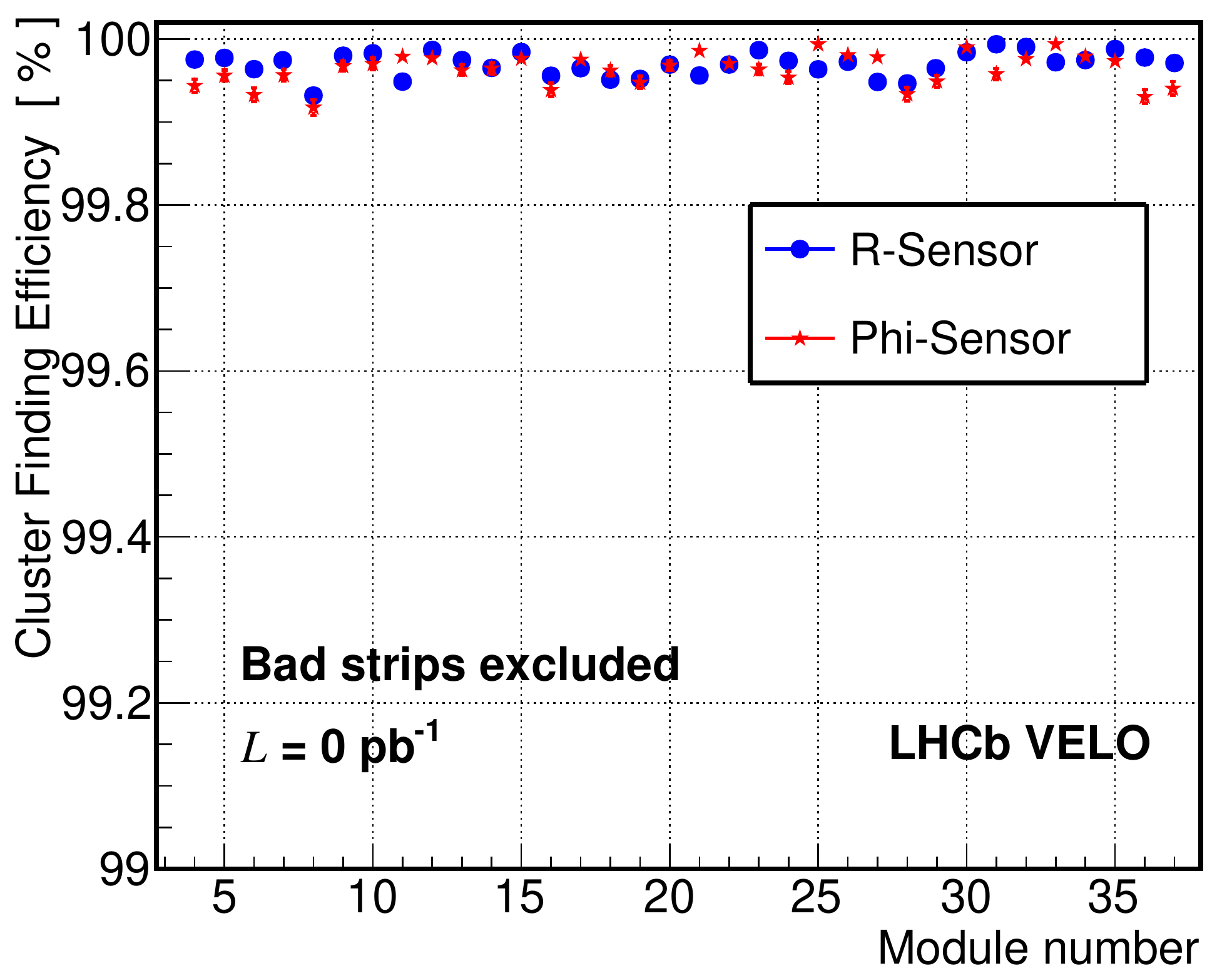}
    }
    \caption{\small The cluster finding efficiency for each sensor with
    all strips (left) and when identified bad strips are
    excluded (right).
    }
    \label{fig:CFE}
  \end{center}
\end{figure}

The region boundary between the inner and outer strips of the \PhiSens sensor and the
middle region boundary of the \RSens sensor both contain a 79\mum wide region with the HV bias resistors. This region and the 
region of the guard ring around the sensor are excluded in the efficiency calculation.  Averaging across sensors, cluster finding efficiencies of $99.45\%$ sensors were obtained at the start of operation as shown in Fig.~\ref{fig:CFE}.  Bad strips are identified with the cluster finding efficiency analysis described below. When these bad strips are neglected the cluster finding efficiency rises to $99.97\%$. The system contains only a single front-end ASIC that is not functioning, the failure of which occurred after production and before the start of physics operations, this is visible in the reduced efficiency of the \PhiSens sensor of module~21 in Fig.~\ref{fig:CFE} (left).

During the production of the modules faulty channels were identified through the application of three techniques: high resolution visual microscope inspections,  the response of the modules to scans of a laser,  and the analysis of noise data. A total of only 0.6\% of the \velo sensor strips were found to be faulty, see Table~\ref{tab:badStrips}. The analysis of noise data can also be performed on the installed experiment. The measured noise on a functioning channel will decrease once it is biased above its depletion voltage, since its capacitance will decrease: unbonded strips or faulty ASIC channels will not display this dependence. However, it was found that the tuning of the selection was highly detector and condition sensitive and hence this method of faulty channel identification is less accurate for repeated semi-automatic analysis than those discussed below.

\begin{table}
\centering
\caption{\small Fraction of faulty strips classified as dead or noisy. The results are obtained from the methods used at production, occupancy spectrum studies, and a cluster finding efficiency analysis.  Values are given at the time of production or  start of operations in 2010, at the end of 2011, and at the end of 2012.}
\begin{tabular}{l|c|c|c}
\hline
Delivered Int. Luminosity  & 0~\invfb & 1.2~\invfb & 3.3~\invfb\\
\hline
Dead Strips &  &  \\
\hline
Production Tests & 0.6\% %0.55\%  % 955 strips 
& -- &-- \\
Occupancy & 0.7\% %0.69 \% 
& 0.7\% & 0.6\% \\
Cluster Finding Efficiency & 0.8\%  %0.77\% %1227 strips 
& 1.1\% & -- % 1468 strips 
\\
\hline
Noisy Strips &  &  \\
\hline
Production Tests & 0.02\% % 36 strips 
& -- \\
Occupancy & 0.01\% & 0.02\% & 0.02\% \\
\hline
\end{tabular}
\label{tab:badStrips}
\end{table}

\begin{figure}[tb]
  \begin{center}
    \resizebox{\textwidth}{!}{
      \includegraphics{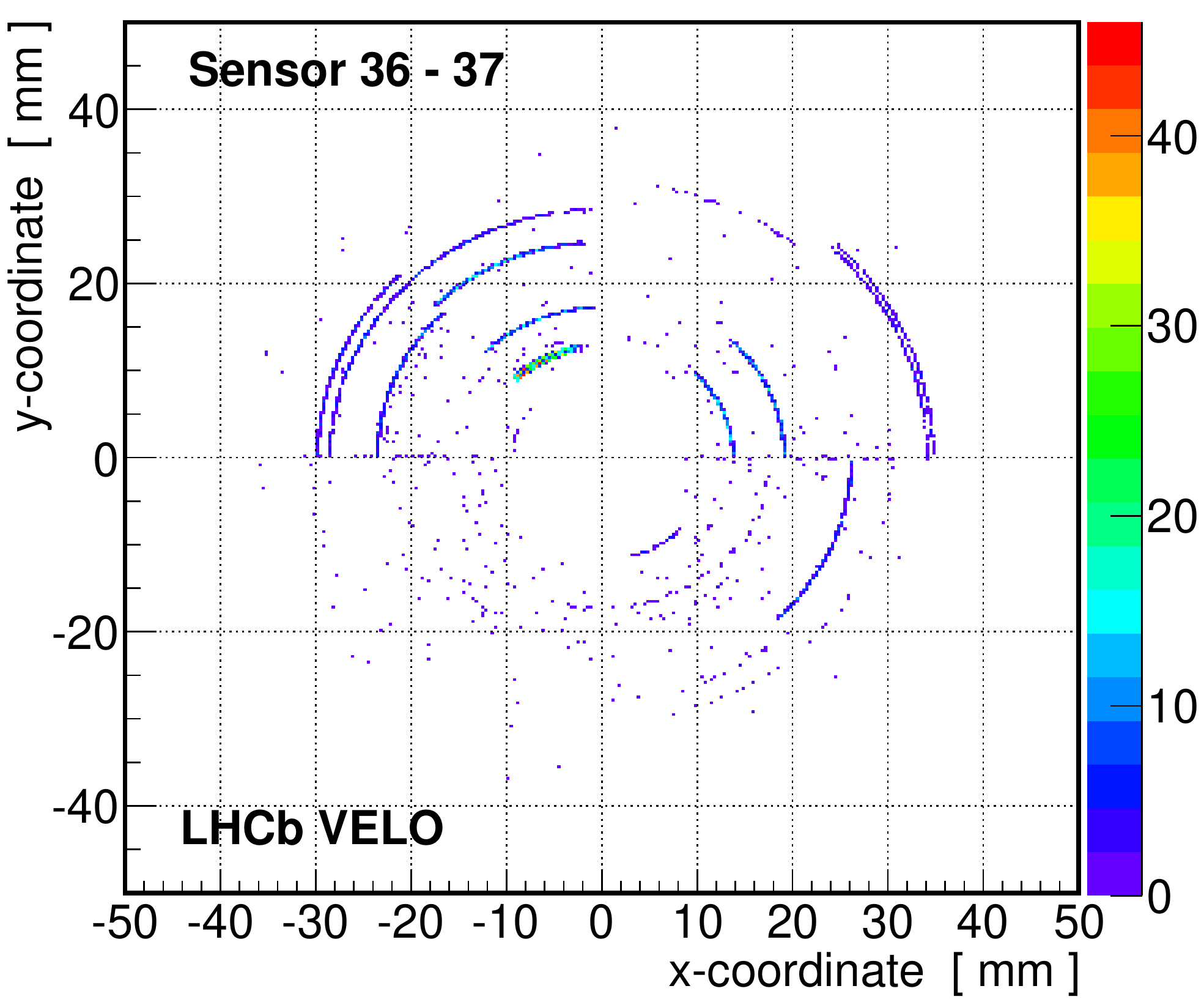}
      \includegraphics{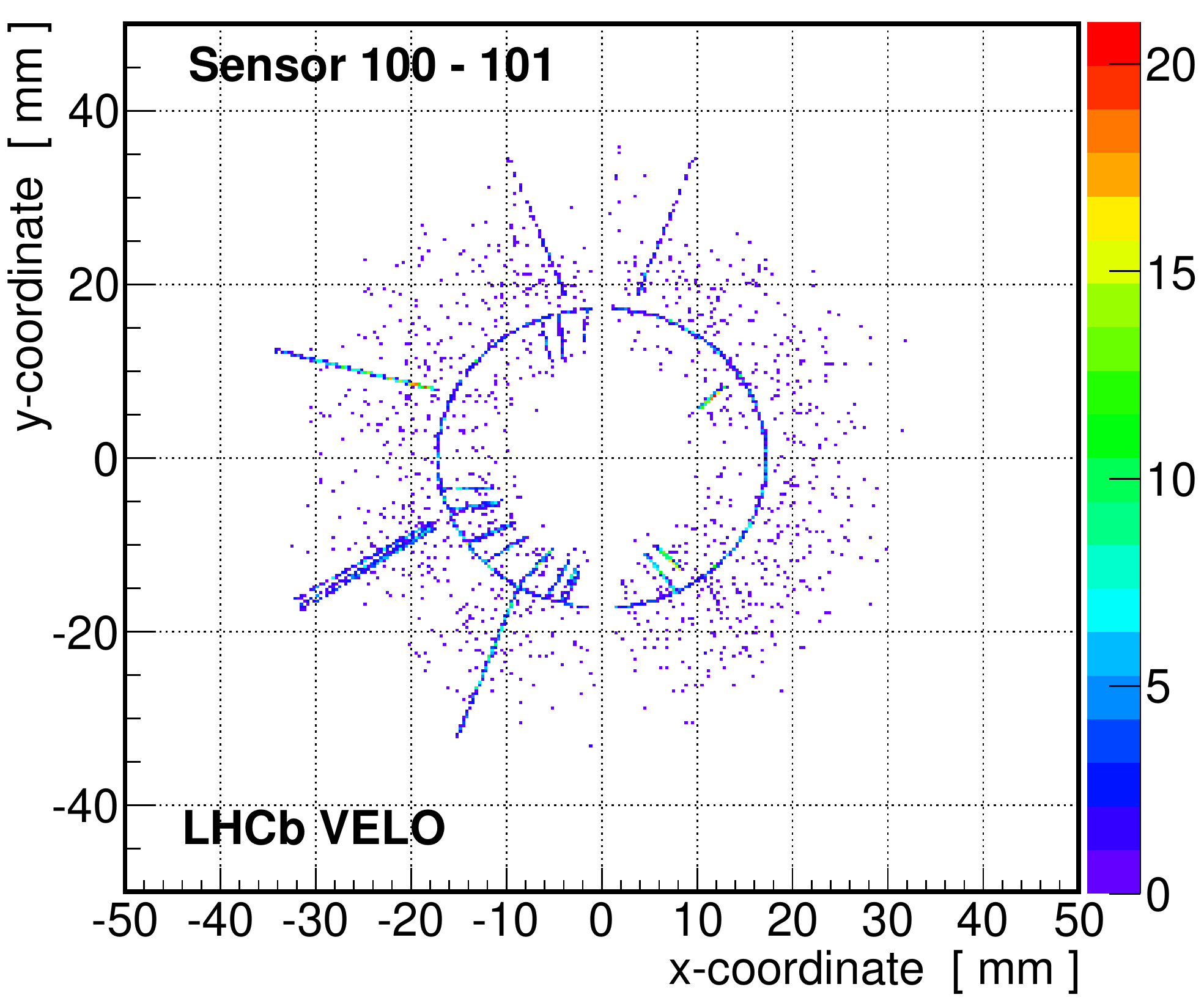}
    }
    \caption{\small The intercept position of an extrapolated track 
    when a cluster is not found for two example \RSens (left) and
    \PhiSens (right) sensors. The bad channels are clearly visible.}
    \label{fig:BadStrips_visual}
  \end{center}
\end{figure}

The cluster finding efficiency analysis can also be used to determine the number of dead channels in the detector.  A track is extrapolated to the detector and clusters searched for in the channels around the intercept point. A channel is identified as dead if tracks extrapolated to this strip show four times more missed clusters than the mean of the $30$ nearest neighbours. Examples of the locations of strips in which clusters are not found are shown in Fig.~\ref{fig:BadStrips_visual}, which shows the boundary region on the \PhiSens sensor between the inner and outer strips and the location of bad strips. This method identifies 89\% of the channels found in the dead channel list made at production. Only 0.8\% of channels were identified as dead at the start of operations by the cluster finding analysis method. At the end of 2011, after two years of operation and a delivered fluence of 1.2\invfb, the fraction had increased to 1.1\% of channels.

Two classes of faulty strips, dead and noisy, are also identified through the analysis of the channel occupancy spectra (see Sect~\ref{sec:Occupancy}). 
 Faulty channels are usually masked in the FPGA data processing algorithms, and hence would not show up in the occupancy spectra. Therefore, the analysis is performed using NZS data and emulating the data processing, but without the faulty channel mask applied. Channels with less than $10\%$ of the average occupancy are identified as dead, and those with greater than three times the average occupancy as noisy. 
%Noisy strips are not masked in the TELL1 algorithms as  they contribute to an increase in the data size and create noise clusters which can be used in the tracking. 
The results are compared with those found at production and with the cluster finding efficiency method in Table~\ref{tab:badStrips}, and seen to be in good agreement. The level of dead and noisy strips is shown to have remained at a low level throughout the operations.

\subsection{Radiation damage studies}
\label{sec:Radiation}

%[This section will be moved to the overall performance section of the paper once that section is edited]

%[Henry - IV, Adam - CCE vs V, Noise vs V - Justin, total 3 pages] 

The proximity of the silicon sensors to the LHC beam results in a high particle fluence of up to $5 \times 10^{13}$~1\mev neutron equivalents / $\cm^2$ (\neutroneq) per \invfb of delivered integrated luminosity for the most irradiated sensor regions, varying by approximately a factor of two as a function of the position of the sensors along $z$. The \velo geometry, with the sensors placed perpendicular to the beam,  gives rise to a highly non-uniform radiation dose across the sensors with the fluence falling off with radius $r$ as approximately $r^{-1.9}$. The estimated fluence is obtained from the \lhcb simulation and using measured values of  displacement damage in silicon. Particle irradiation gives rise to both bulk and surface damage effects in the silicon.  The bulk radiation damage is primarily caused by the displacement of atoms in the silicon sensors from their lattice sites, and induces changes in the leakage current and effective doping concentration of the material.  Consequently, radiation damage monitoring was put in place from the start of operations. The leakage currents, noise and charge collection efficiency (CCE) of the sensors are studied regularly, with dedicated data taking scan procedures having been developed. Detailed studies of radiation damage in the \velo based on 2010 and 2011 data taking have been reported in Ref.~\cite{VeloRadiationDamage}. Further results, based on the 2012 and 2013 data taking periods, are summarised here. 

\subsubsection{Current measurements}

Sensor currents are studied as a function of both voltage and temperature.  Current-voltage scans are taken with an automated procedure on a weekly basis. The shapes are analysed to look for signs of gradient changes that could indicate the onset of breakdown, and the values at the operational voltage of 150\unit{V} are compared to the expected currents. Current-temperature scans are taken a few times a year by controlling the temperature of the cooling system. The current-temperature scans allow the surface and bulk current components to be measured separately.
% as the bulk damage component has an e^{kT}. 
More details of these studies are available in Refs.~\cite{VeloNoteIV, VeloNoteIT}. 

\begin{figure}[tb]
\begin{center}
%\resizebox{\textwidth}{!}{
\includegraphics[width=1.0\textwidth]{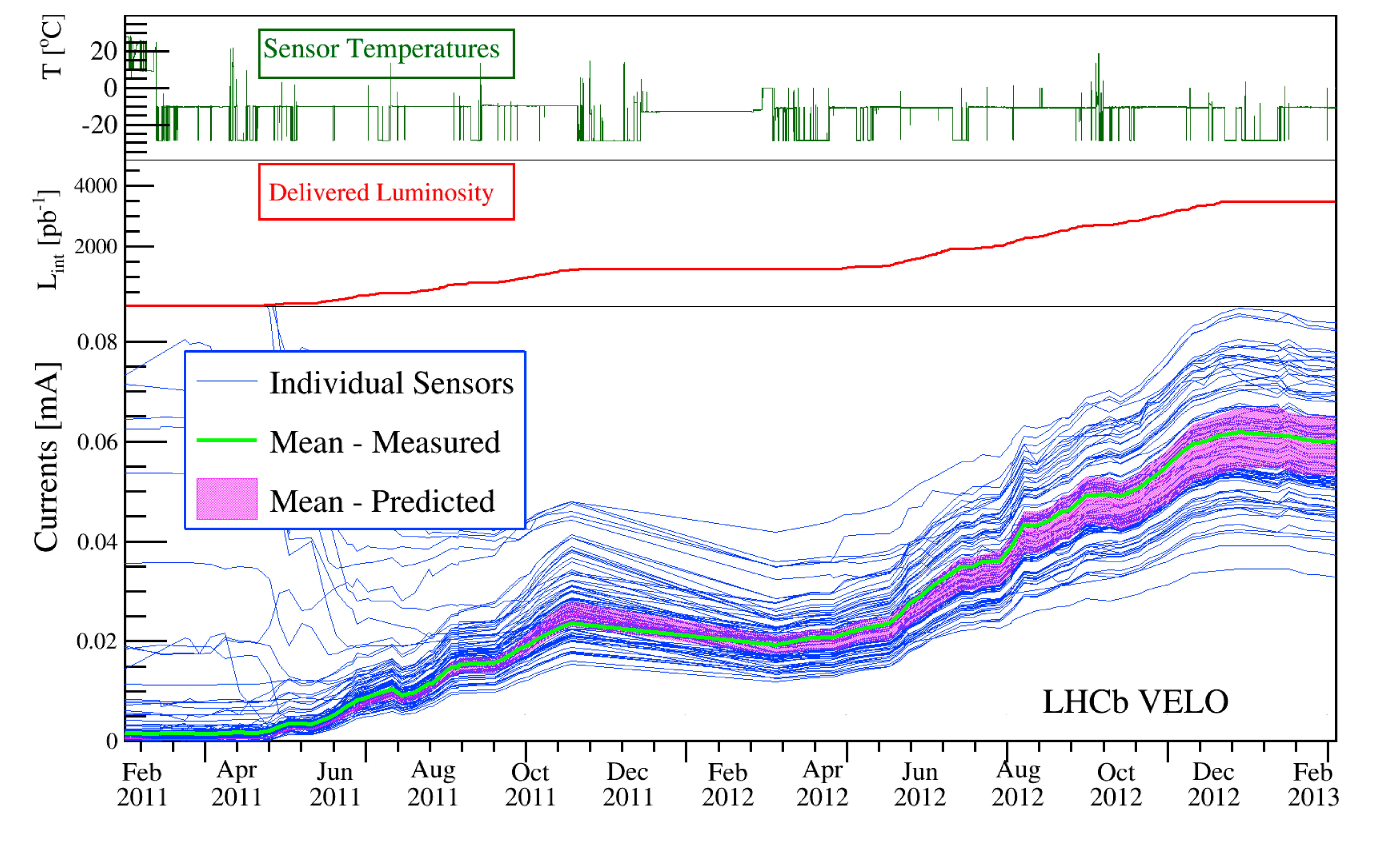}
%\includegraphics{figs/}
%}
\caption{\small 
%(left) Predicted fluence per \invfb of delivered luminosity as a function of module number in radial bins, corresponding to those used in radiation damage studies. 
%Comparison of the predicted and measured sensor currents , scaled to a temperature of 0\degreesC, after a delivered integrated luminosity of 2.7\invfb.
Currents measured for each sensor as a function of time (bottom). The
integrated luminosity delivered to LHCb and the average sensor temperature is shown over the same time
scale (middle and top). Increases in the delivered luminosity are matched by increases in the
sensor currents. The evolution of the mean measured current agrees well with the prediction from simulation. 
The mean measured value excludes sensors that are surface-current dominated.
%3.4~\invfb.
}
\label{fig:fluence}
\label{fig:currentCompared}
\end{center}
\end{figure}

The expected increase in the bulk current in each sensor due to radiation damage is calculated using the predicted fluences and the measured temperature history of the sensor and the relation \cite{MollThesis}

\[  \Delta I = \alpha \phi \rm{V_{Si}}, \]
where $\alpha$ is the annealing parameter in units of A/\cm, which depends on the temperature history, $\phi$ is the fluence in particles per $\cm^2$ and $\rm{V_{Si}}$ is the silicon volume in $\cm^3$. The evolution of the observed currents in the sensors with delivered integrated luminosity are in good agreement with the expectation (see Fig.~\ref{fig:currentCompared}). 

\subsubsection{Effective doping concentration}
The $n$-bulk sensors undergo space-charge sign inversion under irradiation, and hence their depletion voltage initially reduces with irradiation. This continues until type inversion occurs, after which it increases with further irradiation.
%This is a result of the effective doping concentration of the sensors changing with radiation, which subsequently leads to changes in the depletion voltages of the sensors. 
In order for the charge collection efficiency of the sensors to remain reasonably high, the sensors must be close-to or fully depleted during operation. 
As all of the \velo sensors are operated at a constant voltage over long periods, monitoring the sensor depletion voltages is a useful experimental technique 
for ensuring that the CCE for a particular sensor does not decrease significantly due to the sensor being under-depleted. In practice, this is achieved by monitoring the effective depletion voltage (EDV), which is derived using the following method. 
Here we report results for the $n$-type sensors, but note that one $p$-type module is also installed in the \velo and is studied in Ref.~\cite{VeloRadiationDamage}.

\label{sec:CCE}
\label{sec:nonp}

Sensors are grouped into a 1--in--5 pattern, where four of the sensors are operated at the nominal operation voltage (150\unit{V} throughout 2010--2013) whilst a single sensor, referred to as the test sensor, has a range of bias voltages applied to it.
Using only the sensors at the nominal operation voltage, a track is then fitted and extrapolated to the test sensor, where the amount of collected charge at the intercept is determined.
The distribution of collected charge arising from repeating this process for many tracks is fitted with a Landau distribution convolved with a Gaussian distribution to obtain the MPV of the Landau distribution. The EDV is then determined as the voltage at which the MPV of the charge distribution is equal to $80\%$ of the MPV
obtained at the maximum test voltage (currently 200\unit{V}). This value of $80\%$ provides good agreement between the depletion voltage measured from this method prior to irradiation and that obtained from the capacitance-voltage measurements performed on the sensors at production. It also represents a voltage above which the sensors must be operated to ensure significant signal charge is extracted. This process then repeats via an automated procedure for a range of patterns such that all of the sensors are tested. As this procedure requires beam time that would otherwise be used for physics data taking, it is currently only performed around three times per year.

\begin{figure}[tb]
\centering
\includegraphics[width=0.96\textwidth]{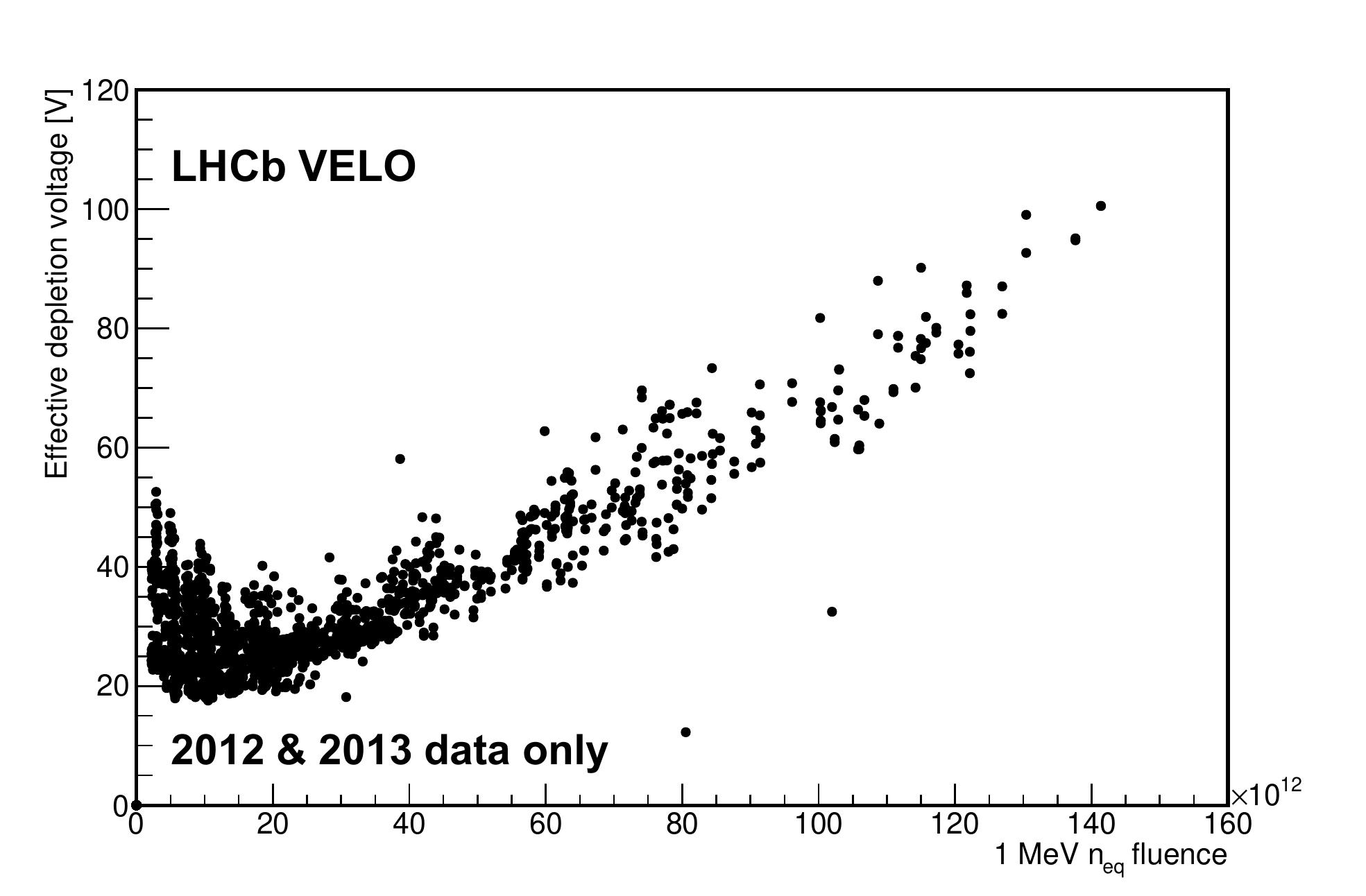}

\caption{\small{The effective depletion voltage versus fluence for all \velo sensors up to 3.4\invfb delivered integrated luminosity.}}

\label{fig:hvratio}
\label{fig:cce}
\end{figure}

Figure~\ref{fig:cce} shows the measured EDV values using all the \velo sensors, and dividing them into radial regions with reasonably constant fluence. The delivered integrated luminosity at each CCE scan has been combined with the expected fluence per \invfb to give the fluence for each radial region of each sensor. The sensors exhibit an overall decrease in EDV with fluence followed by an increase in EDV until the most recent CCE scan (after a delivered integrated luminosity of 3.4\invfb). As a result of the radiation damage that had occurred by the start of 2013, many of the sensors now have EDVs significantly higher than before irradiation. On average, type inversion of the silicon occurs around $20 \times 10^{12}$~\neutroneq, after which the EDV increases linearly with fluence. This behaviour can therefore be used to predict the operational voltages required to give maximal CCE for a given fluence and hence a delivered integrated luminosity. It can also be noted that a minimum EDV of $\sim20$\unit{V} is observed for all sensors, this is assumed to be a result of the finite integration time of the front-end electronics.
%; the actual depletion voltages are expected to approach 0 V around type inversion.

\subsubsection{Charge loss to second metal layer}

The cluster finding efficiency in the sensors is studied using the same technique of track extrapolation as that used to study the CCE (see Sect.~\ref{sec:CCE}). These studies have shown an unexpected radiation induced charge loss due to the presence of the second metal layer on the sensors. The second metal layer on the sensors is used to route signals from inner strips to the outer radius of the sensor. Particles that pass close to a second metal layer trace, but away from the shielding effect of the first metal layer on the strip, have a reduced charge collected on this strip and additionally induce charge on the second metal layer. This gives rise to a reduction in cluster finding efficiency on this strip, and additional noise clusters located on the strip connected to the second metal layer track. This degradation has been observed to continue throughout the 2012 data taking period, although there is some evidence to suggest the rate of change with fluence is decreasing. Currently studies of the \velo tracking efficiency show no degradation associated with this effect within the errors of $\pm 0.3\%$. This effect is discussed in detail in Ref.~\cite{VeloRadiationDamage}.

\section{Physics performance}
\label{sec:Physics}
%[Paula to oversee, max. 12 pages]
The performance of the \velo on more directly physics analysis related
parameters is reviewed in this section. The performance of the pattern
recognition, track reconstruction and the alignment of the detector are
discussed. The impact parameter, primary vertex (PV) resolution and decay time resolution of the detector are then presented.

\subsection{Pattern recognition and tracking}
\label{sec:Tracking}

%[David - 3 pages]

Obtaining a high efficiency for reconstructing the trajectories of
charged particles is particularly important for the analysis of
many-body final states since the selection efficiency scales with the
track finding efficiency to the power of the number of tracks. An accurate knowledge of the
tracking efficiency is also important in many other physics analyses,
especially those that aim to measure a production cross-section or a
branching fraction. The track finding algorithms in \lhcb start with a
search in the \velo detector for straight lines. These tracks are
then combined with hits in the tracking stations to produce the ``long"
tracks that are used for most physics analyses. The track fit is
performed using a Kalman fit. In the trigger a
simplified model of the detector material and a single direction Kalman
fit are used in order to reduce CPU consumption. A full
bi-directional Kalman fit is performed offline using a detailed model of the
detector material.

The \velo pattern recognition algorithm requires a minimum of three \RSens
sensor and three \PhiSens sensor clusters to reconstruct a trajectory. The
basic algorithm collects sets of \RSens clusters consistent with being
on a straight line from the interaction point, then looks for a set of compatible \PhiSens sensor
clusters to confirm the trajectory.  The alternating stereo angles of
the \PhiSens sensor strips resolves the stereo ambiguities when combining the \RSens
and \PhiSens sensor clusters.  A second pass is made combining the unused
clusters to find tracks not from the interaction region. The number of
clusters on a \velo track ranges from six (the minimum requirement is three on \RSens and three on
\PhiSens sensors)  up to the full 42, with an average of 11.  An additional
algorithm, that only runs when the detector is closing, makes 3D space
points from the \RSens and \PhiSens sensor clusters in a module and forms
tracks from these having no assumptions about the track directions.

%In the original design \cite{LHCbVELO_TDR:2001hf} the plan was to use
%the R sensors only to make 2D RZ tracks, however these did not have the
%resolution required to measure the impact parameters of tracks to the
%primary vertices with only 2D information, so a full 3D pattern
%recognition is always performed.

The efficiency of the track reconstruction in the \velo has been
measured using a tag-and-probe method. Samples of \jpsi decays into two muons are
used where the track of one of the muons is fully
reconstructed (tag-muon), while the other muon is only partially
reconstructed using hits in the other tracking stations (probe-muon).  The
momenta of the tag- and probe-muon are used to
reconstruct the \jpsi candidate mass and hence confirm the selection. It
is required that the trigger has not selected the event based on the
probe muon track to prevent any potential trigger bias. The \velo
tracking efficiency is obtained by matching the partially
reconstructed probe muon to a long track (which has a \velo track
segment).

\begin{figure}[tb]
\begin{center}
    \resizebox{\textwidth}{!}{
      \includegraphics{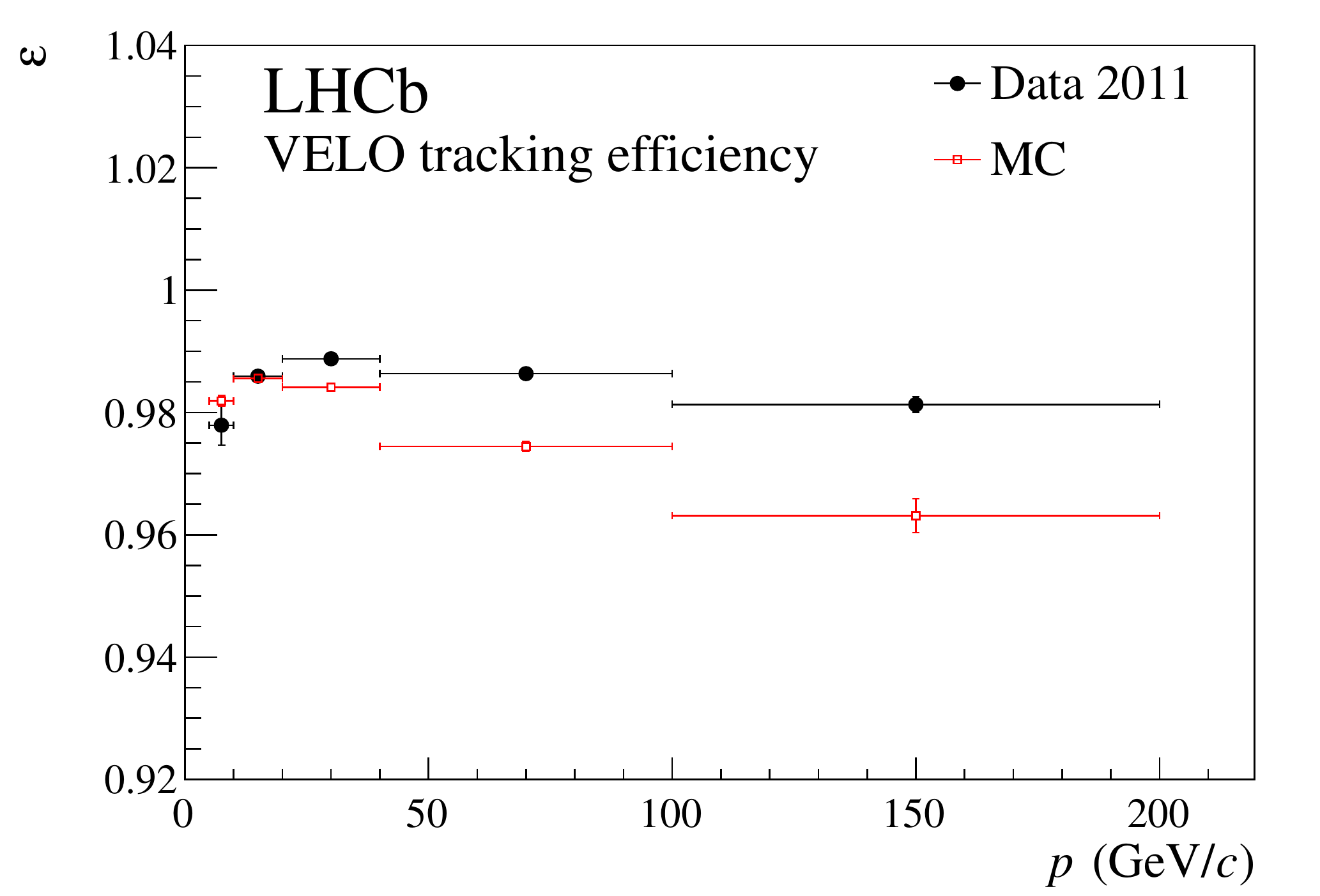}
      \includegraphics{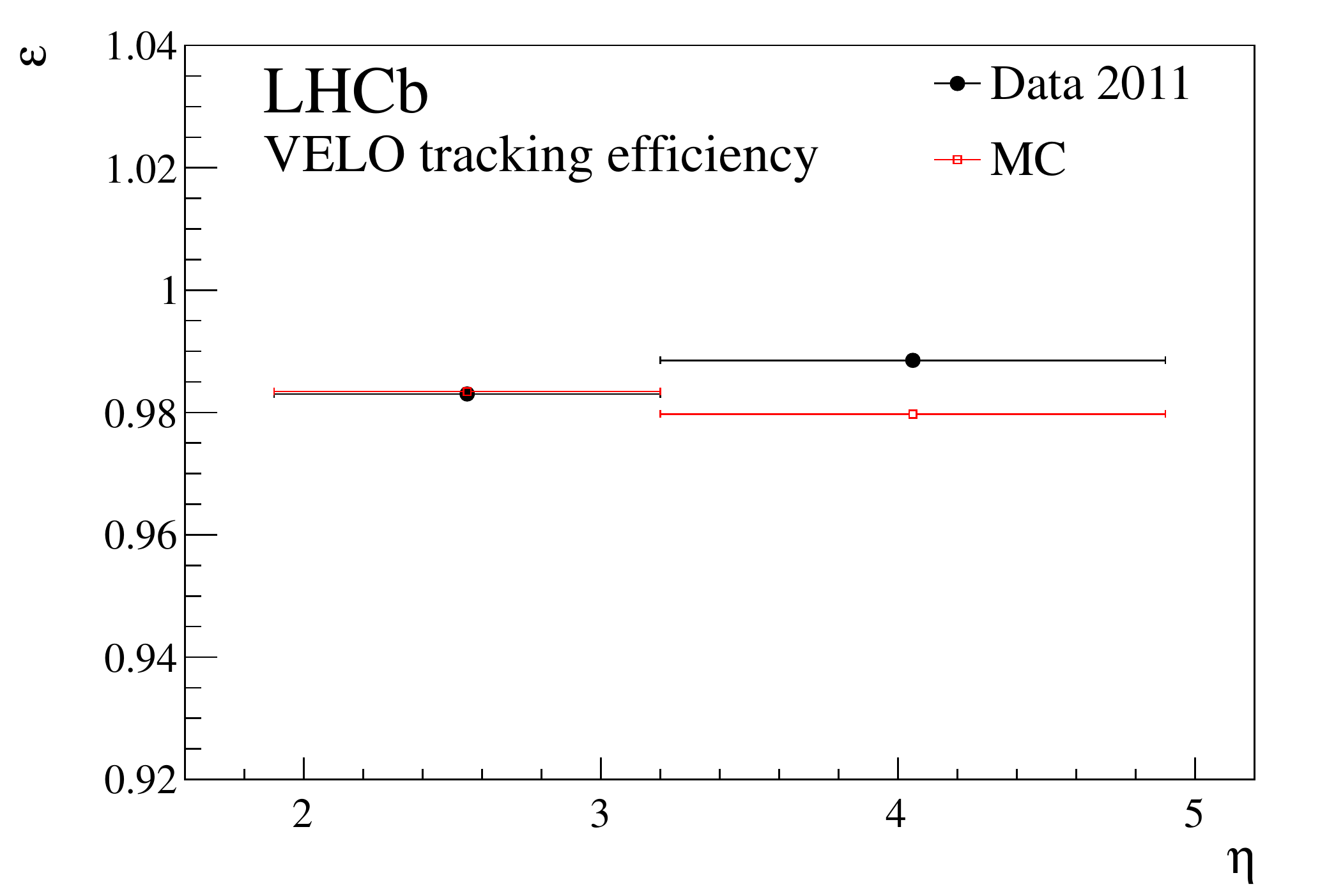}
    }
    \resizebox{\textwidth}{!}{
      \includegraphics{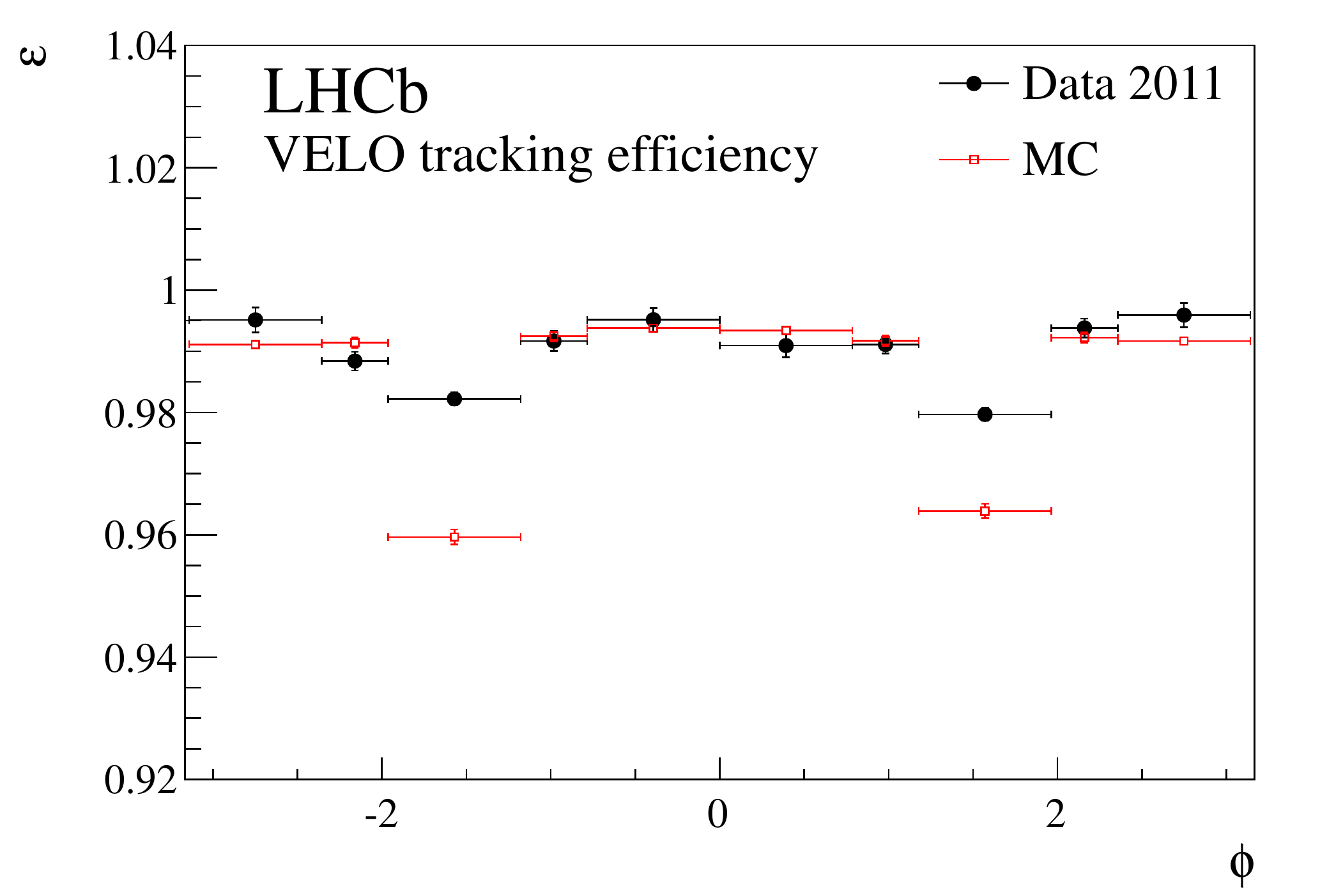}
      \includegraphics{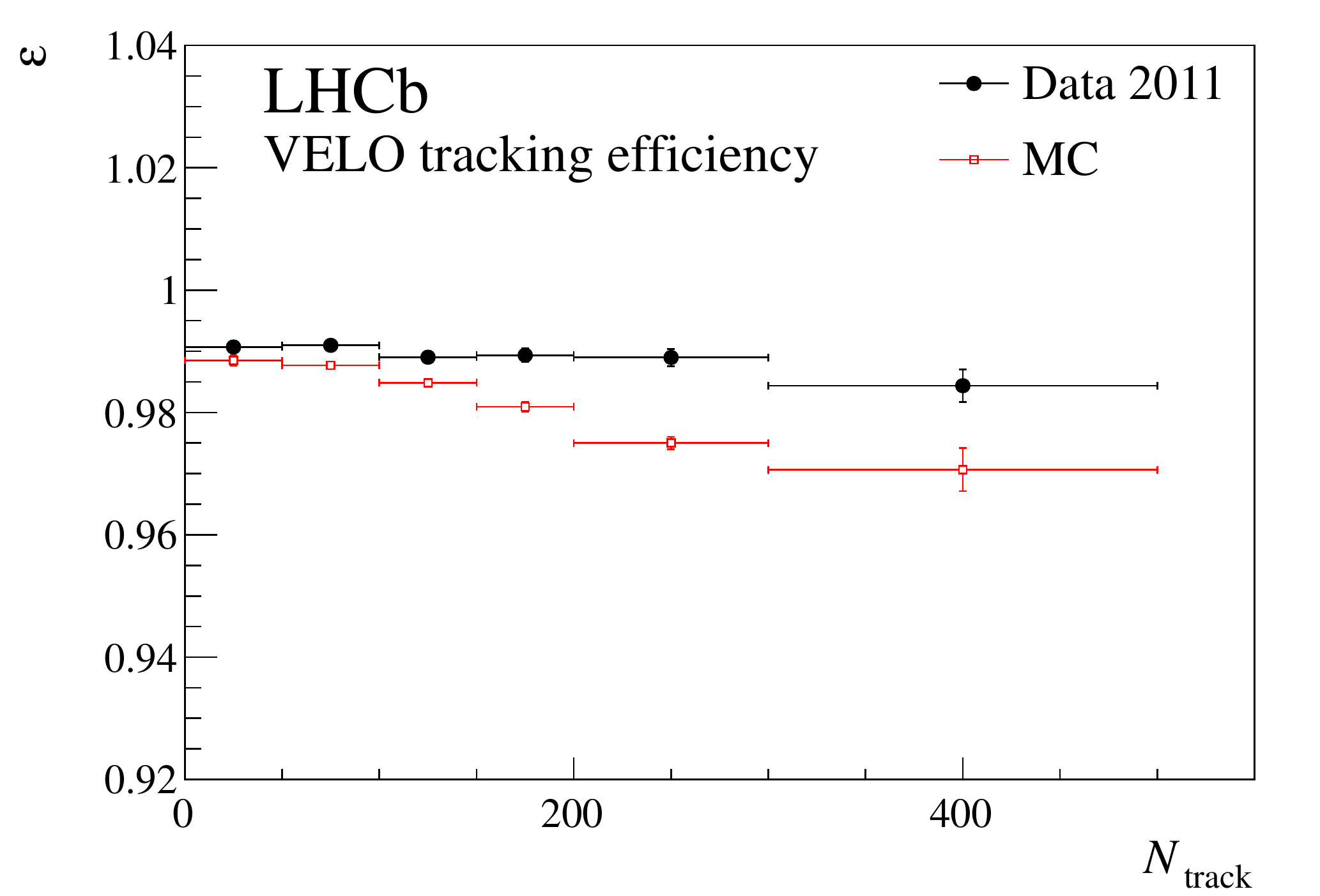}
    }
\caption{\small Tracking efficiency for the 2011 data and
  simulation for the \velo as a function of the momentum, p (top
  left), the pseudorapidity, $\eta$ (top right), the azimuthal
  angle $\phi$ (bottom left) and the total number of tracks in the
  event, $N_{track}$ (bottom right). The simulation has been reweighted
  to the number of tracks observed in data for the p,$\eta$ and $\phi$
  plots. The error bars indicate the statistical uncertainty.}
\label{fig:trackEff} 
\end{center}
\end{figure}

The measured \velo tracking efficiency for long tracks is shown in
Fig.~\ref{fig:trackEff} for data and simulation, and is typically 98\%
or higher in the data. The simulation is weighted by the number of
tracks observed in the data. The efficiency for the \velo pattern
recognition is weakly dependent on the multiplicity of the event, as the
pattern recognition becomes more complex in a higher occupancy
environment. The dips in efficiency at $|\phi|\approx \pi/2$ are caused
by the extra multiple scattering from the material of the RF foil in the
vertical plane, see Fig~\ref{fig:etaphiscanfmp}~(left). 
The discrepancy between data and simulation in these bins is
partially explained by the effect of assumptions made in the tracking: the beam is
centred in the coordinate system in the simulation while it is offset by 0.4\mm in data; and the
distance between the two halves when the system is fully closed differ at the 150\mum
level between data and simulation. The
discrepancy is taken into account in physics analyses by applying a 
reweighting procedure that ensures data and simulation agreement, and
a procedure for determining the systematic uncertainties due to this
correction is also in place.

%The efficiency of the pattern recognition on Monte Carlo events is show
%in fig. \ref{fig:Eff_in_MC}. Tracks are considered to be reconstructable
%if they have at least 3 R and 3 Phi measurements, the minimum required
%for the pattern recognition, and a momentum of $1 \gevc$ and also travel
%through the TT station after RICH1. Additional a selection that the
%track originated from $r<100\mum$ and $-100<z(\mm)<100$ requires that
%they start from the luminous region. When the efficiency verse $p$, $r$
%and $z$ is plotted the cut on the variable is removed, the $p>1\gevc$ is
%still applied for the plot of efficiency verse \pt.

%\begin{figure}[htb]
%  \begin{center}
%    \resizebox{\textwidth}{!}{
%      \includegraphics{figs/Eff_v_p_withZoom.pdf}
%      \includegraphics{figs/Eff_v_pt_withZoom.pdf}
%    }
%    \resizebox{\textwidth}{!}{
%      \includegraphics{figs/Eff_v_R.pdf}
%      \includegraphics{figs/Eff_v_phi.pdf}
%    }
%    \resizebox{\textwidth}{!}{
%      \includegraphics{figs/Eff_v_z.pdf} 
%      \includegraphics{figs/Eff_v_eta.pdf} 
%    }
%    \caption{Pattern recognition efficiencies in Monte Carlo for
%      reconstructable tracks from the interaction point with $p>1
%      \gevc$. Top left: with momentum (inset is a zoom at low $p$), top
%      right: with \pt, with the selection $p>1 \gevc$ (inset is a zoom
%      at low \pt). Middle left: with radius of origin vertex, middle
 %     right: with azimuth of the track direction. Bottom left: with $z$
%      of track origin vertex, bottom right: with pseudo-efficiency of
%      the track.}
%    \label{fig:Eff_in_MC}
%  \end{center}
%\end{figure}

%, both in absolute occupancy and as a function
%of the number of primary vertices. There are three categories of tracks
%to reconstruct: those traversing the whole detector and reaching the
%calorimeter face called long tracks, those reaching TT but not the T
%stations, called upstream tracks and those that are not in the nominal
%LHCb acceptance, including those traveling in the $-z$ direction,
%called VELO only tracks. Plots of the efficiencies are shown in fig.
%\ref{fig:Eff_in_occ}.

\begin{figure}[tb]
  \begin{center}
    \resizebox{\textwidth}{!}{
      \includegraphics{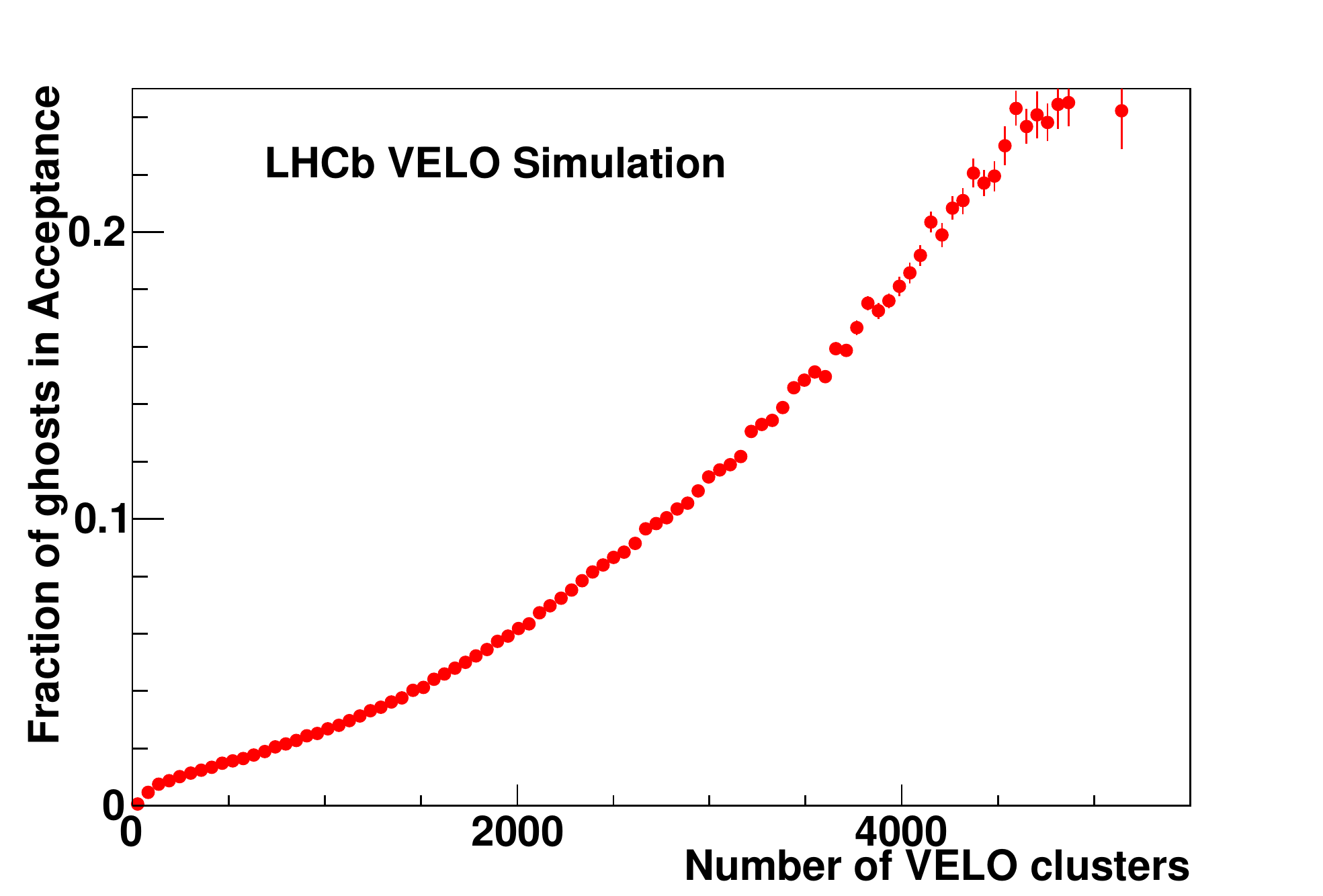}
       \includegraphics{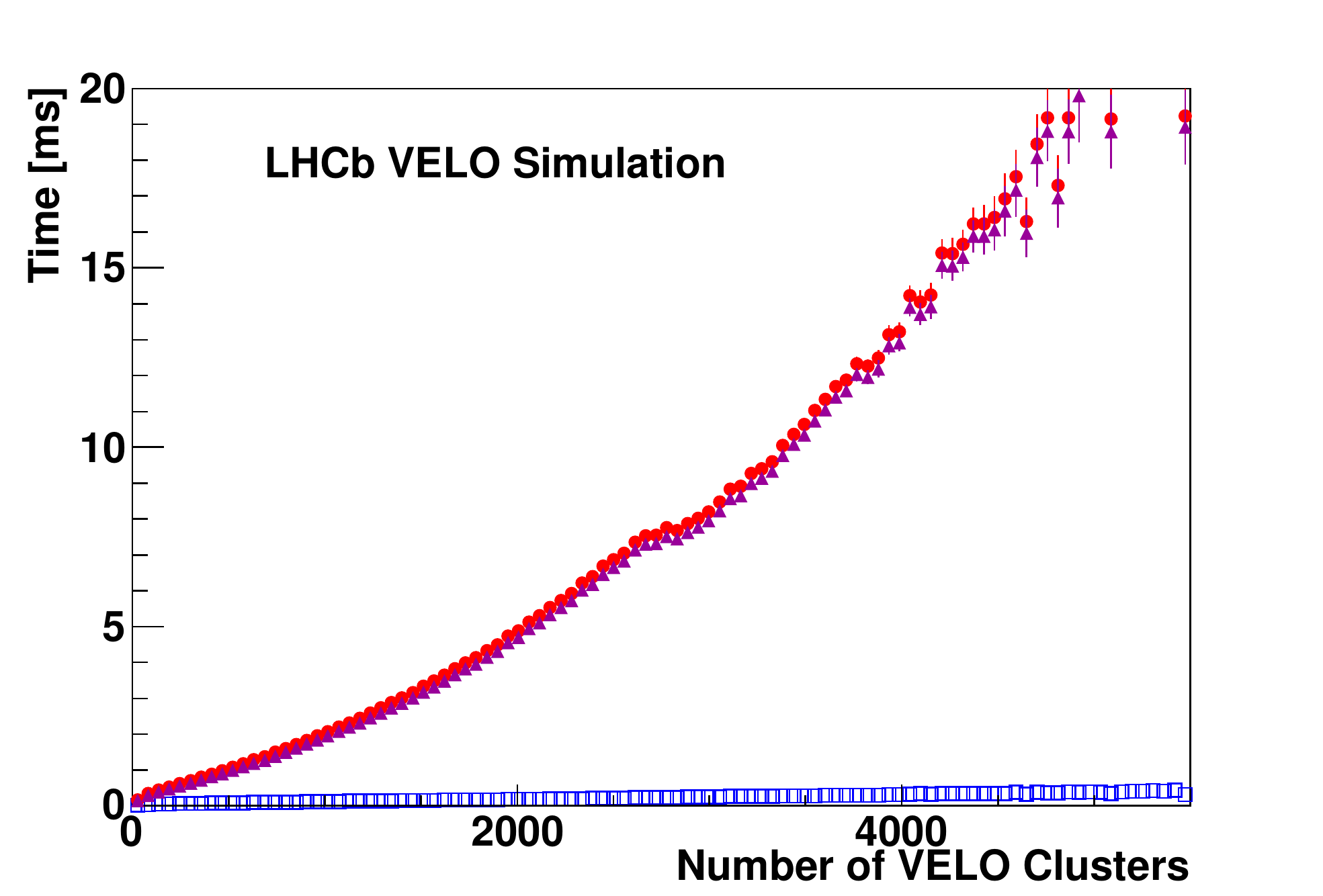}
    }
    \caption{\small (left) Fraction of ghost tracks versus
      number of \velo clusters in simulation.  (right) \velo\ pattern recognition
      timing versus number of clusters, the pattern recognition alone (magenta triangles), the raw data decoding (blue squares) and the combined
      (red circles) timings are shown. The times are scaled to a
      2.8\ghz Xeon processor.  }
    \label{fig:Ghost_with_occ}
    \label{fig:Trigger_timing}
  \end{center}
\end{figure}

Another important measure of the tracking performance is the number of
poor quality or ``ghost" tracks that are produced.  Defining a ghost
track as one in which less than 70\% of the \velo clusters on the track
are from the same simulated particle, the fraction of ghost tracks is
shown in Fig.~\ref{fig:Ghost_with_occ} (left) as a function of the
total number of \velo clusters. These tracks become more frequent as the
detector occupancy rises, with typical occupancies being 0.5\% for
randomly triggered events and 1\% for HLT triggered events (see Sect.~\ref{sec:Occupancy}).

%\begin{figure}[htb]
%  \begin{center}
%    \resizebox{\textwidth}{!}{
%      \includegraphics{figs/Eff_v_nPV.pdf}
%      \includegraphics{figs/Eff_v_Occ.pdf}
%    }
%    \caption{
%      Left: Pattern recognition efficiency verse number of
%      primary vertices for reconstructable tracks separately for long
%      (red circles), upstream (blue squares) and VELO only tracks
%      (magenta triangles). 
%      Right: Pattern recognition efficiency verse number of Velo
%      clusters, same symbols.
%}
%    \label{fig:Eff_in_occ}
%  \end{center}
%\end{figure}

% long tracks with less than 70\% of \velo clusters arising from the same simulated particle 

%The MC performance has been validated by reconstructing
%$J/\psi\to\mu^+\mu^-$ decays with one muons reconstructed using all of
%the tracking detectors and one reconstructed ignoring the VELO
%information, for more details see [Note no paper to reference
%yet]\cite{Eff_data_MC}. The efficiency for track finding in the VELO is
%compared between data and MC in fig. \ref{fig:data_MC_eff_comp}.

%\begin{figure}[htb]
%  \begin{center}
%    \resizebox{0.5\textwidth}{!}{
%      \includegraphics{figs/VeloDataMCEffComp.pdf}
%    }
%    \caption{
%      Efficiency of the pattern recognition to find a muon from a
%      $J/\psi$ decay track in the VELO, for both data and MC samples.
%    }
%    \label{fig:data_MC_eff_comp}
%  \end{center}
%\end{figure}

The CPU resources consumed by the \velo tracking in the high level
trigger are substantial and the algorithms have been optimised for speed. 
For example, to minimise the time taken in decoding the clusters, the 3-bit cluster centre position computed in the TELL1 FPGAs is used (see Sect.~\ref{sec:TELL1}).
The time taken for the decoding and pattern recognition algorithms is shown in Fig.~\ref{fig:Trigger_timing} (right).

%The VELO is used as part of the trigger to identify primary vertices and
%to evaluate the impact parameters of objects triggering at Level-0
%\cite{LHCb_Trigger_TDR}. The pattern recognition has relatively wide
%windows and does not require the ultimate precision from the detector.
%To minimize the time taken a decoding is performed to {\tt
 % VeloLiteClusters} which do not contain the ADC values for the strips,
%then the pattern recognition collects clusters compatible with a track
%and does a linear fit to the clusters. The time taken for this is shown
%in fig. \ref{fig:Trigger_timing}.

%\begin{figure}[htb]
%  \begin{center}
%    \resizebox{\textwidth}{!}{
%      \includegraphics{figs/timePV.pdf}
%      \includegraphics{figs/time_clus.pdf}
%    }
%    \caption{VELO pattern recognition timing verse number of primary
%      vertices for just the pattern recognition (red circles), the raw
%      data decoding (blue squares) and combined (magenta triangles).
%      Right: Pattern recognition timing verse number of Velo clusters,
%      same symbols.  In all cases the times are scaled to a notional 2.8
%      GHz Xeon processor. }
%    \label{fig:Trigger_timing}
%  \end{center}
%\end{figure}

\subsection{Alignment}
\label{sec:Alignment}

%\subsubsection{Introduction}
The \velo has extremely stringent alignment requirements to ensure
that the intrinsic hit resolution of the detector (see
Sect.~\ref{sec:Resolution}) and the impact parameter and decay time
resolution (see below) of the experiment are not adversely affected. Furthermore, as described in~Sect.~\ref{sec:Motion},
the \velo halves are inserted and centred around the beams in each
\lhc fill. Consequently  the high level trigger requires an immediate
update of the alignment parameters. These parameters are also required
for the offline reconstruction. The alignment of the detector is thus
separated into two elements: the underlying system alignment, and the
updates required in each fill.

The underlying alignment of the \velo relies on three components: the
precise construction and assembly of the detector, the mechanical and
optical survey of each part of the detector, and the software
alignment of the system using tracks. The updates for each fill are
then added to this: mechanical measurements of the position to which the
detector is closed are used to determine these.

\subsubsection{Optical and mechanical measurements}

Each component of the detector was surveyed at the various stages of
the assembly using a smart-scope for the individual modules, and a
coordinate measuring machine with optical and touch probe heads for
the modules and for each half of the assembled system. The relative position of the \PhiSens sensor with respect to the \RSens 
sensor in each module was measured with an accuracy of about 3\mum for the $x$ and $y$ translation and with an accuracy of about 20\murad for
 rotations around the  $x$ and $y$ axis. 
The relative module positions within each half of the detector were measured with a precision of about 10\mum for the translations along $x$ and $y$. 
Measurements of the mounting frame were made with a coordinate measuring machine during production,
and then prior to, and after, installation measurements were made using photogrammetry, theodolites, mechanical gauges and levelling instruments.
The position of the two \velo halves was determined with an accuracy of 100\mum
for the translations and 100\murad for the rotations. These survey
measurements are used 
as a starting point for the track-based software alignment, and remain
important for the final alignment quality as some degrees of freedom
(see below) are difficult to align with tracks.

\subsubsection{Track-based alignment methods}

The track-based alignment relies on minimising the residuals between
the fitted tracks and the measured cluster positions. The alignment
can be considered in terms of three different stages:
\begin{enumerate}
\item the relative alignment of each \PhiSens sensor with respect to the \RSens sensor in the
same module. This allows the $x$ and $y$
translations of the sensors to be determined; 
\item the relative alignment of the modules within each \velo half.
 This alignment is primarily sensitive to the $x$ and $y$
translations of the modules and their rotations around the $z$ axis. Only the
\PhiSens sensors are sensitive to this rotation due to the strip
geometry and hence this rotation must be determined at the module
level. The misalignment due to the other three degrees of freedom (the $z$ translation and the rotations
around the $x$ and $y$ axis) cause second order effects to which
sensitivity can only be obtained with \RSens sensors from using a large
data sample with a wide range of track angles;
\item the relative alignment of one \velo half with respect to the other
half. The PV position can be used as a constraint since 
tracks reconstructed in each half originate from this common point.
This method is sensitive to the $x$, $y$ and $z$ translations and the
rotations around the $x$ and $y$  axis. In addition, tracks that cross
both halves of the detector are used.  There is a small overlap
between the sensors in the left and right halves of the \velo when the
detector is fully closed, and tracks that traverse the \velo in this
overlap region are particularly important. The alignment using these
tracks is mainly sensitive to the misalignment of $x$ and $y$ translations and the rotation
around the $z$ axis.
\end{enumerate}

Two methods were developed and used for performing the track-based software alignment.
The first method~\cite{Viret:2008jq,Gersabeck:2008jr} performs the \PhiSens
sensor alignment by fitting an analytical form to the
residuals as a function of $\phi$,  and performs the remaining
alignment stages using a matrix inversion method, based
on Millepede \cite{Blobel:2002ax}, which performs a  $\chi^2$
minimisation that depends upon both the track and alignment
parameters. This alignment method is fast but the implementation fits straight line tracks in
the \velo. The second method uses a global $\chi^2$ minimisation based
on Kalman track fit residuals \cite{Hulsbergen:2008yv}. This method
uses an iterative procedure, that could be time consuming in the case
of a large misalignment. However, the Kalman filter fit takes into
account the corrections for multiple scattering, energy loss effects and the weak magnetic field in the \velo region. The results from both methods are in agreement \cite{VeloTedArticle}.

The alignment is performed using a specially selected data sample
which improves the sensitivity to all degrees of freedom and better
constrains the upstream and downstream of the PV regions of the \velo. This sample
includes a mixture of tracks from collisions with a wide range of angles, and tracks from beam-gas interactions.  The beam-gas interactions are roughly parallel with the beam-axis, giving tracks that cross many modules.

%However, the track-based alignment is not equally sensitive to all degrees of freedom of alignment.
Correlated module misalignments which are poorly constrained in the
track-based alignment, known as weak modes, can bias physics quantities like the impact parameter or the invariant mass.
The most important weak mode in the \velo is the twist of the modules around the $z$-axis.
This misalignment can distort the impact parameter measurement by several tens of microns as a function of the azimuthal direction of the track.
The effect is limited by using the alignment track sample with a range
of track types, and by applying constraints in the alignment procedure
from the survey measurements.

\subsubsection{Mechanical measurement of closing}

The \velo halves are moved independently horizontally ($x$) and
together vertically ($y$) during the closing procedure and the motion
is measured to an accuracy of a few\mum with mechanical position sensors. These measure the
revolutions of the spindle that controls the displacement. The
position of the two \velo halves is updated for each fill using these
mechanical system measurements in $x$ and $y$ by adding these changes
to the underlying system alignment. 

Performing the track-based alignment
of the \velo halves using data collected at different opening
positions (distance between the two halves of 0, $\pm$5\mm,  $\pm$10\mm,  $\pm$29\mm) allows a calibration of the motion system
measurements to be performed. This shows a calibration accuracy of
$0.6\%$ of the motion system along the $x$ direction is
achieved. 

\subsubsection{Alignment performance}

The track-based alignment results are in good agreement with the lower
precision survey results, within the uncertainties, apart from the
effect due to the change in temperature of the system. The system was
surveyed at room temperature, but is normally operated with the
cooling system operated at -30\degreesC. Comparing track-based
alignments performed with the cooling temperature set to +8\degreesC and -30\degreesC a change of distance between the two halves of about 170\mum
was measured. Laboratory measurements were also made by heating individual modules and an
expansion of approximately 1\mum per degree along the $x$ direction
of the carbon fibre that supports the modules was found. Hence, even when using
materials with a low coefficient of thermal expansion it is important to control temperature changes when aiming for~\mum level precision. This is achieved in the \velo by maintaining the mounting base plate of each \velo half at 20\degreesC and operating with a stable cooling system temperature. 

\begin{figure}[tb]
\begin{center}
  \includegraphics[width=0.9\textwidth]{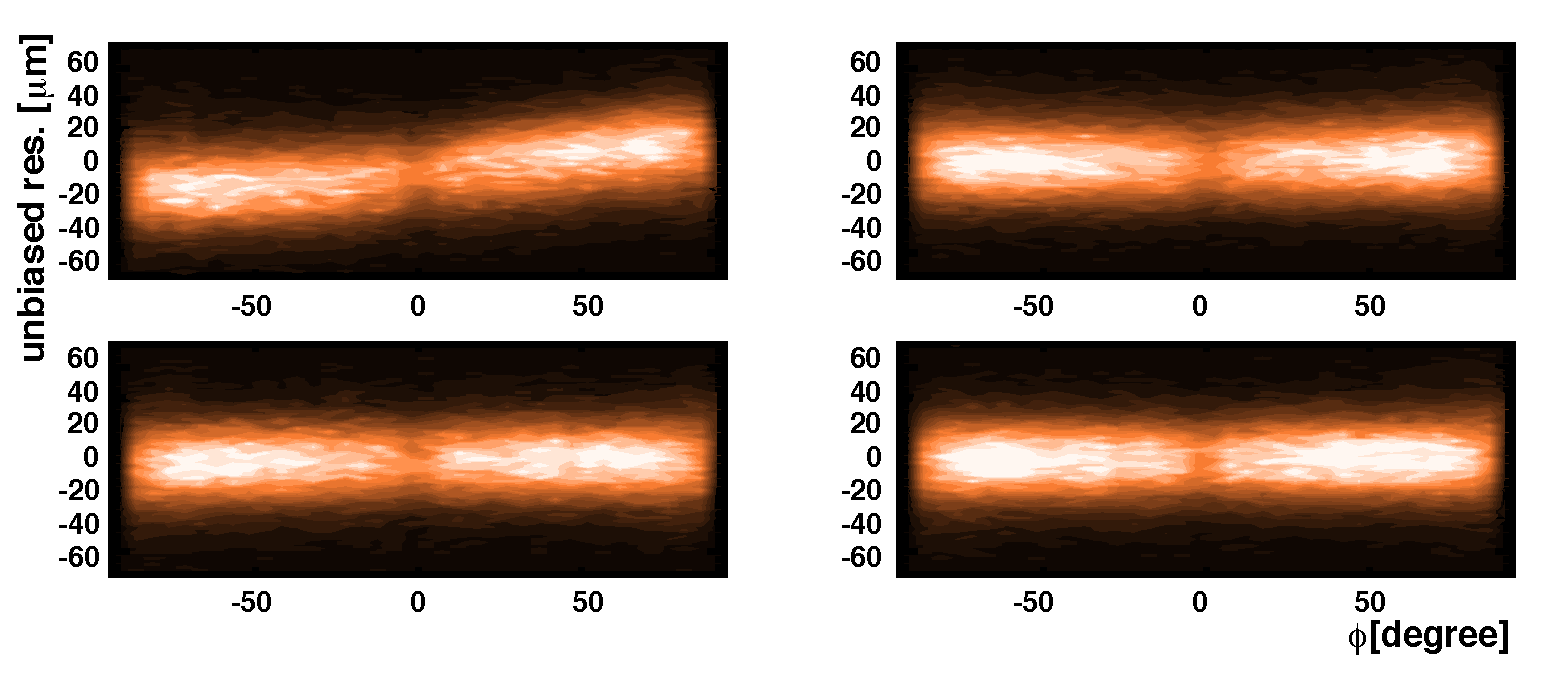}
  \caption{\small Example unbiased sensor residuals as a function of the $\phi$ coordinate using only the survey 
    information (left) and using the track-based software alignment
    (right). Results are given for two different example
    sensors. (top) A significant improvement in the residuals is seen
    in this sensor with the
    track-based alignment. (bottom) In this sensor the alignment quality using the
    survey information is already good.
  }
  \label{fig:alignresphi}
  \end{center}
\end{figure}

The improvement obtained after the track-based alignment procedure over the precision of the survey is illustrated in Fig.~\ref{fig:alignresphi}. The shapes of the residual distributions as a function of the azimuthal coordinate are characteristic of particular sensor misalignments.
The precision obtained after the track-based alignment is
better than the best hit resolution of the detector with an alignment at the few \mum level obtained for the $x$ and
$y$ module translations.

\begin{figure}[tb]
\begin{center}
  \includegraphics[width=0.7\textwidth]{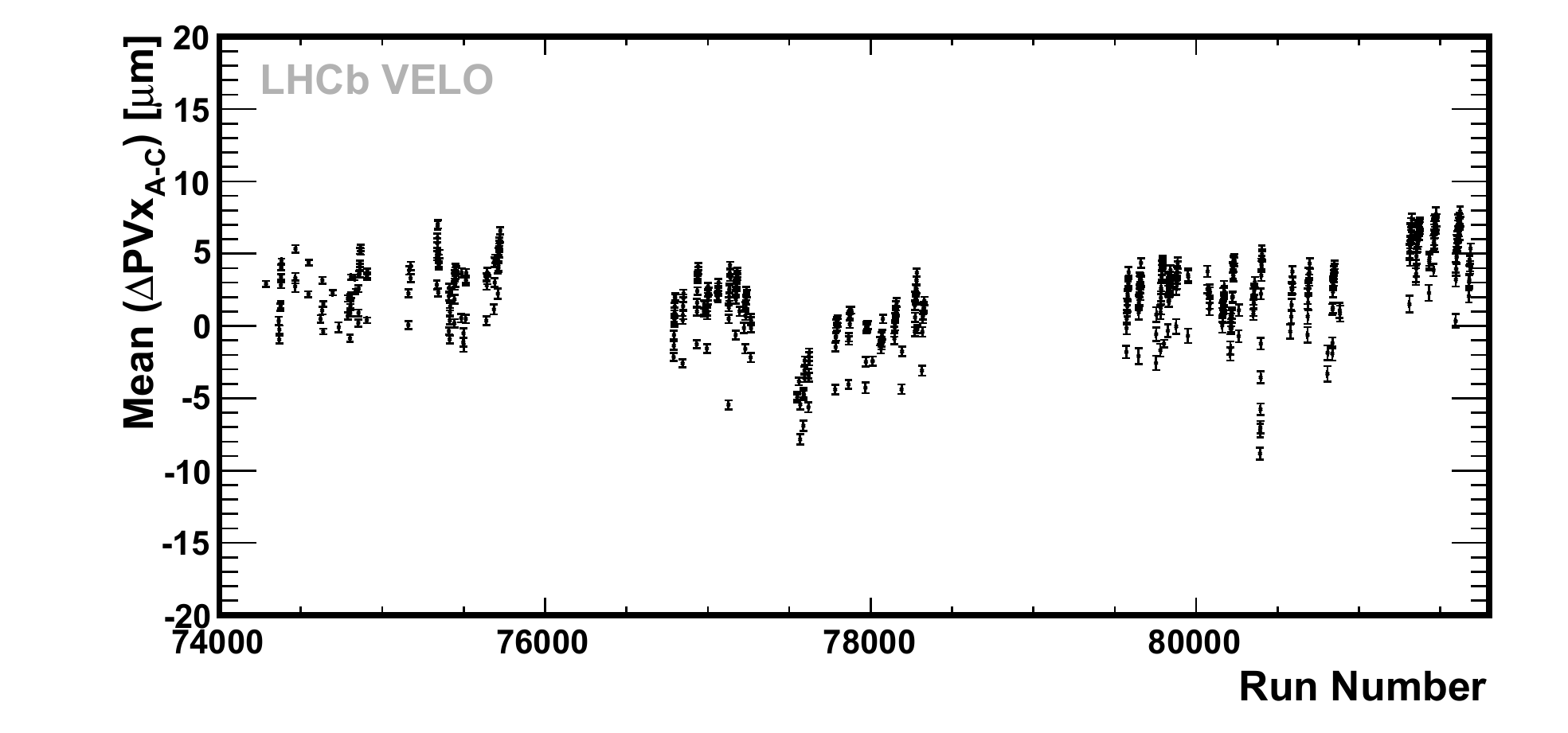}
  \caption{\small Misalignment between the two \velo halves in each
    run, evaluated by fitting the PV separately with tracks in the two halves of the \velo. 
  The run numbers shown here span the period of the last four months
  of operations in 2010.
  }
  \label{fig:alignstability}
  \end{center}
\end{figure}

Fitting the position of the PV separately with tracks in the two
halves of the VELO allows the misalignment between the two halves to
be determined.  This is shown in Fig.~\ref{fig:alignstability} over a
period of four months of operation. The variation between runs shows
the accuracy with which the position in each fill is measured. The
excellent stability of the alignment of the system is also clear.

\subsection{Primary vertex resolution}
\label{sec:Vertex}

%
%==================
%

%(1) In this paragraph it says that you wanted to add something about the vertexing algorithm. I agree this should be added, can you send me a few lines.
%    For comparison on level of detail, in the tracking performance section we explain the pattern recog. with:

%The \velo pattern recognition algorithm requires a minimum of three R
%sensor and three $\Phi$ sensor clusters to reconstruct a trajectory. The basic algorithm
%collects sets of R clusters consistent with being from the interaction point, then looks 
%for a set of compatible $\Phi$ sensor clusters to confirm the trajectory. 
%The alternating stereo angles of the $\Phi$ sensors resolves the stereo ambiguities when combining the R and $\Phi$ sensor clusters. 
%A second pass is made combining the unused clusters to find tracks not from
%the interaction region. An additional algorithm, that only runs
%when the detector is closing, makes 3D space points from the R and $\Phi$ sensor
%clusters in a module and has no assumptions about track directions. 

%(2) We quote the vertex resolution for 25 tracks. A useful addition would be the average number of tracks (say in a minimum bias and heavy flavour event with 1 PV ?). 

%(3) Another useful addition would be to add a sentence on degradation with number of PVs.

%
%==================
%

% why vertex resolution is important
The accurate measurement of decay lifetimes is required for the primary physics aims of the \lhcb experiment in CP violation and rare decay studies.
Precise vertex reconstruction is therefore of
fundamental importance, in order to resolve production and decay
vertices.

%LHCb has been designed to study CP violation and rare decays in the
%B-meson sector. This requires the accurate measurement of decay
%lifetimes and impact parameters, both for flavour tagging and
%background rejection. Precise vertex reconstruction is therefore of
%fundamental importance, in order to resolve production and decay
%vertices.

% method

The PV resolution is strongly correlated to the number of
tracks $N$ used to reconstruct the vertex. The analysis is
performed on an event-by-event basis. The principle is to reconstruct
the same PV twice, and to determine the difference
between these two PV positions. This is achieved by splitting the track
sample of each event into two and making vertices from each
independent set of tracks. The method was verified in the simulation by comparing the reconstructed and generator level information.

The track splitting is done entirely at random, with no ordering of
tracks and no requirement that the same number of tracks is put into
each set. The vertex reconstruction algorithm is applied to each set of tracks. Vertices are `matched'
between the two sets by requiring that the difference in their $z$
position is $<2$\mm. Then, if the number of tracks making a pair of matched vertices is the
same, the residual is calculated. Repeating for many events yields a
series of histograms of residuals in $(x,y,z)$ for varying track multiplicity.

In practice, the number of tracks making a vertex ranges from $5$ (the
required minimum) to around $100$. However, given the track splitting
method roughly divides the total number of tracks in two, it is
difficult to measure the resolution past $40$ tracks. Each residual
histogram is fitted with a Gaussian distribution. The resolution for each
particular track multiplicity is calculated as the $\sigma$ of the
fitted Gaussian divided by $\sqrt 2$, as there are two uncorrelated
resolution contributions in each residual measurement.

% parametrisation

The resolution is fitted with a function which parametrises it in
terms of $N$ as follows:
\begin{equation}
\label{nTrdep} \sigma_{PV} = \frac{A}{N^{B}} + C,
\end{equation}
where $A,B,C$ are constants.
%$C$ can be thought of as the best resolution possible, given multiple scattering. $A$ is a constant multiplier, and $B$ shows the dependence on track multiplicity.

% results in data

\begin{figure}[tb]
\centering
\includegraphics[width=.49\textwidth]{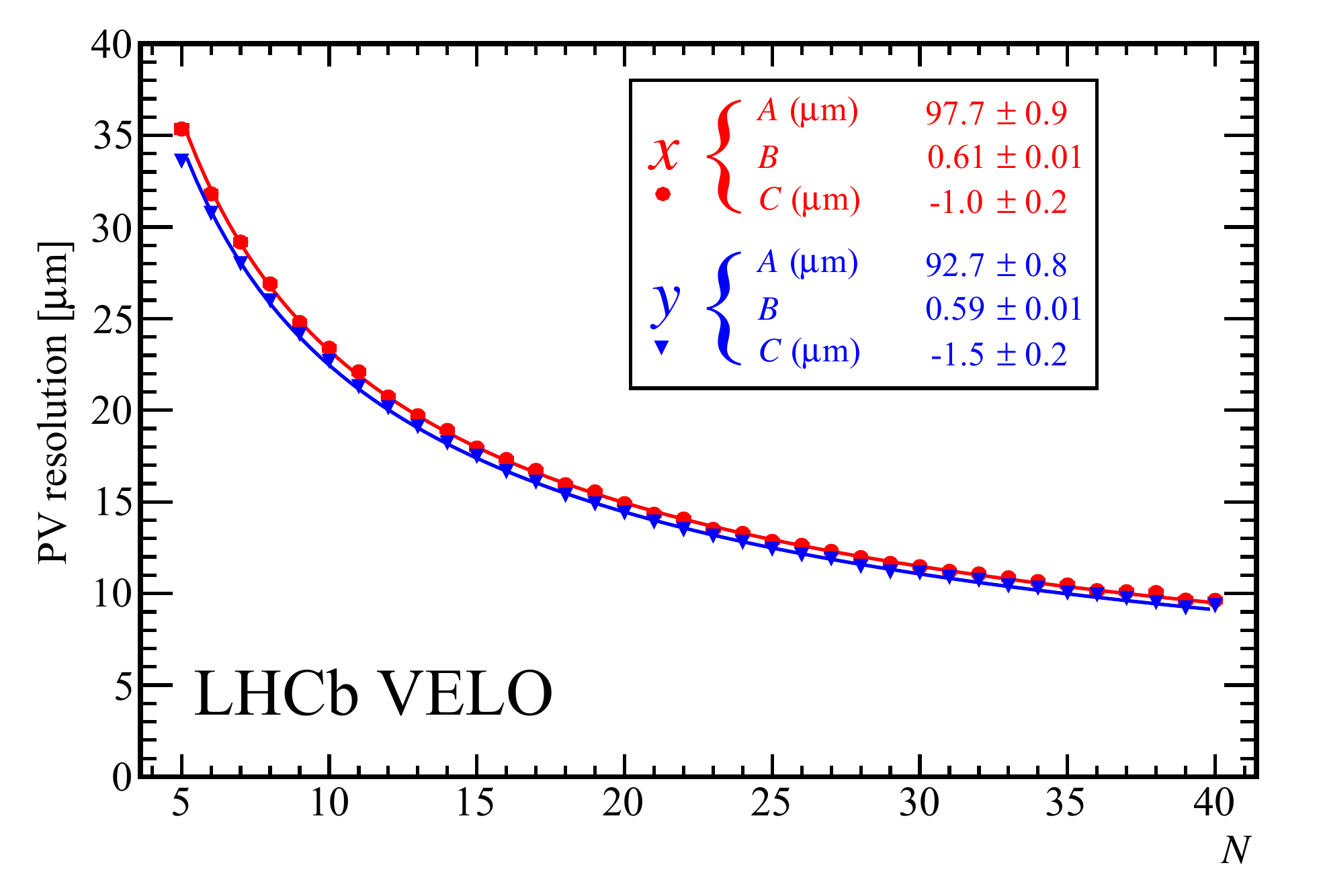}
\hfill
\includegraphics[width=.49\textwidth]{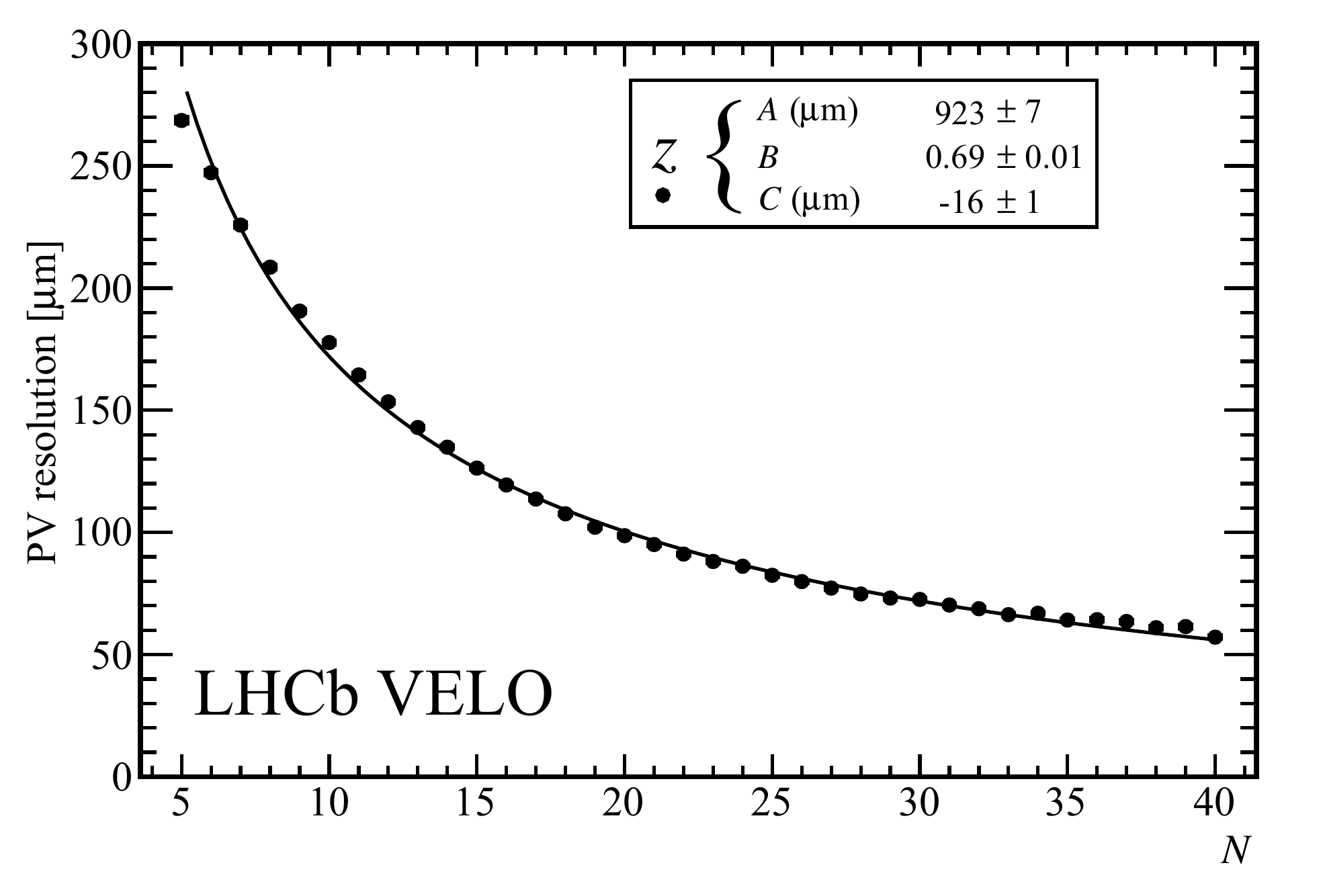}
\caption{\small PV Resolution of events with exactly one PV in 2011
  data as a function of track multiplicity. (left) $x$ (red) and $y$
  (blue) resolution and (right) $z$ resolution. The fit parameters
  $A$, $B$ and $C$ for each coordinate are given.}
\label{fig:data_res}
\end{figure}

In 2011 data it was found that a 25-track vertex has a resolution in
the transverse plane of 13\mum, while the resolution in $z$ is 71\mum,
as shown in Fig.~\ref{fig:data_res}.
The 2011 simulation had a resolution approximately 2\mum better than in the data. 
For data with an average number of visible proton-proton interactions per bunch crossing of around 1.3, the average number of tracks in a minimum bias event containing one PV is 55. The equivalent number in an event in which a candidate $B$ decay has been reconstructed is 120.
As the number of reconstructed PVs in the event increases, the resolution degrades. 
The rate of degradation is approximately 5--10\% per additional
vertex. The vertex resolution results for 2012 data are very similar. 

The stability of the PV position has also been studied. In a single
fill the PV position was found to move in $x$ and $y$ by not more than 4\mum and 2\mum respectively. Based on analysing a three-month
data sample from 2012, the PV position in the LHCb coordinate system
was found to move by a maximum of 50\mum (RMS=16\mum). The VELO 
is centred around the beam with a maximum variation of 20\mum in $x$ and 40\mum in $y$.
For a conservative estimate we apply a safety factor of two to the total
variation, and determine that the beam stability with respect to the
\velo is better than 100\mum.

% results in MC and method check

%The same track split method was applied to MC10 Monte Carlo, see
%Figure~\ref{fig:mc_method_check_res}. In the transverse plane, the
%vertex resolution in Monte Carlo is approximately $2$ microns better
%than in data. The track split method was verified in Monte Carlo using
%truth information. No significant bias was seen. 
%A summary of all
%vertex resolution figures can be found in
%Table~\ref{table:25track_res}.

%\begin{figure}[htb]
%\centering
%\includegraphics[width=.49\textwidth]{figs/DataResX_Offline_1PV_3parfit_MCResX_Offline_1PV_3parfit_TruthResX_Offline_1PV_40_3parfit_comp_v39r4}
%\hfill
%\includegraphics[width=.49\textwidth]{figs/DataResZ_Offline_1PV_3parfit_MCResZ_Offline_1PV_3parfit_TruthResZ_Offline_1PV_40_3parfit_comp_v39r4}
%\caption{Resolution of events with exactly 1PV as a function of track multiplicity. Data is compared with Monte Carlo using the split method (purple) and truth (green).}
%\label{fig:mc_method_check_res}
%\end{figure}

% summary table of results

%\begin{table}[htb]
%\centering
%\begin{tabular}{ | l | c | c | c |}
%\hline			
%25 track resolution & $x$  & $y$  &$z$ \\\hline	
%Data - split & $13.1$ & $12.5$ & $71.1$ \\
%MC - split   & $10.7$ & $10.9$ & $58.1$ \\
%MC - truth  & $10.8$ & $10.8$ & $58.2$ \\
%\hline  
%\end{tabular}
%\caption{25 track resolutions for data and Monte Carlo using the track split method, and using Monte Carlo truth information.}
%\label{table:25track_res}
%\end{table}

%\subsection{dE/dx}
%\label{sec:dEdx}
%[to discuss, Grant, Kazu]

\subsection{Impact parameter resolution}
\label{sec:ImpactParameters}
%[Michael, Stephen - 2 pages]
% - The parametrisation of IP resolutions, showing how it depends on hit resolution, material budget & extrapolation distance, & the origin of the 1/pt dependence.

The impact parameter (IP) of a track is defined as the distance between the track and the PV 
at the track's point of closest approach to the PV. The \B and \D
mesons studied in many \lhcb analyses are long lived particles and
hence their decay vertex is generally displaced from the PV.
%The final state
%tracks from a long lived particle, such as the \B and \D mesons studied in many \lhcb analyses, are
%produced at the decay vertex of the long lived 
%The daughters of a long lived particle,
%such as the \B and \D mesons studied in many \lhcb analyses, are produced at the decay vertex of their mother, which is generally displaced from the primary
%vertex. 
The tracks made by particles coming from the decay of long lived
particles therefore tend to have larger IPs than those made by particles
produced at the PV. Consequently, cuts on the IP are very effective at excluding
prompt backgrounds, and maximising the signal content of a data set. It is thus of great importance
for an experiment like \lhcb to be able to measure IPs to a high precision, and to have a good 
understanding of the effects contributing to the resolution of IP measurements.

The IP resolution is governed by three main factors: multiple scattering of particles by the detector
material; the resolution on the position of hits in the detector from which tracks are reconstructed;
and the distance it is required to extrapolate a track from its first hit in the detector to
the PV. An approximate analytical expression can be derived \cite{Papadelis:Thesis},

%  and IP $\chi^2$
%The angle of deflection of particles, with momentum \ptot, passing through 
%a block of material, of width $x$ and with radiation length \Xrad, follows a Gaussian distribution
%with width
%\cite{Eidelman:multipleScattering}

%\begin{equation}
%  \label{eq:multipleScatteringAngle}
%  \sigma_\theta = \frac{0.0136}{\ptot}\sqrt{x/\Xrad}[1+0.038 \ln(x/\Xrad) ],
%\end{equation}
%for \ptot in \gev.
%When extrapolated over a distance $\Delta_{01}$ between the first hit on a track and the primary
%vertex, this leads to a contribution to the IP resolution of 

%\begin{equation}
%  \label{eq:multpleScatteringIPRes}
%  \sigma_{IP,MS}^2 = \Delta_{01}^2 \sigma_\theta^2.
%\end{equation}
%If the resolution on the first and second hits on the track are $\sigma_1$ and $\sigma_2$, and the distance 
%between the primary vertex and the second hit is $\Delta_{02}$, the contribution to the IP resolution
%from the detector resolution is given by \cite{Papadelis:Thesis}

%\begin{equation}
%  \label{eq:hitResIPRes}
%  \sigma_{IP,hit}^2 = \frac{ \Delta_{02}^2 \sigma_1^2 + \Delta_{01}^2 \sigma_2^2 }{ (\Delta_{02} - \Delta_{01})^2 }.
%\end{equation}
%Thus, the total IP resolution is given by

\begin{eqnarray}
  \label{eq:IPResFull}
%  \sigma_{IP}^2 & = & \sigma_{IP,MS}^2 + \sigma_{IP,hit}^2 \nonumber \\
%  & = & \left( \frac{0.0136}{\ptot}\sqrt{x/\Xrad}[1+0.038 \ln(x/\Xrad) ] \right)^2 \Delta_{01}^2 
%  + \frac{ \Delta_{02}^2 \sigma_1^2 + \Delta_{01}^2 \sigma_2^2 }{ (\Delta_{02} - \Delta_{01})^2 } \nonumber \\
 \sigma_{IP}^2  & = & \left( \frac{0.0136}{\pt}\sqrt{x/\Xrad}[1+0.038 \ln(x/\Xrad) ] \right)^2 r_1^2 
  + \frac{ \Delta_{02}^2 \sigma_1^2 + \Delta_{01}^2 \sigma_2^2 }{ (\Delta_{02} - \Delta_{01})^2 },
\end{eqnarray}
where the first term is due to multiple scattering and the second term due to the detector resolution. 
The particle has a transverse momentum \pt and it passes through  a piece of material, of thickness $x$, and with radiation length \Xrad.
The track is extrapolated over a distance $\Delta_{01}$ between the first hit on a track and the PV, and the distance from  the PV to the second hit is $\Delta_{02}$.
The resolution of the first and second hits on the track are
$\sigma_1$ and $\sigma_2$. This expression applies to a 1D IP
measurement. Parametrising the resolution in the plane
perpendicular to the beam, $IP_x$ and $IP_y$, as a function of \invpt
one expects identical, roughly linear distributions with an $x$ or
$y$-intercept  dependent on the detector resolution, and a gradient proportional to the material budget. 
%using $\Delta_{01}/\ptot = r_1/\pt$, where $r_1$ is the radius of the first hit. 
%This applies to a 1D IP measurement.
In 3D geometry an IP has two degrees of freedom: three as it is a distance in 3D space, minus one from the requirement of being calculated at the point of closest approach to the PV. 
%The two underlying variables have identical Gaussian distributions with $\sigma$ given by equation \ref{eq:IPResFull}, and so the measured IP resolution is decoupled into its 1D $x$ and $y$ components. 
Due to the forward geometry of \lhcb the $z$ component is negligible, and the IP measurement in 3D space is thus simply the sum in quadrature of its $x$ and $y$ components.
%, $\sqrt{IP_x^2 + IP_y^2}$. 
The mean offset of such a measurement from its true value is given by the resolution on the 1D components multiplied by $\sqrt{\pi/2}$. 

% - A brief description of how the plots are made:
%    - Use only high quality Long tracks & events with 1 reconstructed PV.
%    - Refit the PVs to exclude each track in turn.
%    - Extrapolate the track to the same z as the PV & plot the difference in x & y between the track & PV in bins of 1/pt (or phi).
%    - Perform single Gaussian fits to each bin & plot the sigma.

The vast majority of tracks reconstructed at \lhcb are made by
particles produced promptly at the PV.
The measured IP of such tracks is non-zero only due to the measurement resolution. Thus, the IP
resolution can be measured by examining the width of the $IP_x$ and $IP_y$ distributions for all tracks. To do this,
only good quality long tracks 
%(tracks reconstructed in both the \velo and the downstream tracking
%stations, for the best momentum resolution) 
from events with only one reconstructed PV
are used. The PV is required to have at least 25 tracks included in its fit to minimise
the contribution of the vertex resolution to the measured IP. Furthermore, the PV is refitted
excluding each track in turn before its IP is calculated so the track has no influence on the 
PV position. The $IP_x$ and $IP_y$ are then plotted in bins of the variable of interest, 
such as \invpt, and a fit of a Gaussian function performed in each bin. The width of the fitted Gaussian is taken as the resolution.

% - That the first plots are of the expected shape, and show that the IP resolution is very good 
%   & the VELO is performing as well as expected in this respect.

\begin{figure}[tb]
  \centering
 % \subfloat[]{
    \label{fig:IPX_Vs_PT}
    \includegraphics[width=0.49\textwidth]{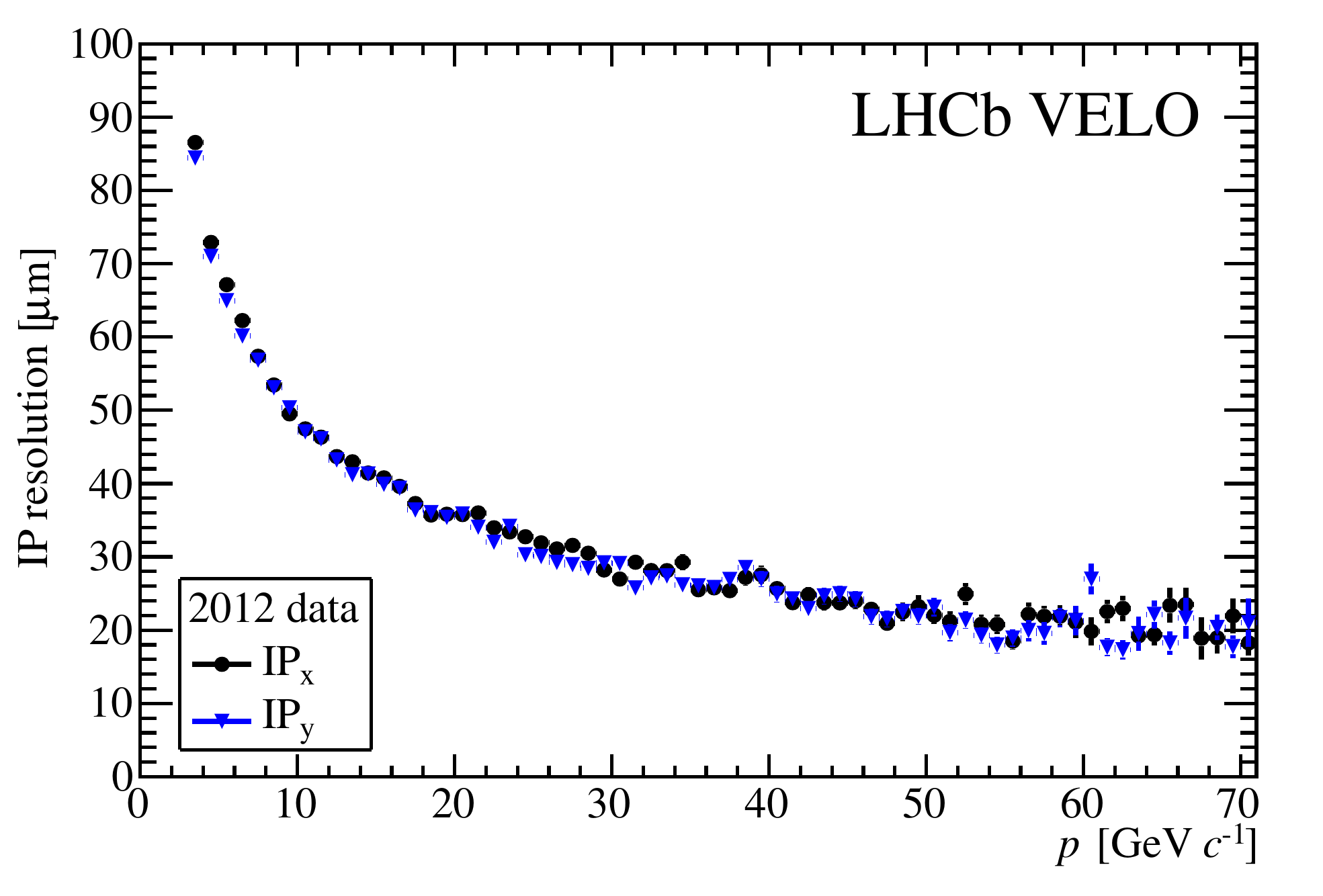}
  %}
%  \subfloat[]{
%    \label{fig:IPY_Vs_PT}
%    \includegraphics[width=0.45\textwidth]{figs/IPY-Resolution-Vs-PT-2011Data}
%  } \\
  %\subfloat[]{
    \label{fig:IPX_Vs_InversePT}
    \includegraphics[width=0.49\textwidth]{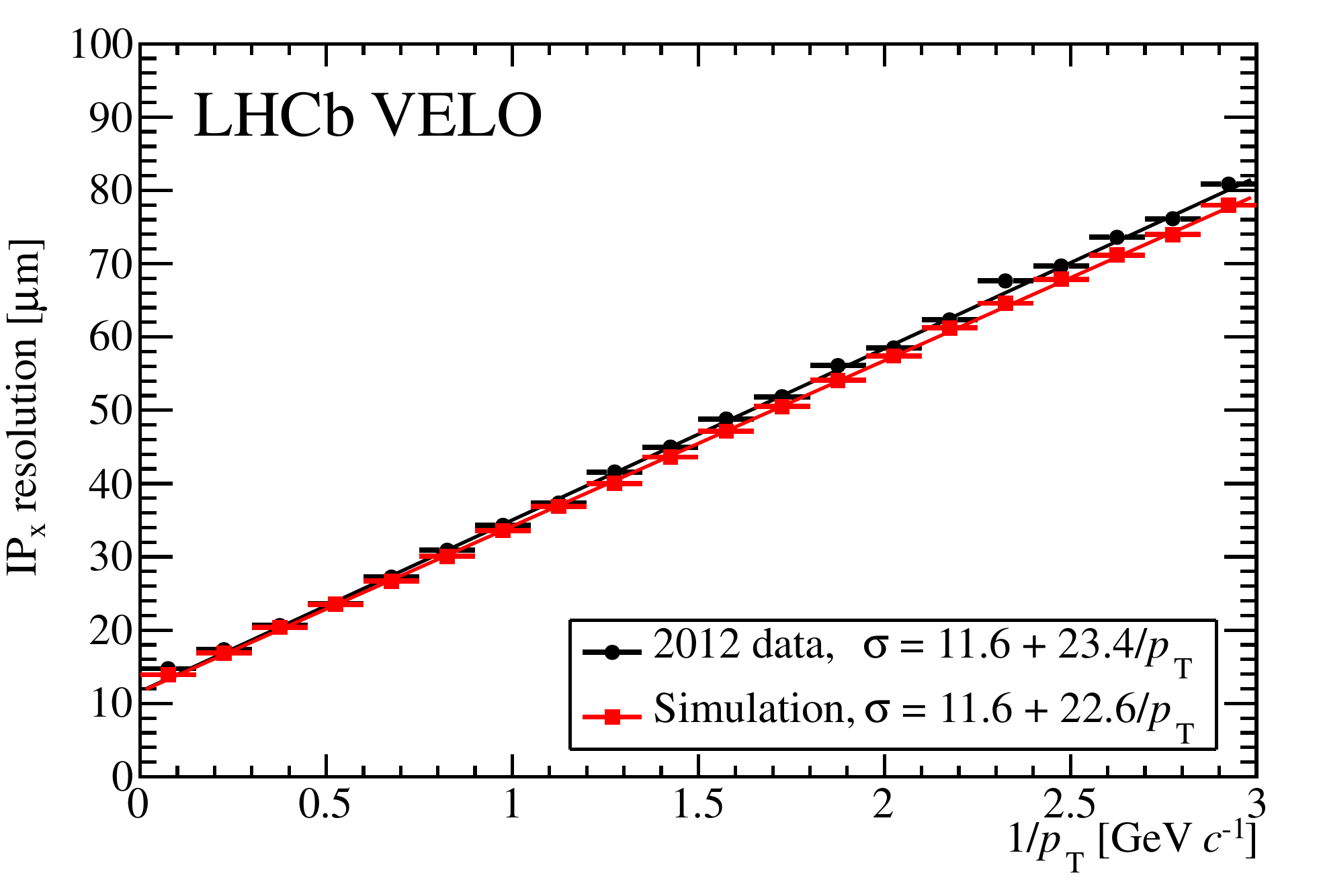}
 % }
%  \subfloat[]{
%    \label{fig:IPY_Vs_InversePT}
%    \includegraphics[width=0.45\textwidth]{figs/IPY-Resolution-Vs-InversePT-2011Data}
%  }
  \caption[IP resolution as a function of momentum and \invpt.]{ \small
    $IP_x$ and $IP_y$ resolution as a function of momentum (left) and $IP_x$ as a function of
    \invpt and compared with simulation (right). Determined with 2012 data.}
  \label{fig:IPRes_Vs_InversePT}
\end{figure}

Figure \ref{fig:IPRes_Vs_InversePT} shows plots of the IP resolution 
versus momentum and \invpt. The resolution of $IP_x$ and $IP_y$ is almost identical. They are asymptotic 
at high \pt, tending to $\sim$12\mum, and depend roughly linearly on \invpt. The performance of the \velo in
this respect is excellent, achieving IP resolutions of $<$35\mum for particles with \pt$>$1\gevc.

% - The relationship between alignment quality & hit resolution, and how this is reflected in the 2nd plot.

\begin{figure}[tb]
  \centering
%  \begin{minipage}{0.47\textwidth}
%  \includegraphics[width=\textwidth]{figs/IPX-Resolution-Vs-InversePT-CompareAlignments-2010Data-edit}
%  \caption[]{ \small $IP_x$ resolution as a function of \invpt, comparing different qualities of alignment, measured on 2010 data.}
%  \label{fig:IPXRes_CompareAlignments}
%  \end{minipage}\qquad
%  \begin{minipage}{0.47\textwidth}
  \includegraphics[width=0.6\textwidth]{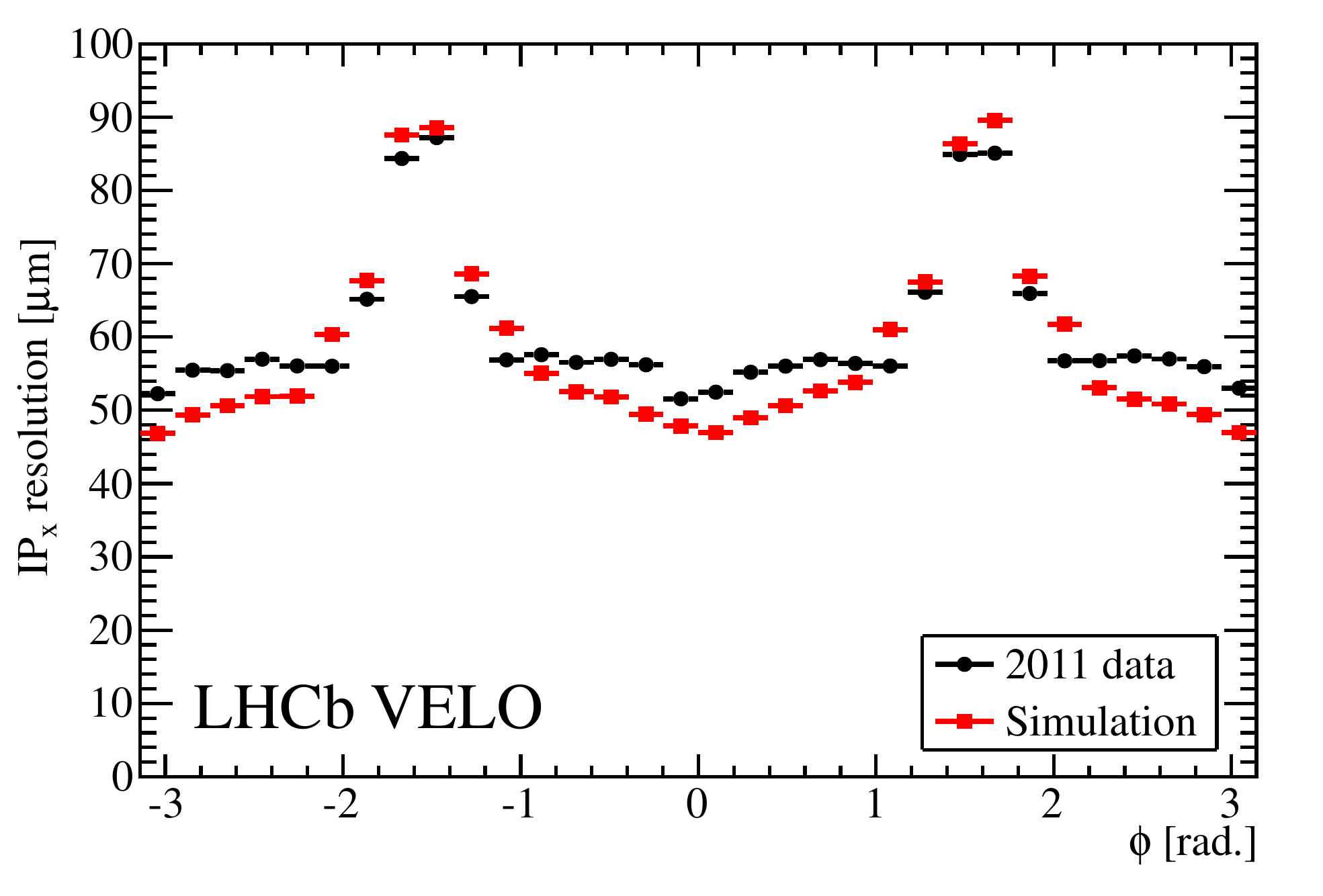}
  \caption[]{\small $IP_x$ resolution as a function of azimuthal angle
    $\phi$, measured on 2011 data and compared to simulation.}
  \label{fig:IPXRes_Vs_Phi}
 % \end{minipage}
\end{figure}

%To examine the dependence of IP resolutions on detector resolution,
%the same study was made using detector
%alignment data from early in the 2010 data taking period, when the alignment was less well known, as is shown in 
%Fig.~\ref{fig:IPXRes_CompareAlignments}. Three alignments of different
%qualities are compared: the first was determined prior to the 2010 run; the second includes the alignment of the \velo halves re-evaluated with 2010 data; and 
%the third also has the module and sensor alignment evaluated using 2010 collision data. Poorer alignment 
%means poorer hit resolution, and so, as expected the $y$-intercept of the IP resolution distribution reduces as the 
%alignment improves, while the gradient remains largely unchanged. 

% - How the material budget varies within in the VELO, and how this variation is reflected in the 3rd plot.

The distribution of material in the \velo is non-uniform, and this
also affects IP resolutions. 
The excellent agreement with simulation shown in
Fig.~\ref{fig:IPRes_Vs_InversePT} was only obtained after
careful study of the material distribution in the RF foil.
In the region in which the two halves of the \velo overlap the two sides of the 
RF foil also overlap, greatly increasing the material density. This can be seen by measuring 
the IP resolution in bins of the azimuthal angle $\phi$, as is shown in Fig.~\ref{fig:IPXRes_Vs_Phi}.
The increase in material is reflected in the increase in IP resolution about $\phi = \pm \pi/2$,
\ie\ in the overlap region. 
%The IP resolution in 2011 simulation was
%approximately in agreement with the data for high momentum tracks
%(where it is dominated by the hit resolution) but was 15\% better than
%the data on the gradient term dominated by the material (see Sect.~\ref{sec:Material}).

% - That, as evidenced by these plots, our understanding of the behaviour of IP resolutions is very good.

Thus, it can be seen that the \velo provides accurate IP measurements
on which the \lhcb physics programme relies for the rejection of
prompt backgrounds to long-lived heavy flavour hadron decays. The IP
resolution behaves as expected, with a roughly linear dependence on
\invpt, and a clear 
dependence on both the hit resolution and the distribution of material.

\subsection{Decay time resolution}
\label{sec:Propertime}

The reconstructed decay time of strange, charm and beauty
hadrons is used in offline event selections and in precise measurements of
lifetimes. However, the most stringent demands on the decay time
resolution stem from the requirement to resolve the fast oscillations
induced by \Bs--\Bsb{} mixing. Consequently, we illustrate the
performance of the \velo with an analysis of the decay time resolution
of \BsToJPsiPhi{} decays.

The reconstructed decay time in the rest frame of the decaying
particle can be expressed in terms of the reconstructed decay length
$l$, momentum $p$ and mass $m$ of the particle in the LHCb frame as
\begin{equation}
t \; = \; \frac{m \: l}{p}.
\end{equation}
The decay time is computed with a vertex fit that constrains the
decaying particle to originate from the PV.  The decay
time uncertainty is a function of the actual decay time. For small
decay times it is dominated by the resolution on the decay length $l$,
which in turn is dominated by the secondary vertex resolution of the \velo.
%which in turn is fully determined by the resolution obtained on the
%track parameters by the \velo. 
For large decay times, it is dominated by the momentum, which
is mostly determined by the tracking stations before and after the LHCb
magnet.  The two contributions are approximately equal at several
times the $B$ hadron lifetime. Therefore, the momentum resolution
plays only a small role.

%The decay time resolution can be written schematically in terms of
%decay time and momentum as
%\[
%\sigma_t \; = \; \sqrt{ 
%  \left( \frac{m}{p} \sigma_l\right)^2 +
%  2 \frac{m t}{p^2} \rho_{lp} \sigma_{l}\sigma_p +
%  \left( \frac{t}{p} \sigma_p\right)^2 }
%\]
%where the middle term on the right hand side represents a correlation
%between the decay length and momentum errors. 
%Consequently, for small decay length the resolution is dominated by
%the decay length resolution while for large decay length is dominated
%by the momentum resolution. As we shall see below, in the limit of
%small decay time the typical decay time resolution is about $50$~fs.
%With a typical relative momentum resolution of $0.004$ the
%contributions of decay length and resolution are equal for a decay
%time of approximately $7$ times the $B$ lifetime. This illustrates
%that the momentum resolution plays only a marginal role.

Time dependent $CP$ violation effects are measured as the amplitude of
an oscillation in the $B$ decay time distribution. The size of the
observed amplitude is damped by a dilution factor from the finite
decay time resolution. 
% If the decay time resolution is characterised
%by a resolution or `response' function $R(t,t')$, the dilution is
%given by~\cite{Moser:1996xf}
%\begin{equation}
%D \; = \; \int_{-\infty}^{\infty} \: R(t,0) \: \cos( \dm t) \: \mathrm{d}t 
%\label{equ:dilution}
%\end{equation}
%where \dm{} is the mixing frequency. 
For a resolution function that is
a single Gaussian function with RMS $\sigma_t$, the dilution is
\begin{equation}
D \; = \; \exp\left[ - \frac{1}{2} \dm^2 \sigma_t^2 \right] ,
\label{equ:dilutiononegaus}
\end{equation}
where \dm is the mixing frequency.

The sensitivity to the oscillation amplitude is proportional to the
dilution. Therefore, for optimal sensitivity the dilution must be as
close to unity as possible. However, even more important is the
understanding of the dilution itself, since a bias in the estimated
dilution leads to a bias in the measurement of a $CP$ violating
effect.

\begin{figure}[tb]
  \includegraphics[width=0.49\textwidth]{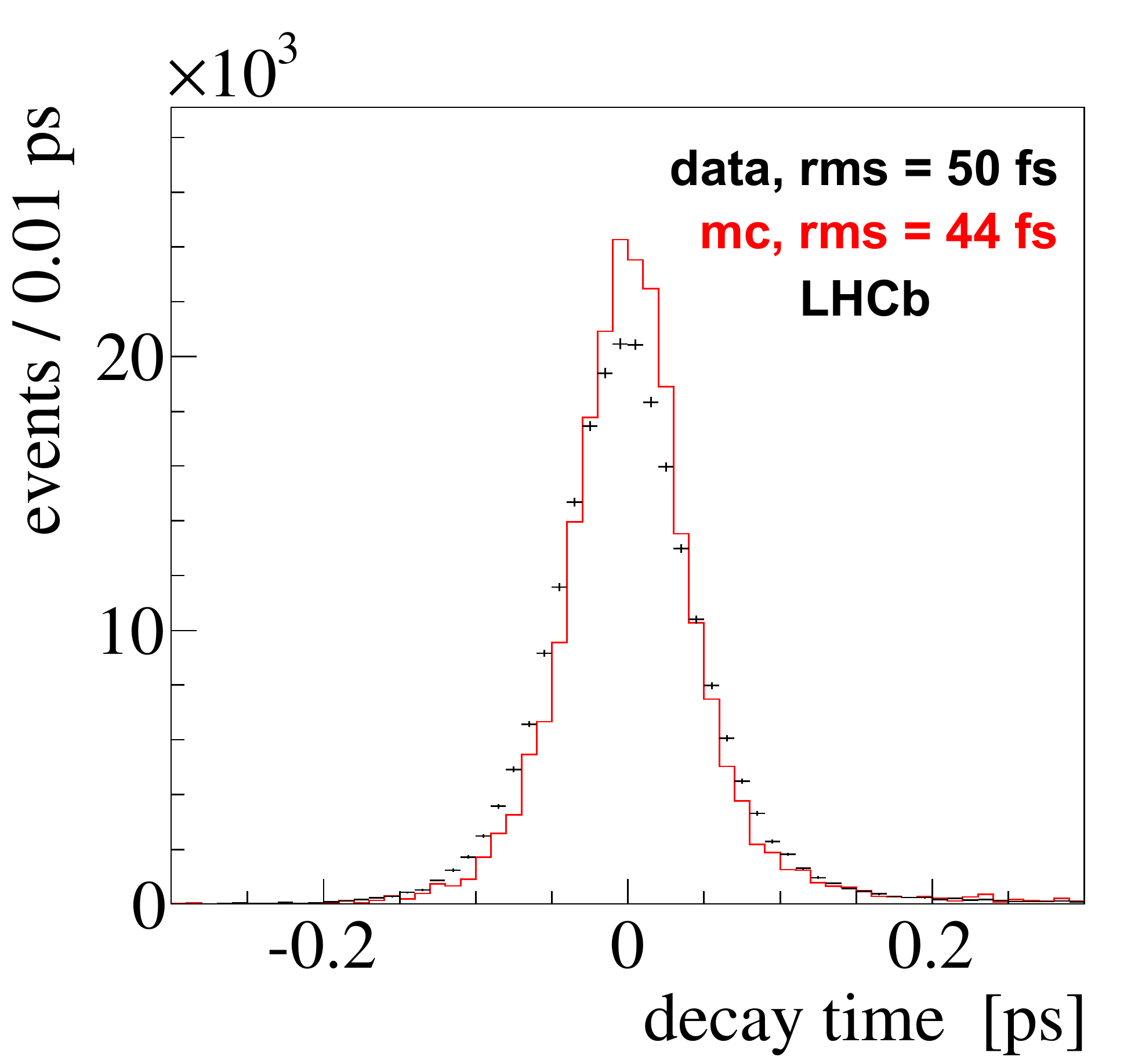}
  \includegraphics[width=0.49\textwidth]{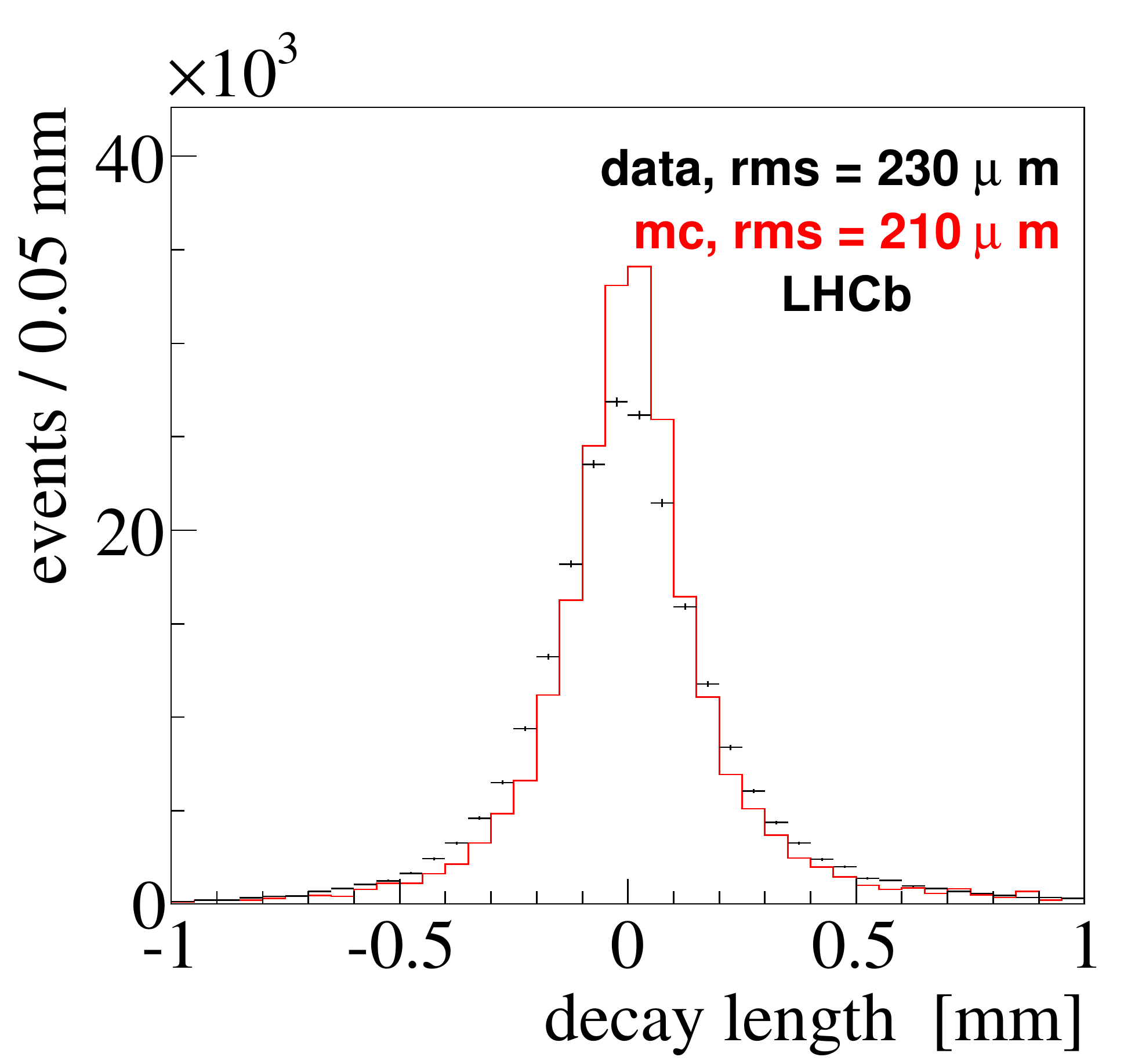}
  \caption{\small Decay time (left) and decay length (right) distribution for
    fake, prompt $\Bs\to\jpsi\phi\to\mumu\Kp\Km$ candidates in 2011 data
    (black points) and simulation (solid red histogram). Only events
    with a single PV are used. The simulated data are
    generated inclusive $\jpsi$ events from which signal
    $\Bs\to\jpsi\phi$ are removed.  In the data contributions from
    non-\jpsi{} di-muon combinations and from true $\Bs\to\jpsi\phi$
    are subtracted using the \sPlot\ technique~\cite{Pivk:2004ty}.}
  \label{fig:promptjpsiphi}
\end{figure}

The decay time resolution depends on the topology of the decay and is
calibrated for each final state on data. The calibration method uses
prompt combinations that fake signal candidates. Ignoring the small
contribution from signal candidates and long-lived background, the
shape of the prompt peak is determined only by the resolution
function. Figure~\ref{fig:promptjpsiphi}  (left) shows the decay time
distribution for fake $\Bs\to\jpsi\phi\to\mumu\Kp\Km$ decays in data
and simulation. The \BsToJPsiPhi{} candidate selection is described
in Ref.~\cite{LHCb:2011aa}.
The contribution from signal decays (which would
result in a tail on the right of the distribution) is removed. The RMS
values quoted in the figure are computed using events with negative decay time
or decay length only, to reduce the sensitivity to a small
contribution from other $B\to\jpsi X$ decays. For a mixing frequency of
17.7\invps{}, the decay time resolution corresponds to a dilution of
about $0.7$.

The resolution in both decay time and decay length is
about 10\% worse in 2011 data than in the corresponding simulated events. Note that the decay time
resolution is more Gaussian shaped than the decay length resolution. This is
largely because, for a fixed opening angle resolution, the decay
length resolution is proportional to the momentum. This momentum
dependence largely cancels in the decay time.

\begin{figure}[htb]
  \includegraphics[width=0.49\textwidth]{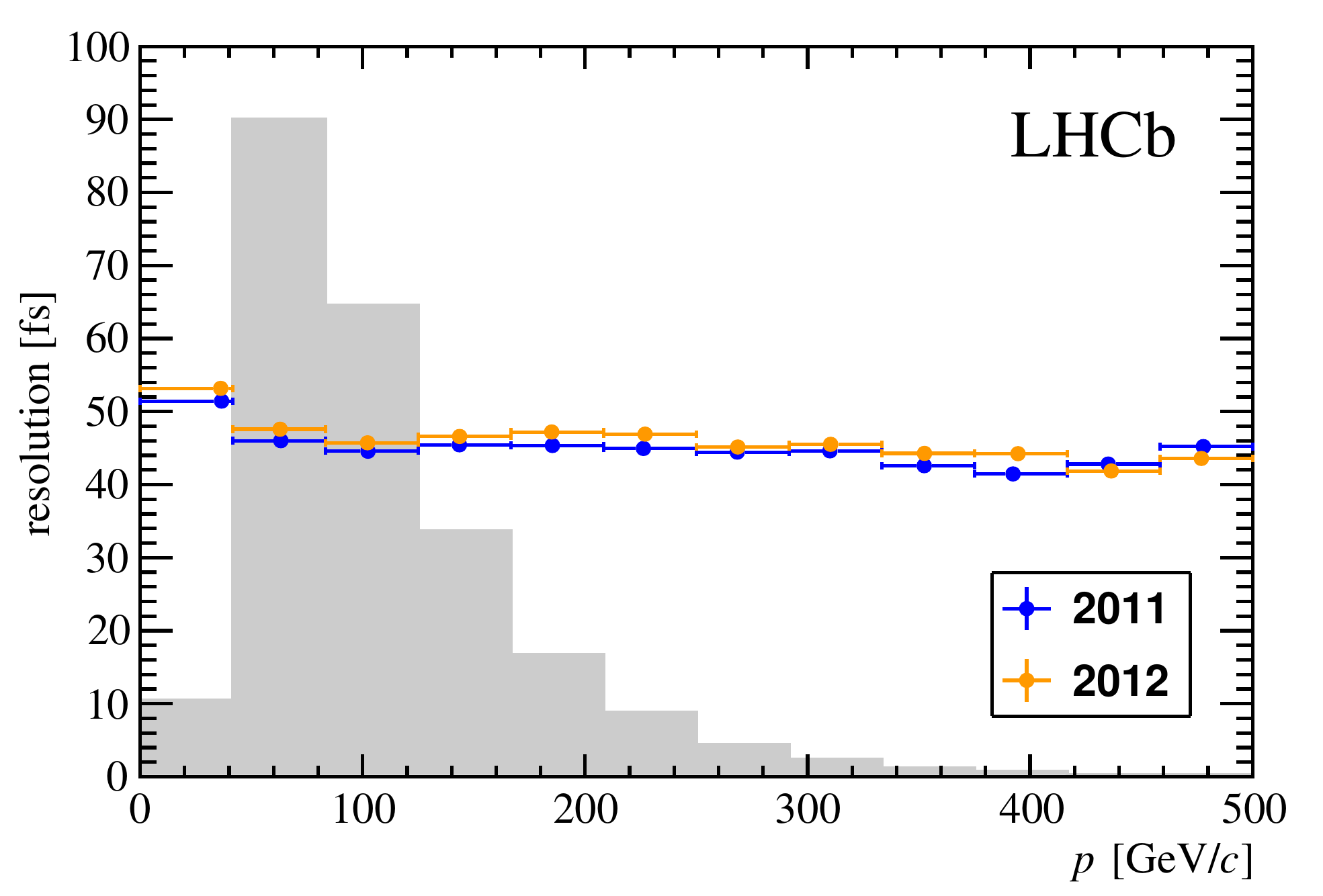}
  \includegraphics[width=0.49\textwidth]{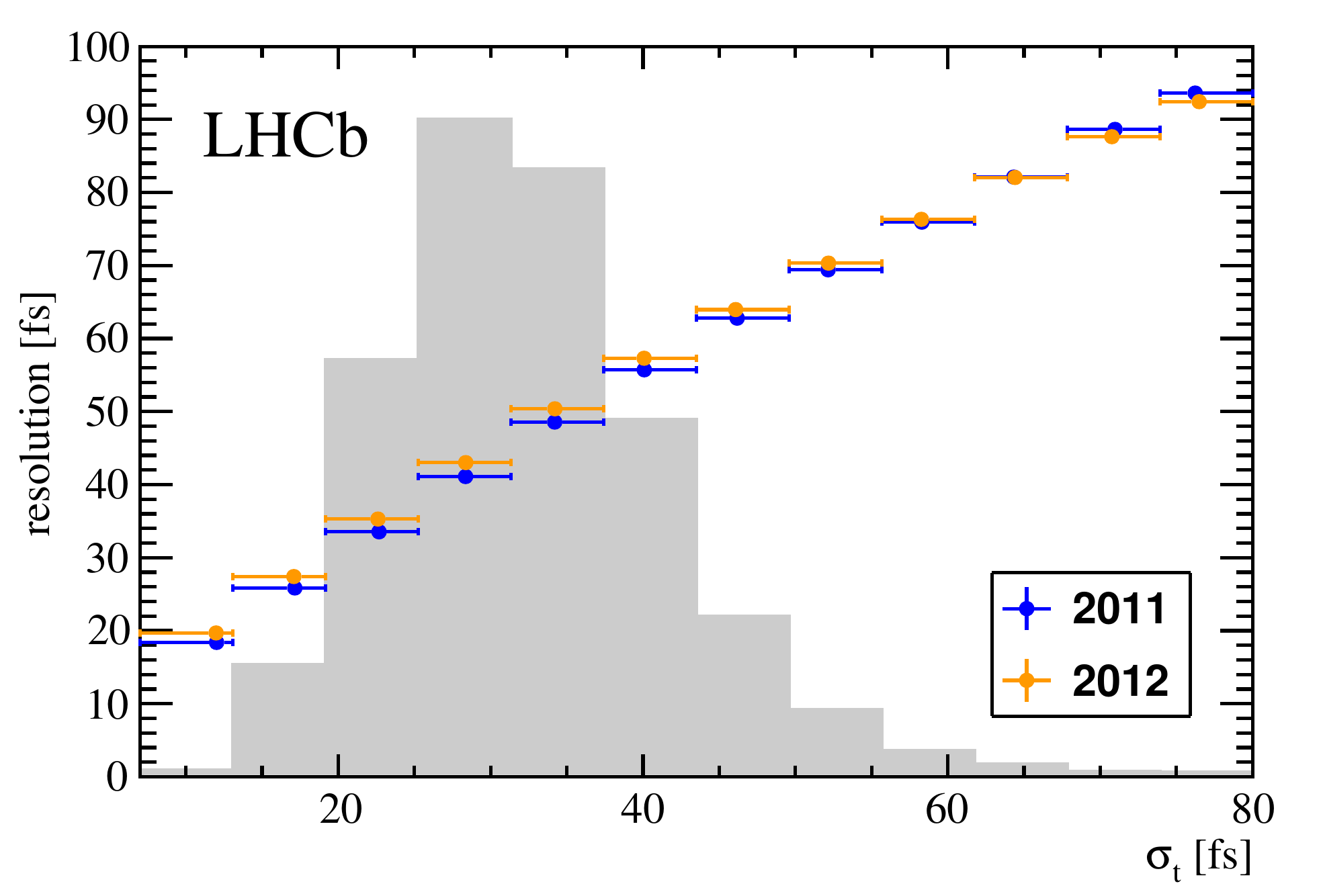}
  \caption{\small Decay time resolution (points) as a function of momentum (left) and
    as a function of the estimated decay time uncertainty (right) of
    fake, prompt $\Bs\to\jpsi\phi\to\mumu\Kp\Km$ candidates in
    data. Only events with a single PV are used.
    Contributions from non-\jpsi{} di-muon combinations and from true
    $\Bs\to\jpsi\phi$ are subtracted using the \sPlot\
    technique. The superimposed histogram shows the
    distribution of momentum (left) and estimated decay time
    uncertainty (right) on an arbitrary scale.}
    %~\cite{Pivk:2004ty}
  \label{fig:decaytimereso}
\end{figure}

This effect is illustrated in Fig.~\ref{fig:decaytimereso} (left) which
shows the resolution as a function of the (fake) $B$ candidate
momentum, where a mixing frequency of  17.7\invps{}  and Eqn.~\ref{equ:dilutiononegaus} have been used to obtain the decay time resolution.
%The resolution was estimated by using
%equation~\ref{equ:dilution} with an oscillation frequency of
%17.7~\invps{} to compute the dilution and
%subsequently~\ref{equ:dilutiononegaus} to turn the dilution in an
%effective decay time resolution. 
Only events with $t<0$ were used in
order to reduce the sensitivity from remaining long-lived candidates.
This procedure gives a resolution that is most relevant for the
time-dependent $CP$ violation measurement in \Bs{} decays.

Finally, Fig.~\ref{fig:decaytimereso} (right) shows the resolution as a
function of the per-event estimated uncertainty in the decay time. The
latter is computed with the same vertex fit that computes the decay
time itself. It is a non-trivial function of the track parameters and
covariance matrices of the four final state particles in the decay. As
expected, the resolution is a linear function of the estimated
uncertainty. However, since track parameter uncertainties are not
perfectly calibrated yet in the data, the slope is $\sim1.2$ rather
than unity. The typical decay time resolution in \lhcb is 50\fs, 
and this resolution plays a crucial role in the sensitivity of many \lhcb physics measurements.

% $Id: introduction.tex 4475 2011-04-05 10:27:28Z uegede $
% ===============================================================================
% Purpose: introduction to the standard template
% Author: Tomasz Skwarnicki
% Created on: 2010-09-24
% ===============================================================================

\section{Conclusions}
\label{sec:Conclusions}
%[Chris to provide. 0.5 page]

The performance of the \lhcb \velo during its first years of operation has been described. The operational experience of key  subsystems has been reviewed. The sensors have been successfully operated in a secondary vacuum at a pressure of $2\times10^{-7}$\mbar, and cooled using a bi-phase $\rm CO_2$ cooling system which has maintained the operating temperature of the sensors at $(-7\pm2)$\degreesC. The sensors are moved in 210~seconds from their fully retracted position to be centred 7\mm from the LHC beam for physics operation to an accuracy of better than 4\mum. The average material budget of the detector for tracks in the \lhcb acceptance is $0.22\,{\rm X_0}$. 

The determination of the sampling time of the front-end pulse-shape and the digitisation time to maximise the signal, minimise spillover to the next and preceding event, and synchronise with the LHC beam collisions is described.  The timing is found to be stable over a period of one year of operation at the nanosecond level. The normalisation of the gain of the system has also been described and this calibration is performed approximately every six months. The digitised data are processed by a series of FPGA algorithms that have been described and perform pedestal subtraction, common mode suppression and clusterisation. Errors in the system are also identified in the FPGA algorithms. A novel technique is applied of emulating the algorithms in bit-perfect C code which is used in the main analysis framework of the experiment to tune the operational parameters. A parameter retuning is required approximately every two months to ensure optimal performance. Single event upsets are observed in the front-end electronics at a rate of around 2.9 register bit-flips for all front-end ASICs combined per \invpb of delivered integrated luminosity. Extensive data quality monitoring has been put in place, including automatic processing of each run, the use of a graphical user interface to display plots, and templated reports in an electronic logbook.

The sensors initially had a signal to noise ratio of approximately 20:1, with the noise depending on the strip capacitance. The hit resolution varies with pitch and track angle; for the optimal track angle of 7--11\degrees it varies from  4\mum in the 40\mum pitch region to 20\mum at 100\mum pitch. This resolution has been achieved using analogue readout for the front-end ASICs and pulse-height weighted cluster position determination. The typical cluster occupancy in the experiment during 2011 operation was around 0.48\% for randomly triggered events in beam-crossings, but rises to 0.93\% for events that have passed the high level trigger.  Faulty strips in the detector have been determined using noise distributions, occupancy spectra and the  cluster finding efficiency. The detector has less than 1\% of faulty strips. Tracks arising from interactions between the LHC beams and gas molecules  are observed and provide a useful sample for alignment and for beam imaging for luminosity studies. Beam backgrounds giving rise to splashes of hits in localised regions of the detector are also seen. The radiation damage in the detector has been studied through the analysis of the currents drawn and charge collection versus voltage. Dedicated beam time is used for the latter study for which an automated procedure has been developed. The inner radius regions of the \nonn sensors are observed to have undergone space-charge sign inversion, which is expected due to their proximity to the LHC beams. Charge loss is also observed to have developed after irradiation due to the presence of the second metal layer that is used for routing out the signals from strips on the sensor.

The track finding efficiency of the \velo has been determined using a tag and probe method using \jpsi decays, and is typically above 98\%.  The modules have been aligned using a track-based software alignment procedure. The position to which the \velo is inserted in each fill is measured by mechanical motion sensors and this is used to update the alignment.  An alignment precision of 1\mum for translations in the plane transverse to the beam, and a 5\mum stability of the relative alignment of the two \velo halves over an operational year is obtained. 

The \lhcb physics selection and analysis of long-lived heavy flavour decays relies on the background rejection and flavour tagging from the impact parameters and vertexing performance of the detector.   The vertex resolution is strongly dependent on the number of tracks in the vertex. A resolution of 13\mum in the transverse plane and 71\mum along the beam axis is achieved for vertices with 25 tracks. A 1D impact parameter resolution of 12\mum in the plane transverse to the beam for high momentum tracks is obtained. It is primarily determined by the detector cluster position resolution and the distance of the sensors for the LHC beams. For lower momentum tracks the impact of multiple scattering in the detector material becomes dominant, and an impact parameter resolution of  35\mum is achieved for particles with transverse momentum of 1\gevc.  A decay time resolution of approximately 50~fs is obtained (evaluated for the  \BsToJPsiPhi{} decay channel) which plays a key role in many \lhcb physics results.
%, including the world's best measurements of \dms, \dgs and \phis in  \Bs\Bsb{} oscillations.

\section {Acknowledgements}
This complex detector could only be constructed with the dedicated effort of many technical collaborators in the
institutes forming the \lhcb \velo Collaboration. A special acknowledgement goes to all our \lhcb collaborators who over the years have
contributed to obtain the results presented in this paper. 
We express our gratitude to our colleagues in the CERN accelerator departments
for the excellent performance of the LHC. We thank the technical and administrative
staff at the LHCb institutes. We acknowledge support from CERN and from the national
agencies: CAPES, CNPq, FAPERJ and FINEP (Brazil); NSFC (China); CNRS/IN2P3
and Region Auvergne (France); BMBF, DFG, HGF and MPG (Germany); SFI (Ireland);
INFN (Italy); FOM and NWO (The Netherlands); SCSR (Poland); MEN/IFA (Romania);
MinES, Rosatom, RFBR and NRC ``Kurchatov Institute'' (Russia); MinECo, XuntaGal
and GENCAT (Spain); SNSF and SER (Switzerland); NAS Ukraine (Ukraine); STFC
(United Kingdom); NSF (USA). We also acknowledge the support received from the ERC
under FP7. The Tier1 computing centres are supported by IN2P3 (France), KIT and
BMBF (Germany), INFN (Italy), NWO and SURF (The Netherlands), PIC (Spain),
GridPP (United Kingdom). We are indebted to the communities behind the multiple open
source software packages we depend on. We are also thankful for the computing resources
and the access to software R\&D tools provided by Yandex LLC (Russia).

% Do not include this in analysis note and conference reports
%\input{acknowledgements}

%\input{appendix}

% This should be taken out in the final paper
%\input{supplementary-app}

\addcontentsline{toc}{section}{References}
\bibliographystyle{LHCb}
\bibliography{main}

\end{document}